\documentclass[12pt]{article}
\usepackage{a4wide,epsfig,psfrag,amsmath,amssymb,cite,scalefnt}
\usepackage[dvipsnames]{xcolor}
\usepackage{amsmath,comment,braket}
\usepackage{placeins}
\usepackage{breqn}
\usepackage{slashed}
\usepackage{comment}
\usepackage{subcaption}

\graphicspath{ {figs_3l_local/} }

\allowdisplaybreaks

\begin{document}

\title{\vskip-3cm{\baselineskip14pt
    \begin{flushleft}
      \normalsize P3H-25-060, TTP25-029, ZU-TH 56/25
    \end{flushleft}} \vskip1.5cm
  Two-loop QCD corrections to $ZH$ and off-shell $Z$ boson pair
  production in gluon fusion
}

\author{
  Joshua Davies$^{a}$,
  Dominik Grau$^{b}$,
  Kay Sch\"onwald$^{c}$,
  \\
  Matthias Steinhauser$^{b}$,
  Daniel Stremmer$^{b}$,
  Marco Vitti$^{b,d}$
  \\[1mm]
  {\small\it (a) Department of Mathematical Sciences, University of
    Liverpool,}
  {\small\it Liverpool, L69 3BX, UK}
  \\
  {\small\it (b) Institut f{\"u}r Theoretische Teilchenphysik}\\
  {\small\it Karlsruhe Institute of Technology (KIT)}\\
  {\small\it Wolfgang-Gaede Stra\ss{}e 1, 76131 Karlsruhe, Germany}
  \\
  {\small\it (c) Physik-Institut, Universit\"at Z\"urich, Winterthurerstrasse 190,}\\
  {\small\it 8057 Z\"urich, Switzerland}
  \\
  {\small\it (d) Institut f{\"u}r Astroteilchenphysik,
    Karlsruhe Institute of Technology (KIT),}\\
  {\small\it Hermann-von-Helmholtz-Platz 1, 76344 Eggenstein-Leopoldshafen, Germany, Germany}
}

\date{}

\maketitle

\thispagestyle{empty}

\begin{abstract}

  We compute two-loop corrections to the associated production of a
  Higgs boson with a $Z$ boson and to off-shell $Z$ boson pair
  production in the gluon fusion channel, mediated by a heavy quark.
  We perform deep expansions in the
  high-energy region and around the forward limit and show that their
  combination covers the whole phase space. 
  Our results constitute the next-to-leading order 
  virtual corrections to these processes.
  Their numerical evaluation is fast and the dependence on
  all parameters is maintained, thus a change of parameter values or renormalization
  scheme is straightforward.
  As a by-product of our calculation, we also obtain the two-loop heavy quark mediated
  virtual corrections to the processes of off-shell di-photon and photon-$Z$ production.

\end{abstract}


\thispagestyle{empty}

\newpage

\section{Introduction}

\begin{figure}[t]
    \centering
  \begin{tabular}{c}
    \includegraphics[width=.9\textwidth]{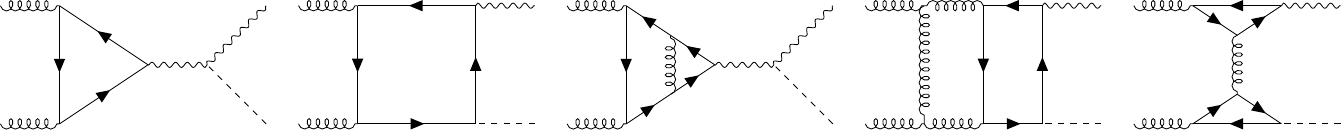} \\
    \\
    \includegraphics[width=.9\textwidth]{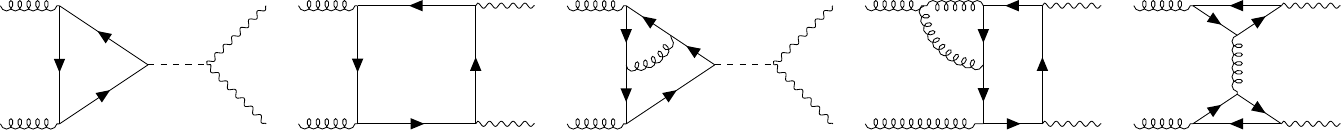}
  \end{tabular}
  \caption{
	\label{fig::FD1}
        Sample Feynman diagrams contributing to 
        $gg\to ZH$ (top) and $gg\to ZZ$ (bottom).
        Straight, wavy, curly and dashed lines refer to
        quarks, $Z$ bosons, gluons and Higgs bosons, respectively.
        }
\end{figure}

The high-luminosity phase of the LHC will require accurate predictions
for the gluon-initiated production of pairs of massive bosons, which
we denote generically as $gg \to XY$. The next-to-leading-order (NLO) QCD corrections
mediated by heavy-quark loops are hard to compute because of the
many scales involved in this class of processes. This is particularly
true for the box-type diagrams
(see Fig.~\ref{fig::FD1}) which appear in the virtual corrections.

In this context the use of analytic expansions is known to be
beneficial, allowing the results to be expressed in terms of simpler
integrals which are relatively fast to evaluate numerically, while
retaining the dependence on all the parameters involved. In this
paper we are interested in one of the most complicated cases, in
which the momenta of the two final-state particles are associated to
different scales. The calculations involve five scales: the top quark
mass $m_t$, the Mandelstam variables $s$ and $t$, and the two (different)
virtualities of the final-state bosons $q_X^2$ and $q_Y^2$.

This class of processes includes $gg\to ZH$ and
off-shell\footnote{Note that in the literature \emph{off-shell}
often refers to the case in which the Higgs boson in the triangle-type
diagrams of Fig.~\ref{fig::FD1} is off-shell, as opposed to resonant
single-Higgs production. Here, we are rather interested in the case in
which the final-state $Z$ bosons are off-shell.} $gg \to ZZ$, denoted as $gg
\to Z^\star Z^\star$. The former is a non-negligible contribution to
associated $ZH$ production at the LHC, and represents one of the main
sources of theoretical uncertainty in current
measurements~\cite{LHCHiggsCrossSectionWorkingGroup:2016ypw, Brein:2012ne, ATLAS:2024yzu, CMS:2023vzh}. The top-mediated corrections to $gg \to ZZ$
are most important because of the interference between the
Higgs-mediated and the non-resonant amplitudes, which has been used to
put indirect constraints on the total decay width of the Higgs
boson~\cite{Kauer:2012hd,Caola:2013yja,Campbell:2013una}.

The LO contribution to $gg \to ZH$ and $gg \to ZZ$ is loop-induced
and has been computed in
Refs.~\cite{Kniehl:1990iva,Dicus:1988yh,Dicus:1987dj,Glover:1988rg}. At
NLO in QCD, expansion results are available for the large-mass expansion
(LME)
\cite{Hasselhuhn:2016rqt,Davies:2020drs,Melnikov:2015laa,Campbell:2016ivq},
the high-energy expansion~\cite{Davies:2020drs,Davies:2020lpf}, the
small-mass expansion~\cite{Wang:2021rxu} and the $p_T$
expansion~\cite{Alasfar:2021ppe,Degrassi:2024fye}. Numerical results
have been presented in
Refs.~\cite{Chen:2020gae,Agarwal:2020dye,Chen:2022rua,Agarwal:2024pod}.
As a first result of this paper, we provide results for the
$t$-expansion~\cite{Davies:2023vmj} of $gg \to ZH$ and top-quark mediated $gg \to
Z^\star Z^\star$ including higher-order terms
than Refs.~\cite{Alasfar:2021ppe,Degrassi:2024fye}, and we provide new
terms in the high-energy expansion. In the case of $gg \to ZZ$ we
allow the $Z$ bosons to be off-shell. 
This is an advantage of the analytic approach; the same methods can be applied in
a straightforward way, although the intermediate and final expressions become larger.
On the other hand, introducing an additional mass scale into
the calculations of
Refs.~\cite{Agarwal:2020dye,Agarwal:2024pod}
would lead to a significant increase in complexity.
The helicity amplitudes for the light-quark mediated contribution to on-shell $gg\to ZZ$
can be found in Refs.~\cite{vonManteuffel:2015msa,Caola:2015ila}.

The possibility to rely only on analytic approximations for the
virtual corrections of similar $gg$-initiated processes has been discussed in
Refs.~\cite{Bellafronte:2022jmo,Davies:2023vmj}, showing that the
expansions in the forward limit and in the high-energy limit
can be combined in such a way that the complete phase space is covered
with an accuracy at the percent level. As a second result of this
paper, we combine the new results for the two approximations following
Ref.~\cite{Davies:2023vmj}.  For $gg \to ZH$, the combination of the
high-energy expansion \cite{Davies:2020drs} and the $p_T$ expansion
\cite{Alasfar:2021ppe} has been presented in
Ref.~\cite{Degrassi:2022mro}. However, this analysis has the
possible shortcoming that only three expansion terms for $p_T\to0$ have been
considered,  and furthermore only 13 expansion terms in the high-energy limit have been included. 
In this work, we improve the description of $gg\to ZH$ in both regions of phase space:
our expansion in the forward limit includes terms up to $t^5$ and $m_{Z,H}^8$,
and we use a deep high-energy expansion which includes terms up to $m_t^{112}$
and $m_{Z,H}^4$.

Our results for $gg\to Z^\star Z^\star$ can be used to obtain
the helicity amplitudes for $gg\to \gamma^\star\gamma^\star$ and 
$gg\to Z^\star \gamma^\star$ by dropping the triangle contributions,
setting the axial-vector coupling of the $Z$ boson to zero, and replacing
either both or one of the $Z\,t$ couplings with the $\gamma\,t$ coupling.
Numerical results for on-shell
di-photon production can be found in Refs.~\cite{Maltoni:2018zvp,Chen:2019fla} 
and recent exact calculations
of the helicity amplitudes are available from~\cite{Becchetti:2025rrz,Ahmed:2025osb}.

The remainder of the paper is structured as follows: in the next section we 
provide technical details common to both processes
and describe the procedure to obtain ultraviolet and infrared finite results.
We then discuss off-shell $Z$ boson pair production in Section~\ref{sec::ggZZ}
and $gg\to ZH$ in Section~\ref{sec::ggZH}.
At one-loop order we compare our approximations
to the exact results. At two loops we show that the expansions in the forward limit and 
the high-energy limit agree for intermediate values of the transverse momentum $p_T$.
For both processes we compare the total NLO virtual finite
corrections to precise numerical results.
We conclude in Section~\ref{sec::concl}. In the Appendix
we present results for the projectors to
one of the helicity amplitudes for $gg\to Z^\star Z^\star$.

\section{Technical details}

\subsection{\label{sub::kin}Kinematics}

We consider the scattering of two gluons in the initial state with
momenta $q_1$ and $q_2$ into two massive particles in the final state,
$gg\to XY$, with momenta $q_3$ and $q_4$. The Mandelstam variables are
then given by
\begin{eqnarray}
  s = (q_1+q_2)^2\,,\qquad t = (q_1+q_3)^2\,,\qquad u = (q_1+q_4)^2\,,
  \label{eq::stu}
\end{eqnarray}
where all momenta are incoming. Furthermore we have
\begin{eqnarray}
  q_1^2=q_2^2=0\,.
\end{eqnarray}
For the production of on-shell particles we have
\begin{eqnarray}
  q_3^2=q_X^2=m_X^2\,,\qquad q_4^2=q_Y^2=m_Y^2\,,
  \label{eq::q3sq4s}
\end{eqnarray}
where in general $m_X$ and $m_Y$ are allowed to be different
and the transverse momentum of the final-state particles is given by
\begin{eqnarray}
  p_T^2 &=&\frac{u\,t-q_3^2 q_4^2}{s}\,.
            \label{eq::pT}
\end{eqnarray}
The (internal) top quark mass is denoted by $m_t$.

\subsection{Amplitude construction}

For the computation of the amplitudes in terms of scalar integrals we use a well-tested and automated setup.
The diagrams required for the computation of the amplitudes 
for $gg \to ZH$ and $gg \to Z^\star Z^\star$ are generated with 
\texttt{qgraf}~\cite{Nogueira:1991ex}.
Afterwards, we use \texttt{tapir}~\cite{Gerlach:2022qnc} and \texttt{exp}~\cite{Harlander:1998cmq,Seidensticker:1999bb} 
to map the diagrams onto integral families which are exact in all kinematic parameters.
The output is converted to \texttt{FORM}~\cite{Ruijl:2017dtg} notation, which is used 
by the in-house ``{\tt calc}'' setup to perform the 
Dirac and color algebra and finally provide the amplitudes
expressed in terms of scalar Feynman integrals.
The output is then further processed to obtain expansions in the 
high-energy and forward limits. 
We discuss the details of these expansions in the following sections.

\subsection{Integral expansions}

In this paper we perform two kinds of expansion: around the forward limit and
for high energies. It has been shown for various $2\to2$ processes that
the combination of these expansions leads to precise results over the whole phase space
region~\cite{Davies:2019dfy,Bellafronte:2022jmo,Degrassi:2022mro,Chen:2022rua,Davies:2023vmj,Degrassi:2024fye}.
We follow the approach of Ref.~\cite{Davies:2023vmj} and compute a Taylor
expansion in the Mandelstam variable $t$. For the high-energy expansion we use
the approach developed in Refs.~\cite{Davies:2018ood,Davies:2018qvx}.

We have two approaches to compute the $t$ expansion (see below for more details):
\begin{itemize}
\item[(i)] expansion of the amplitude before reduction to master integrals, 
\item[(ii)]
expansion of the master integrals after integration-by-parts (IBP) reduction.  
\end{itemize}
We apply both
approaches and use the agreement of the final results as a consistency check.
In approach (i) we perform a simultaneous expansion in $t$ and the final-state
masses whereas in approach (ii) we first perform a Taylor expansion in the
final-state masses and then apply an IBP reduction to obtain
the amplitude in terms of $t$-dependent master integrals, which are subsequently expanded.
We use \texttt{Kira}~\cite{Klappert:2020nbg} and \texttt{FIRE}~\cite{Smirnov:2019qkx,Smirnov:2023yhb} for the IBP reductions.

The high-energy expansion follows
approach (ii) only, since approach (i) would lead to a
complicated asymptotic expansion. 
Note that the amplitude in terms of master integrals, before expansion, is
identical to that of the $t$ expansion in approach (ii).

\subsubsection{\label{subsub::texp_amp}Forward expansion at the amplitude level}

Our amplitudes have an analytic dependence on $t$, allowing us to perform a
straightforward Taylor expansion of the integrands in the variable
$\delta q = q_1+q_3$.
The expansion of the denominators produces numerator scalar products of the form
$\delta q\cdot q_i$, where $q_i$ is
one of the external momenta. Here we use the kinematics given in
Eqs.~(\ref{eq::stu}) to~(\ref{eq::q3sq4s}) in order to express them in terms
of invariants.\footnote{The use of the ``full'' kinematics with massive final-state
particles is different from the approach suggested in
Ref.~\cite{Davies:2023vmj}. There an
expansion in $q_3^2$ and $q_4^2$ is performed first, followed by a $\delta q$
expansion in massless kinematics (where $q_3^2=q_4^2=0$).

Expanding the scalar integrals first
in the final-state masses generates spurious negative
powers of $t$, which only cancel after
the expansion in $\delta q$. Thus, as compared to the approach used
in this paper, a deeper expansion in $\delta q$ is necessary to
arrive at the desired order in $t$ in the final expression.}
This produces a simultaneous expansion\footnote{A similar approach has been introduced in Ref.~\cite{Bonciani:2018omm}.} in $t$, $q_3^2$ and
$q_4^2$.

The expansion in $\delta q$ also leads to scalar products $\delta q\cdot p_i$,
where $p_i$ is a loop momentum.  Here we perform a tensor reduction which
eliminates $q_3$ from the numerator. We use the program {\tt
  OPITeR}~\cite{Goode:2024cfy} in order to generate {\tt FORM}~\cite{Ruijl:2017dtg}
code which we can include in our setup.
Since the expansion in $\delta q$ eliminates $q_3$ from the denominators of the
integrands, the resulting integrals depend only on $s/m_t^2$ and not on $t$,
$q_3^2$ nor $q_4^2$.

We manage to compute expansion terms up to $t^{n_t} (q_3^2)^{n_3} (q_4^2)^{n_4}$ with $n_t+n_3+n_4\leq4$ for both processes.
 This leads to tensor integrals up to rank $14$, which constitutes one of the main bottlenecks of this approach.
 In this expansion of the helicity amplitude $\mathcal{A}_{++00}$ of $gg\to Z^\star Z^\star$
 (see Section~\ref{sec::ggZZ}) we encounter tensor integrals only up to rank $10$, which allows us to extend the expansion to also include the terms $n_t+n_3+n_4=5$ in this particular case.

\subsubsection{\label{subsub::texp_MI}Forward expansion of the master integrals}

In the second approach to obtain the $t$ expansion we do not expand the
integrand but first perform a reduction to master integrals and then expand
them. To simplify the reduction problem we expand each scalar integral
in the final-state masses $q_3^2$ and $q_4^2$ using the program {\tt
  LiteRed}~\cite{Lee:2013mka}. This leads to a reduction problem which only
depends on $s$, $t$ and $m_t^2$. It has already been studied in
Refs.~\cite{Davies:2018ood,Davies:2018qvx}, where 161 two-loop master
integrals were identified.  The $t$ expansion of these master integrals
was constructed in Ref.~\cite{Davies:2023vmj} with the help of the
semi-analytic ``expand and match'' approach~\cite{Fael:2021kyg,Fael:2021xdp,Fael:2022miw}.

The bottleneck in this approach are the huge intermediate expressions which
are generated after the expansion in $q_3^2$ and $q_4^2$ and the insertion of
the IBP reduction tables. Once the $t$ expansions of the master integrals are
inserted, much more compact expressions are obtained.  We manage to compute quartic
terms in $q_3$ and $q_4$ and an expansion up to
$t^5$ for $gg\to ZH$ and $t^{10}$
for $gg\to Z^\star Z^\star$.  In all expansion
terms which overlap with those obtained in Section~\ref{subsub::texp_amp} we
find agreement.

Our final results for the helicity amplitudes use a combination of both
$t$-expansion approaches. The contributions have the form
$t^{n_t} (q_3^2)^{n_3} (q_4^2)^{n_4}$. For $n_3+n_4=0,1,2$ we include all
terms up to $n_t\leq5$ ($ZH$) and $n_t\leq10$ ($Z^\star Z^\star$), using
approach (ii). ``Above'' these we additionally include terms with
$n_t+n_3+n_4\leq4$, computed using approach (i).\footnote{For the helicity amplitude $\mathcal{A}_{++00}$ of $gg\to Z^\star Z^\star$
we include terms to $n_t+n_3+n_4\leq5$, as discussed in Section \ref{subsub::texp_MI}.}

\subsubsection{High-energy expansion of master integrals}

The starting point for the high-energy expansion is the same amplitude in
terms of the 161 two-loop master integrals as for the $t$ expansion.
Thus, the agreement of the $t$ expansions obtained in
Sections~\ref{subsub::texp_amp} and~\ref{subsub::texp_MI} imposes a strong
check on our high-energy results.  After expanding the coefficients for
$m_t\to 0$ and inserting the high-energy expansion of the master integrals
from Refs.~\cite{Davies:2018ood,Davies:2018qvx,Mishima:2018olh,Davies:2023vmj} we
obtain results for the form factors and helicity amplitudes
expanded up to $(q_3^2)^{n_3} (q_4^2)^{n_4}$ with $n_3+n_4\leq2$ and up to $m_t^{112}$.

We use these deep expansions to construct Pad\'e approximants in the variable $m_t$. This is achieved by first inserting
numerical values for all kinematic variables and for the $\log(m_t)$ terms, which leads to a polynomial in $m_t$ with the highest exponent $N$. Afterwards we compute
a set of
Pad\'e approximants 
where $N$ is varied in a given range. Their combination provides,
for each phase-space point, a central value and an uncertainty estimate. For more details we refer to Ref.~\cite{Davies:2020lpf}.
Note that it is important to include expansion terms up to about $m_t^{80}$; including more terms
only leads to a marginal change of the numerical results. For practical reasons we include for
$gg\to Z^\star Z^\star$ terms up to $m_t^{100}$ and for $gg\to ZH$ terms up to $m_t^{112}$.

\subsection{\label{sub::UVIR}Ultraviolet renormalization and infrared subtraction}

To treat the ultraviolet divergences we work in the six-flavour theory and
renormalize the top quark mass and the gluon wave function on shell and the
strong coupling $\alpha_s$ in the $\overline{\rm MS}$ scheme.  For the process
$gg\to ZH$ we have additional (finite) renormalization constants due to our
treatment of $\gamma_5$. They are different for axial-vector (``A'') currents, and
pseudo-scalar (``P'') currents which appear due to $s$-channel Goldstone bosons in the triangle diagrams, and are given by~\cite{Larin:1993tq}:
\begin{eqnarray*}
  Z_{5A} &=& 1 - \frac{\alpha_s}{\pi} C_F + {\cal O}(\epsilon)\,,\nonumber\\
  Z_{5P} &=& 1 - 2 \frac{\alpha_s}{\pi} C_F + {\cal O}(\epsilon)\,,
\end{eqnarray*}
with $C_F=4/3$.
At this point all remaining poles in $\epsilon$ are of infrared nature.
Since they are usually treated in the five-flavour theory
we switch from $\alpha_s^{(6)}$ to $\alpha_s^{(5)}$.

For the subtraction of the infrared poles we follow Ref.~\cite{Catani:1998bh}
and compute\footnote{We present the formula for form factors; the
  application to helicity amplitudes is in complete analogy.}
\begin{eqnarray}
  F^{(1)} &=& F^{(1),\rm IR} - K_g^{(1)} F^{(0)} \,.
  \label{eq::FIRsub}
\end{eqnarray}
Here the superscripts ``(0)'' and ``(1)'' indicate the LO and NLO
contributions to the form factors. $F^{(1),\rm IR}$ is ultraviolet
renormalized but still infrared divergent and $F^{(1)}$ on the left-hand
side is finite. In our conventions $K_g^{(1)}$ is given by
\begin{eqnarray}
  K_g^{(1)} &=& - \left(\frac{\mu^2}{-s-i\delta}\right)^\epsilon 
  \frac{e^{\epsilon\gamma_E}}{ 2 \Gamma(1-\epsilon)}
  \left[\frac{C_A}{\epsilon^2} +
    \frac{1}{\epsilon}\left( \frac{11}{6}C_A-\frac{1}{3}n_f \right)
  \right]
  \,,
  \label{eq::Ig1}
\end{eqnarray}
where $C_A=3, n_f=5$ and $\gamma_E$ is Euler's constant.

\section{\label{sec::ggZZ}$gg \to Z^\star Z^\star$}

\subsection{Amplitudes and projectors}

The amplitude for $gg\to Z^\star Z^\star$ can be written as
\begin{align}
\mathcal{A}_{\lambda_1\lambda_2\lambda_3\lambda_4}(q_1,q_2,q_3,q_4) = \mathcal{A}_{\mu\nu\rho\sigma}(q_1,q_2,q_3,q_4)\epsilon_{\lambda_1}^\mu(q_1)\epsilon_{\lambda_2}^\nu(q_2)\epsilon_{\lambda_3}^{*\rho}(q_3)\epsilon_{\lambda_4}^{*\sigma}(q_4)\,,
\end{align}
where the polarization indices take the values $\{+,-\}$ for the gluons and 
$\{+,-,0\}$ for the $Z$ bosons.
The most general form consists of $138$ parity-even tensor structures, but this number can be reduced to $20$, see Refs.~\cite{Caola:2015ila,vonManteuffel:2015msa,Davies:2020lpf}, by taking into account the transversality of the external particles ($\varepsilon(q_i)\cdot q_i=0$) and by fixing the gauge of the gluon polarization vectors. We consider an axial gauge with the reference momentum $q_2$ for $\varepsilon(q_1)$ and $q_1$ for $\varepsilon(q_2)$, so that the two gauge conditions are given by $\varepsilon(q_1)\cdot q_2=0$ and $\varepsilon(q_2)\cdot q_1=0$. The polarization sums are given by
\begin{align}
\sum_\lambda \varepsilon_\lambda^\mu(q_1)\varepsilon_\lambda^{*\nu}(q_1)=\sum_\lambda\varepsilon_\lambda^\mu(q_2)\varepsilon_\lambda^{*\nu}(q_2)=-g^{\mu\nu}+\frac{q_1^\mu q_2^\nu+q_2^\mu q_1^\nu}{q_1\cdot q_2}.
\end{align}
On the other hand, the polarization sums of the two $Z$ bosons lead to
\begin{align}
\sum_\lambda \varepsilon_\lambda^\mu(q_i)\varepsilon_\lambda^{*\nu}(q_i)=-g^{\mu\nu}+\frac{q_i^\mu q_i^\nu}{q_i^2}
\end{align}
for $i=3,4$.
We write the amplitude as
\begin{align}
\mathcal{A}^{\mu\nu\rho\sigma}(q_1,q_2,q_3,q_4)=\sum_{i=1}^{20} F_i(s,t,q_3^2,q_4^2)\,T_i^{\mu\nu\rho\sigma},
\end{align}
where the $20$ tensor structures are given by
\begin{alignat}{4}
  &T_{1}^{\mu\nu\rho\sigma}=s^2g^{\mu\nu}g^{\rho\sigma},\quad
  &&T_{2}^{\mu\nu\rho\sigma}=s^2g^{\mu\rho}g^{\nu\sigma},\quad
  &&T_{3}^{\mu\nu\rho\sigma}=s^2g^{\mu\sigma}g^{\nu\rho},\quad
  &&T_{4}^{\mu\nu\rho\sigma}=sg^{\mu\nu}q_1^{\rho}q_1^{\sigma},\quad
  \nonumber\\
  &T_{5}^{\mu\nu\rho\sigma}=sg^{\mu\nu}q_1^{\rho}q_2^{\sigma},\quad
  &&T_{6}^{\mu\nu\rho\sigma}=sg^{\mu\nu}q_1^{\sigma}q_2^{\rho},\quad
  &&T_{7}^{\mu\nu\rho\sigma}=sg^{\mu\nu}q_2^{\rho}q_2^{\sigma},\quad
  &&T_{8}^{\mu\nu\rho\sigma}=sg^{\mu\rho}q_1^{\sigma}q_3^{\nu},\quad
  \nonumber\\
  &T_{9}^{\mu\nu\rho\sigma}=sg^{\mu\rho}q_2^{\sigma}q_3^{\nu},\quad
  &&T_{10}^{\mu\nu\rho\sigma}=sg^{\mu\sigma}q_1^{\rho}q_3^{\nu},\quad
  &&T_{11}^{\mu\nu\rho\sigma}=sg^{\mu\sigma}q_2^{\rho}q_3^{\nu},\quad
  &&T_{12}^{\mu\nu\rho\sigma}=sg^{\nu\rho}q_1^{\sigma}q_3^{\mu},\quad
  \nonumber\\
  &T_{13}^{\mu\nu\rho\sigma}=sg^{\nu\rho}q_2^{\sigma}q_3^{\mu},\quad
  &&T_{14}^{\mu\nu\rho\sigma}=sg^{\nu\sigma}q_1^{\rho}q_3^{\mu},\quad 
  &&T_{15}^{\mu\nu\rho\sigma}=sg^{\nu\sigma}q_2^{\rho}q_3^{\mu},\quad
  &&T_{16}^{\mu\nu\rho\sigma}=sg^{\rho\sigma}q_3^{\mu}q_3^{\nu},\quad
  \nonumber\\
  &T_{17}^{\mu\nu\rho\sigma}=q_1^{\rho}q_1^{\sigma}q_3^{\mu}q_3^{\nu},\quad
  &&T_{18}^{\mu\nu\rho\sigma}=q_1^{\rho}q_2^{\sigma}q_3^{\mu}q_3^{\nu},\quad
  &&T_{19}^{\mu\nu\rho\sigma}=q_1^{\sigma}q_2^{\rho}q_3^{\mu}q_3^{\nu},\quad
  &&T_{20}^{\mu\nu\rho\sigma}=q_2^{\rho}q_2^{\sigma}q_3^{\mu}q_3^{\nu}.
  \label{eq::Ti}
\end{alignat}
The ordering of the tensor structure is the same as in Ref. \cite{Agarwal:2020dye} but with additional factors of $s$ such that all tensor structures have the same mass dimension.
Note that the amplitude $\mathcal{A}^{\mu\nu\rho\sigma}$ is dimensionless
and that the form factors have mass dimension $-4$.

We introduce the projectors $P_i^{\mu\nu\rho\sigma}$ for the form factors
$F_i$ via the ansatz
\begin{align}
P_i^{\mu\nu\rho\sigma}=\sum_{j=1}^{20} c_{ij}\,T_{ j}^{\mu\nu\rho\sigma},
\end{align}
and determine the coefficients $c_{ij}(s,t,q_3^2,q_4^2)$ by requiring
\begin{align}
F_i=P_{i}^{\mu\nu\rho\sigma}\sum_{\lambda_1,\lambda_2,\lambda_3,\lambda_4}
\varepsilon_{1,\mu}^*\varepsilon_{1,\mu'}\varepsilon_{2,\nu}^*\varepsilon_{2,\nu'}\varepsilon_{3,\rho}\varepsilon_{3,\rho'}^*\varepsilon_{4,\sigma}\varepsilon_{4,\sigma'}^*\mathcal{A}^{\mu'\nu'\rho'\sigma'}\,,
\label{eq::Fi}
\end{align}
with $\varepsilon_{i,\mu}^{(*)} \equiv \varepsilon_{\lambda_i,\mu}^{(*)}(q_i)$.
The summation over the polarizations as introduced in Eq.~(\ref{eq::Fi}) guarantees
that only the 20 tensor structures from Eq.~(\ref{eq::Ti}), and not all possible 138 tensor structures, have to be considered.

In our practical calculation we first compute the projectors
$P_{i}^{\mu\nu\rho\sigma}$ and use them to construct
the projectors for the helicity amplitudes $\mathcal{A}_{\lambda_1\lambda_2\lambda_3\lambda_4}$
which are given by
\begin{align}
\mathcal{A}_{\lambda_1\lambda_2\lambda_3\lambda_4}=\mathcal{P}_{\lambda_1\lambda_2\lambda_3\lambda_4}^{\mu\nu\rho\sigma}
\sum_{\lambda_1^\prime,\lambda_2^\prime,\lambda_3^\prime,\lambda_4^\prime}
\varepsilon_{1,\mu}^*\varepsilon_{1,\mu'}\varepsilon_{2,\nu}^*\varepsilon_{2,\nu'}\varepsilon_{3,\rho}\varepsilon_{3,\rho'}^*\varepsilon_{4,\sigma}\varepsilon_{4,\sigma'}^*\mathcal{A}^{\mu'\nu'\rho'\sigma'}\,.
\end{align}
The projectors $\mathcal{P}_{\lambda_1\lambda_2\lambda_3\lambda_4}^{\mu\nu\rho\sigma}$ are obtained from the multiplication of the $20$ projectors $P_i^{\mu\nu\rho\sigma}$ by the contraction of the tensor structures $T_i^{\mu\nu\rho\sigma}$ with the polarization vectors of the gluons and $Z$ bosons as
\begin{align} \label{eq::polproj}
\mathcal{P}_{\lambda_1\lambda_2\lambda_3\lambda_4}^{\mu\nu\rho\sigma}=\sum_{i=1}^{20}P_i^{\mu\nu\rho\sigma}\left(\epsilon_{\lambda_1}^{\mu'}(q_1)\epsilon_{\lambda_2}^{\nu'}(q_2)\epsilon_{\lambda_3}^{*\rho'}(q_3)\epsilon_{\lambda_4}^{*\sigma'}(q_4)T_{i,\mu'\nu'\rho'\sigma'}\right)=\sum_{i=1}^{20}a_{\lambda_1\lambda_2\lambda_3\lambda_4}^{(i)}T_i^{\mu\nu\rho\sigma}.
\end{align}
The projectors for the helicity amplitudes are written as linear combinations of 
the $20$ tensor structures $T_i^{\mu\nu\rho\sigma}$ and coefficients 
$a^{(i)}_{\lambda_1\lambda_2\lambda_3\lambda_4}(s,t,q_3^2,q_4^2)$,
see right-hand side of Eq.~(\ref{eq::polproj}).
Explicit results for the $a^{(i)}_{\lambda_1\lambda_2\lambda_3\lambda_4}$
are obtained after specifying the polarization vectors as given below.
For illustration, we provide explicit results for 
$\mathcal{A}_{++00}$
in Appendix~\ref{app::proj};
results for all helicity amplitudes can be found
in the supplementary material of this paper~\cite{progdata}.

We choose the same polarization vectors as in Ref. \cite{Agarwal:2020dye}, extended to the case of two off-shell $Z$ bosons, where we parametrize the four momenta as
\begin{alignat}{2}
&q_1=\frac{\sqrt{s}}{2}\left(\begin{array}{c}1\\ 0\\ 0\\ 1\end{array}\right)
\!,\quad
&&q_3=\frac{\sqrt{s}}{2}
\left(\begin{array}{c}-\left(1+\frac{q_3^2}{s}-\frac{q_4^2}{s}\right)\\ -\beta \sin \theta \\ 0\\ -\beta \cos \theta \end{array}\right)
\!,\,\,
\nonumber\\
&q_2=\frac{\sqrt{s}}{2}\left(\begin{array}{c}1\\ 0\\ 0\\ -1\end{array}\right)
\!,\quad
&&q_4=\frac{\sqrt{s}}{2}
\left(\begin{array}{c}-\left(1-\frac{q_3^2}{s}+\frac{q_4^2}{s}\right)\\ \beta \sin \theta \\ 0\\ \beta \cos \theta \end{array}\right)\,,
\end{alignat}
with
\begin{align}
\beta=\sqrt{1-2\frac{q_3^2+q_4^2}{s}+\frac{(q_3^2-q_4^2)^2}{s^2}},
\end{align}
and the polarization vectors are given by 
\begin{alignat}{2}
\nonumber&\varepsilon_\pm(q_1)=\varepsilon_\mp(q_2)=\frac{1}{\sqrt{2}}\left(\begin{array}{c}0\\ \mp 1\\-i\\0\end{array}\right)\!,\quad
&&\varepsilon_0(q_3)=\frac{\sqrt{s}}{2\sqrt{q_3^2}}\left(\begin{array}{c}\beta\\ \sin\theta\,\left(1+\frac{q_3^2}{s}-\frac{q_4^2}{s}\right)\\0\\ \cos\theta\,\left(1+\frac{q_3^2}{s}-\frac{q_4^2}{s}\right)\end{array}\right)\!,\\
&\varepsilon_\pm(q_3)=\varepsilon_\mp(q_4)=\frac{1}{\sqrt{2}}\left(\begin{array}{c}0\\ \mp\cos\theta\\-i\\ \pm\sin\theta\end{array}\right)\!\,,
&&\,\,\varepsilon_0(q_4)=\varepsilon_0(q_3)\Big|_{\theta\to\theta+\pi,\,q_3^2\leftrightarrow q_4^2}\,.
\end{alignat}

In practice we write the coefficients $a_{\lambda_1\lambda_2\lambda_3\lambda_4}^{(i)}$ in Eq.~\eqref{eq::polproj} as
\begin{align} \label{eq::projcoeff}
a_{\lambda_1\lambda_2\lambda_3\lambda_4}^{(i)}=\frac{p_T^{b_{\lambda_1\lambda_2\lambda_3\lambda_4}}}{\beta^2\sqrt{q_3^2}^{\delta_{\lambda_3 0}}\sqrt{q_4^2}^{\delta_{\lambda_4 0}}}\left(a_{\lambda_1\lambda_2\lambda_3\lambda_4}^{(i,\beta^0)}+\beta\,a_{\lambda_1\lambda_2\lambda_3\lambda_4}^{(i,\beta^1)}\right),
\end{align}
where $b_{\lambda_1\lambda_2\lambda_3\lambda_4}$ is either $0$, $1$ or $2$, depending on the helicity amplitude. The coefficients are written in such a way that the normalization factor and the factor $\beta$ in the brackets in Eq.~\eqref{eq::projcoeff} are hidden during the forward and high-energy expansions so that the helicity amplitudes are given in both cases by
\begin{align}
\mathcal{A}_{\lambda_1\lambda_2\lambda_3\lambda_4}=\frac{p_T^{b_{\lambda_1\lambda_2\lambda_3\lambda_4}}}{\beta^2\sqrt{q_3^2}^{\delta_{\lambda_3 0}}\sqrt{q_4^2}^{\delta_{\lambda_4 0}}}\left(\mathcal{A}^{(\beta^0)}_{\lambda_1\lambda_2\lambda_3\lambda_4}+\beta\,\mathcal{A}^{(\beta^1)}_{\lambda_1\lambda_2\lambda_3\lambda_4}\right),
\end{align}
where the latter term is zero for the helicity amplitudes $\mathcal{A}_{\lambda_1\lambda_2 00}$.
The choice of the normalization factors ensures that the 
first expansion term of the forward expansion of 
$\mathcal{A}_{\lambda_1\lambda_2\lambda_3\lambda_4}^{(\beta^0)}$ 
and 
$\mathcal{A}_{\lambda_1\lambda_2\lambda_3\lambda_4}^{(\beta^1)}$ 
is independent of $t$, $q_3^2$ and $q_4^2$. In addition, we 
have found that not expanding the factors of $\beta$ 
drastically improves the convergence of the expansion in 
$q_3^2$ and $q_4^2$, especially for small values of 
$\sqrt{s}$.

Due to several symmetries between the helicity amplitudes, see also Ref. \cite{Agarwal:2020dye}, we find that just eight are independent which we choose as
\begin{eqnarray} \label{eq::ggZZ_hel}
    \mathcal{A}_{++++},\, \mathcal{A}_{+++-},\, \mathcal{A}_{++-0},\, \mathcal{A}_{++00},\, 
    \mathcal{A}_{+-++},\, \mathcal{A}_{+-+-},\, \mathcal{A}_{+--0},\, \mathcal{A}_{+-00}\,.
\end{eqnarray}
All helicity amplitudes with $\lambda_1=-$ can be obtained from those with $\lambda_1=+$ with the relation
\begin{equation}
\mathcal{A}_{\lambda_1\lambda_2\lambda_3\lambda_4}=(-1)^{\delta_{\lambda_30}+\lambda_40}\mathcal{A}_{-\lambda_1 -\lambda_2 -\lambda_3 -\lambda_4}.
\end{equation}
In addition, we use the following relations for the replacement $q_3^2 \leftrightarrow q_4^2$
\begin{equation}
\begin{array}{cclcccl}
\mathcal{A}_{++0-} &=& \mathcal{A}_{++-0}\Big|_{q_3^2 \leftrightarrow q_4^2},
&\quad&
\mathcal{A}_{+-0-} &=& -\mathcal{A}_{+-+0}\Big|_{q_3^2 \leftrightarrow q_4^2},
\end{array}
\end{equation}
and the replacement $\beta \to -\beta$ provides relations
\begin{equation}
\begin{array}{cclcccl}
\mathcal{A}_{++--} &=& \mathcal{A}_{++++}\Big|_{\beta\to -\beta},
&\quad&
\mathcal{A}_{++-+} &=& \mathcal{A}_{+++-}\Big|_{\beta\to -\beta},\\
\mathcal{A}_{+--+} &=& \mathcal{A}_{+-+-}\Big|_{\beta\to -\beta},
&\quad&
\mathcal{A}_{+---} &=& \mathcal{A}_{+-++}\Big|_{\beta\to -\beta},\\
\mathcal{A}_{+++0} &=& -\mathcal{A}_{++-0}\Big|_{\beta\to -\beta},
&\quad&
\mathcal{A}_{++0+} &=& -\mathcal{A}_{++0-}\Big|_{\beta\to -\beta},\\
\mathcal{A}_{+-+0} &=& -\mathcal{A}_{+--0}\Big|_{\beta\to -\beta},
&\quad&
\mathcal{A}_{+-0+} &=& -\mathcal{A}_{+-0-}\Big|_{\beta\to -\beta},
\end{array}
\label{eq:ZZhelicity}
\end{equation}
to obtain the remaining helicity amplitudes. As an additional cross-check we have also calculated the helicity amplitudes $\mathcal{A}_{++0-}$ and $\mathcal{A}_{+-0-}$, so that all remaining helicity amplitudes could in principle be obtained from only the relations coming from $\beta\to -\beta$. As another cross-check we have verified the following relations up to quadratic order in $q_3$ and $q_4$ in our results
\begin{equation}
\begin{array}{cclcccl}
\mathcal{A}_{++++} &=& \mathcal{A}_{++++}\Big|_{q_3^2 \leftrightarrow q_4^2},
&\quad&
\mathcal{A}_{+-+-} &=& \mathcal{A}_{+-+-}\Big|_{q_3^2 \leftrightarrow q_4^2},\\
\mathcal{A}_{++00} &=& \mathcal{A}_{++00}\Big|_{q_3^2 \leftrightarrow q_4^2},
&\quad&
\mathcal{A}_{+-00} &=& \mathcal{A}_{+-00}\Big|_{q_3^2 \leftrightarrow q_4^2},\\
\mathcal{A}_{+++-} &=& \mathcal{A}_{+++-}\Big|_{q_3^2 \leftrightarrow q_4^2,\,\beta\to -\beta},
&\quad&
\mathcal{A}_{+-++} &=&\mathcal{A}_{+-++}\Big|_{q_3^2 \leftrightarrow q_4^2,\,\beta\to -\beta},
\end{array}
\label{eq:ZZsymmetry}
\end{equation}

We define the perturbative expansion of the helicity amplitudes as
\begin{equation}
\mathcal{A}_{\lambda_1\lambda_2\lambda_3\lambda_4} = \frac{\alpha_s}{2\pi}\mathcal{A}^{(0)}_{\lambda_1\lambda_2\lambda_3\lambda_4}+\left(\frac{\alpha_s}{2\pi}\right)^2\mathcal{A}^{(1)}_{\lambda_1\lambda_2\lambda_3\lambda_4}
\end{equation}
and decompose them at each loop order according to the
vector, axial-vector, triangle and double-triangle contribution
\begin{equation}
\mathcal{A}^{(i)}_{\lambda_1\lambda_2\lambda_3\lambda_4} = v_t^2\mathcal{A}^{(i),{\rm v_t^2}}_{\lambda_1\lambda_2\lambda_3\lambda_4} +a_t^2\mathcal{A}^{(i),{\rm a_t^2}}_{\lambda_1\lambda_2\lambda_3\lambda_4}+\frac{e^2}{s_w c_w}\frac{s}{s-m_h^2}\mathcal{A}^{(i),{\rm tri}}_{\lambda_1\lambda_2\lambda_3\lambda_4}+\mathcal{A}^{(i),{\rm dt}}_{\lambda_1\lambda_2\lambda_3\lambda_4}\,,
\end{equation}
where $v_t$ and $a_t$ are given by
\begin{equation}
    v_t = \frac{e}{2s_wc_w}\left(I^3_t-2Q_ts_w^2\right), \quad\quad\quad\quad\quad\quad a_t = \frac{e}{2s_wc_w}I^3_t\,,
\end{equation}
with $e = \sqrt{4\pi\alpha}$ where $\alpha$ is the fine structure constant,
$Q_t=2/3$ is the (fractional) electric charge of the top quark,
$I_t^3=1/2$ is the third component of the weak isospin, and 
$s_w^2=1-c_w^2$ with $c_w = \cos(\theta_W)$ being the cosine of the weak mixing angle.
For the numerical results shown below we use
$c_w=m_W/m_Z$ and $\sqrt{2}G_F = \pi\alpha/(m_W^2 s_w^2)$
and use the Fermi constant $G_F$ and the $W$ and $Z$ boson masses
as input.

In Sections~\ref{sec::ggZZ1l}, ~\ref{sec::ggZZ2l} and \ref{sec::ggZZ2lVfin}
we only discuss the results for the box contributions
$\mathcal{A}^{(i),{\rm v_t^2}}_{\lambda_1\lambda_2\lambda_3\lambda_4}$
and $\mathcal{A}^{(i),{\rm a_t^2}}_{\lambda_1\lambda_2\lambda_3\lambda_4}$.
We have checked that our high-energy expansions converge very quickly for the
triangle contribution $\mathcal{A}^{(i),{\rm tri}}_{\lambda_1\lambda_2\lambda_3\lambda_4}$. 
Analytic results for the
double-triangle contribution can be found in Ref.~\cite{Campbell:2016ivq} for
on-shell $Z$ bosons. To our knowledge, the off-shell
result is not yet known. Note that the leading term of the $t$ expansion corresponds to the
exact result for the triangle form factors and helicity amplitudes.

In Ref.~\cite{Davies:2020lpf} the high-energy expansion of the form
factors has been computed including terms up to $m_t^{32}$; in this work
we compute the eight helicity amplitudes in Eq.~(\ref{eq::ggZZ_hel})
including one hundred expansion terms in $m_t$.

\subsection{\label{sec::ggZZ1l}One-loop results to $gg \to Z^\star Z^\star$}

\begin{figure}[t]
    \begin{center}
    \begin{tabular}{cc}
    \includegraphics[width=.45\textwidth]{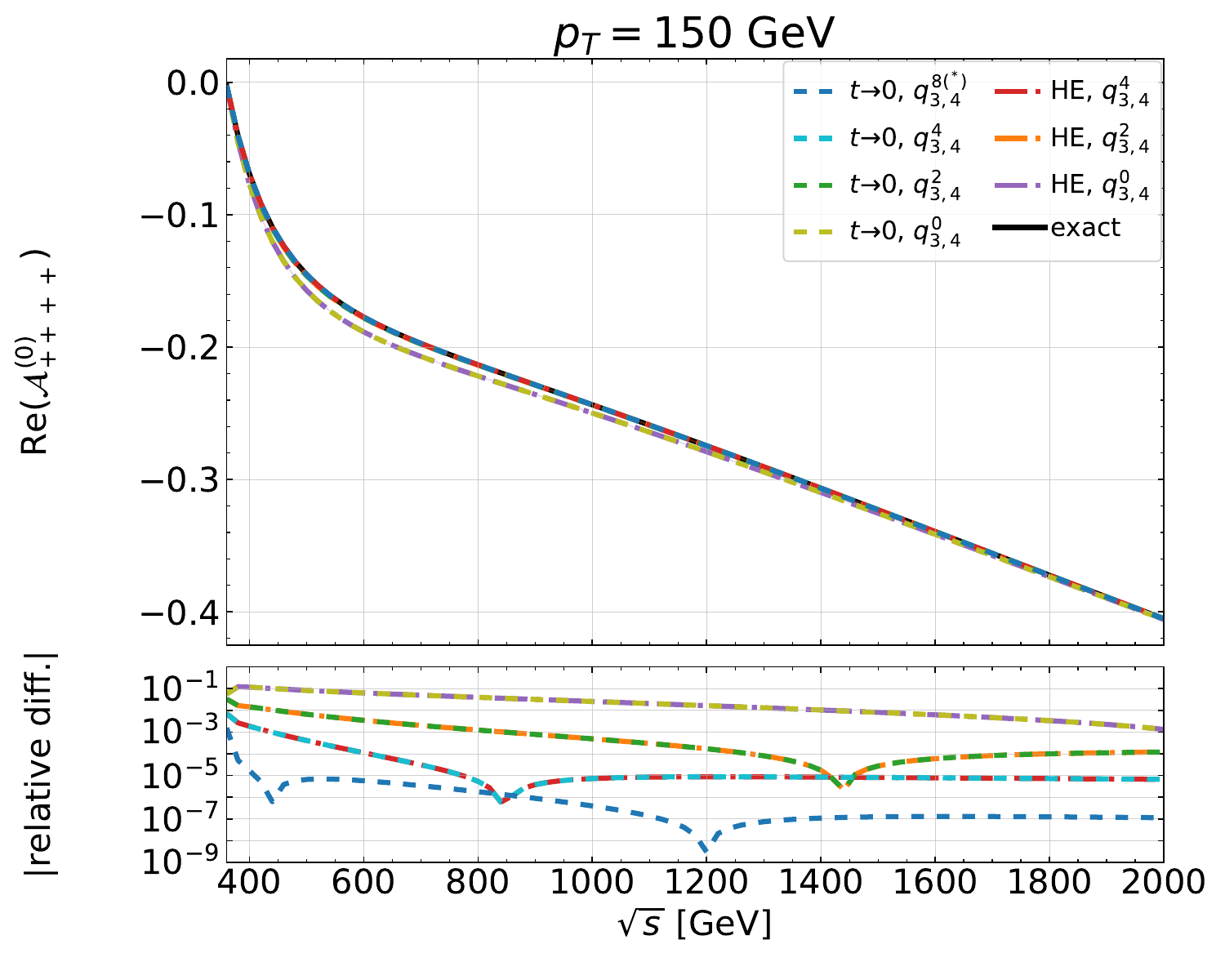} &
    \includegraphics[width=.45\textwidth]{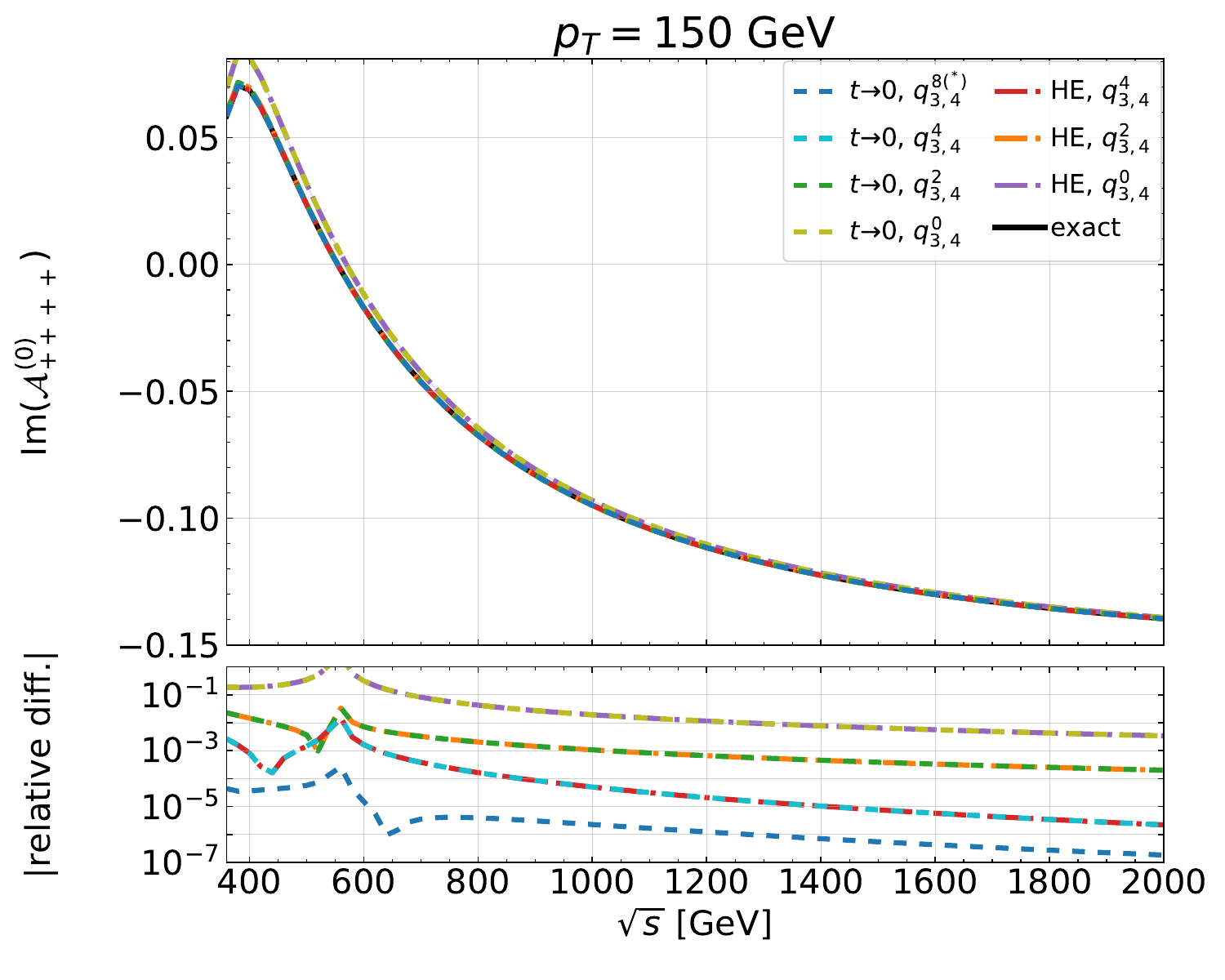} \\
    \includegraphics[width=.45\textwidth]{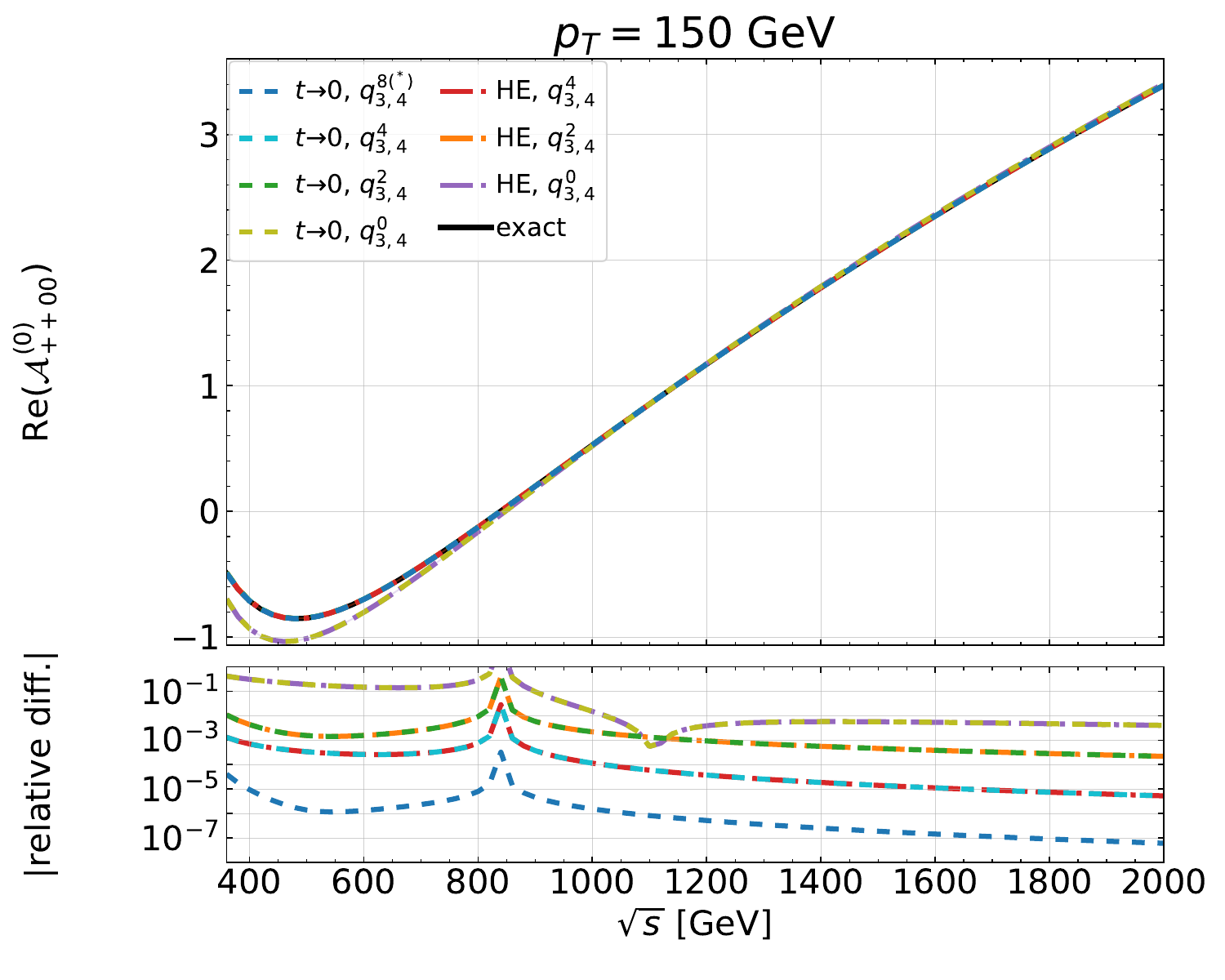} &
    \includegraphics[width=.45\textwidth]{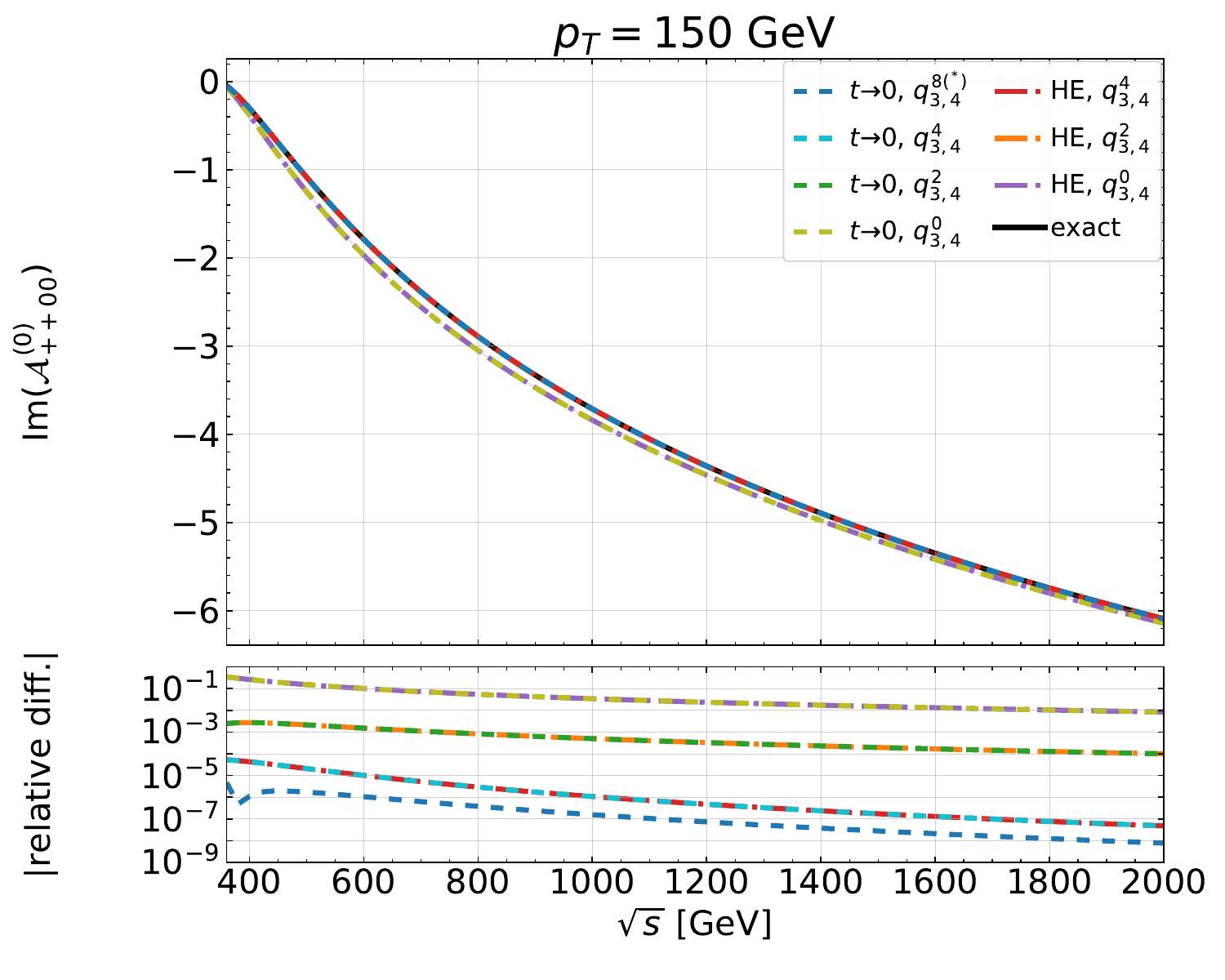}
    \end{tabular}
    \end{center}
  \caption{\label{fig::ggzz1l_pt150}
  Real and imaginary parts of ${\cal A}^{(0)}_{++++}$ and ${\cal A}^{(0)}_{++00}$ as a function of $\sqrt{s}$ for $p_T=150$~GeV and $q_3^2 = q_4^2 = m_Z^2$. High-energy and $t\to 0$ expansions are shown, including mass corrections up to $q_{3,4}^{\{0,2,4\}}$. Also shown are higher mass corrections up to $q_{3,4}^8$ for the $t\to 0$ expansion with fewer expansion terms in $t$ according to $t^{n_t}(q_3^2)^{n_3}(q_4^2)^{n_4}$ with $n_t+n_3+n_4\leq 4$ denoted by $\star$. Lower panels display the relative difference with respect to the exact result.}
\end{figure}

In this section we consider the one-loop 
helicity amplitudes and compare our approximations
to the exact results.
The exact one-loop results have been calculated with the \texttt{calc} setup without 
applying any expansions and are expressed in terms of scalar one-loop functions. 
We use \texttt{OneLOop}~\cite{vanHameren:2010cp} for their numerical evaluation.
We use the same number of expansion terms at one and two loops. For the high-energy expansion we have mass corrections up to $q_{3,4}^{\{0,2,4\}}$ and $m_t$ terms up to $100$. In addition, we construct a Pad\'e approximation following Ref.~\cite{Davies:2020lpf}, where we use expansions in $m_t$ between $m_t^{88}$ and $m_t^{100}$. For the $t\to 0$ expansion we consider mass corrections up to $q_{3,4}^{\{0,2,4\}}$ and terms up to $t^{10}$. In addition, we also include higher mass corrections up to $q_{3,4}^8$ with fewer expansion terms in $t$, that are obtained from the forward expansion at the amplitude level (``approach (i)''); this includes terms of the form $t^{n_t}(q_3^2)^{n_3}(q_4^2)^{n_4}$ with $n_t+n_3+n_4\leq 4$.

In Fig.~\ref{fig::ggzz1l_pt150}
we show the real and imaginary parts of the two helicity amplitudes ${\cal A}^{(0)}_{++++}$ and ${\cal A}^{(0)}_{++00}$
for $p_T=150$~GeV and on-shell $Z$ boson production,
i.e., $q_3^2=q_4^2=m_Z^2$,
as a function of $\sqrt{s}$.
The upper panels show the helicity amplitudes in various approximations and the lower
panels the relative deviation to the exact results, i.e.~$10^{-2}$
on the $y$ axis corresponds to a 1\% deviation.
In each panel we show the exact result as a solid black line.
The results from the $t$ 
expansion are shown as dashed lines. Here we show four curves which correspond to
$m_Z^2=0$, the inclusion of quadratic terms in $m_Z$, the inclusion
of quartic terms, and the inclusion of the additional higher order
terms in $m_Z$.
We show three dash-dotted curves for the high-energy approximation which include either
no $m_Z^2$ term, an expansion up to $m_Z^2$ or 
up to $m_Z^4$.
From Fig.~\ref{fig::ggzz1l_pt150} we observe good agreement between the forward and high-energy approximation over the whole range of $\sqrt{s}$.  For $m_Z=0$ the agreement with the
exact result is at the few-percent level. This improves significantly after including the
quadratic mass corrections and is well below the percent level once quartic terms are included.
The additional $m_Z$ terms which are available for the $t$ expansion provide
small additional improvements for the kinematic range chosen in Fig.~\ref{fig::ggzz1l_pt150}.
For lower energies they become more important.

\begin{figure}[t]
    \begin{center}
    \begin{tabular}{cc}
    \includegraphics[width=.45\textwidth]{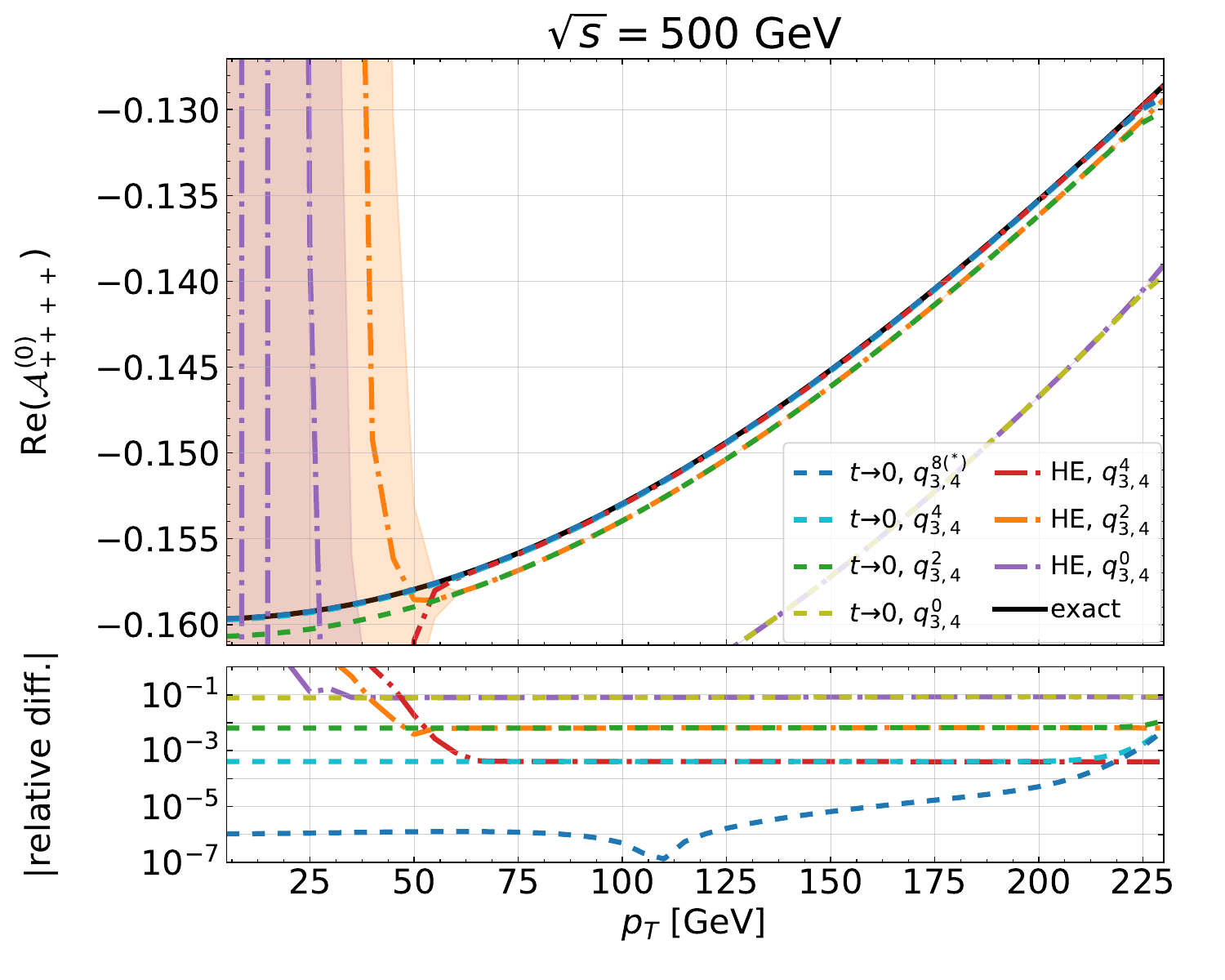} &
    \includegraphics[width=.45\textwidth]{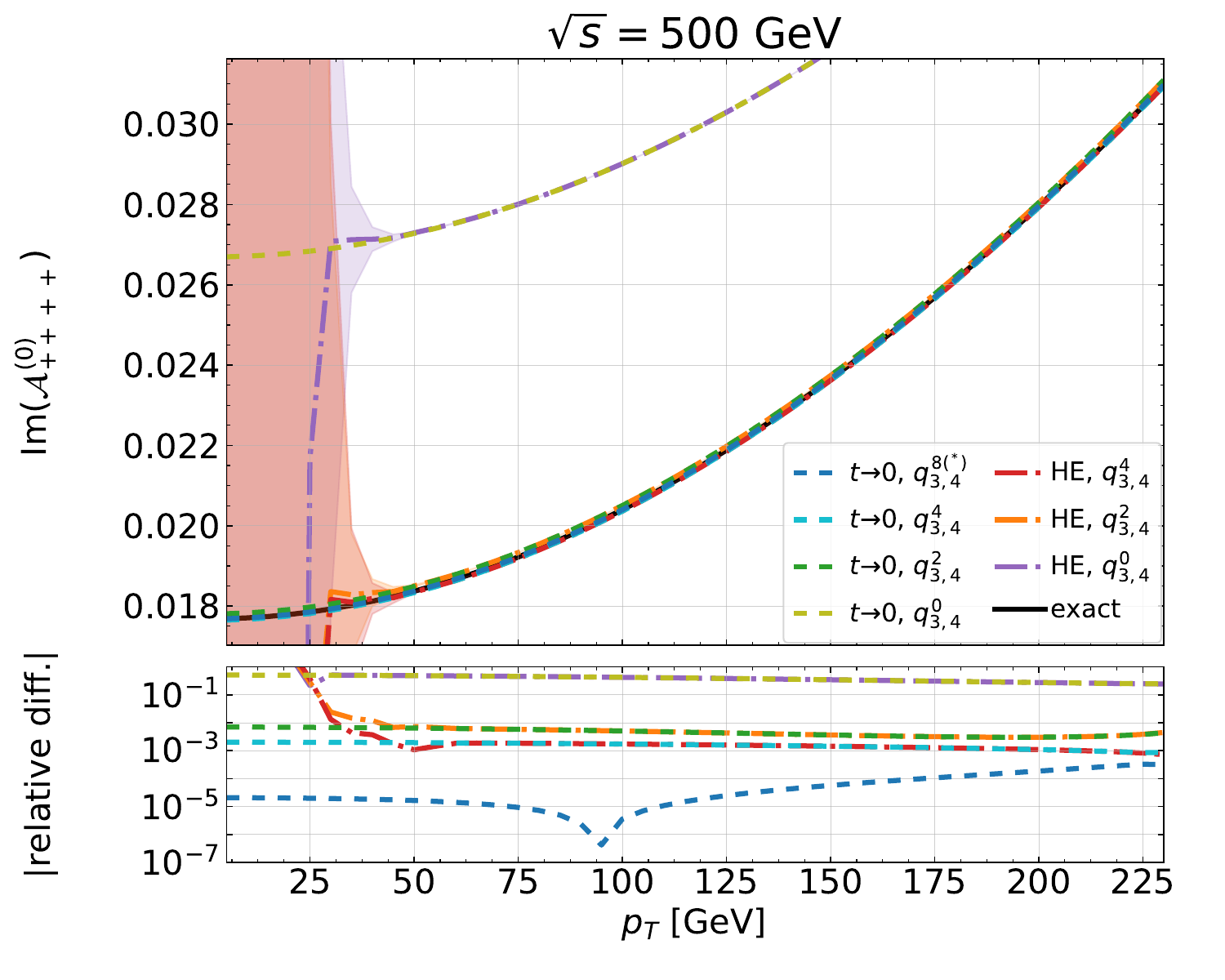} \\
    \includegraphics[width=.45\textwidth]{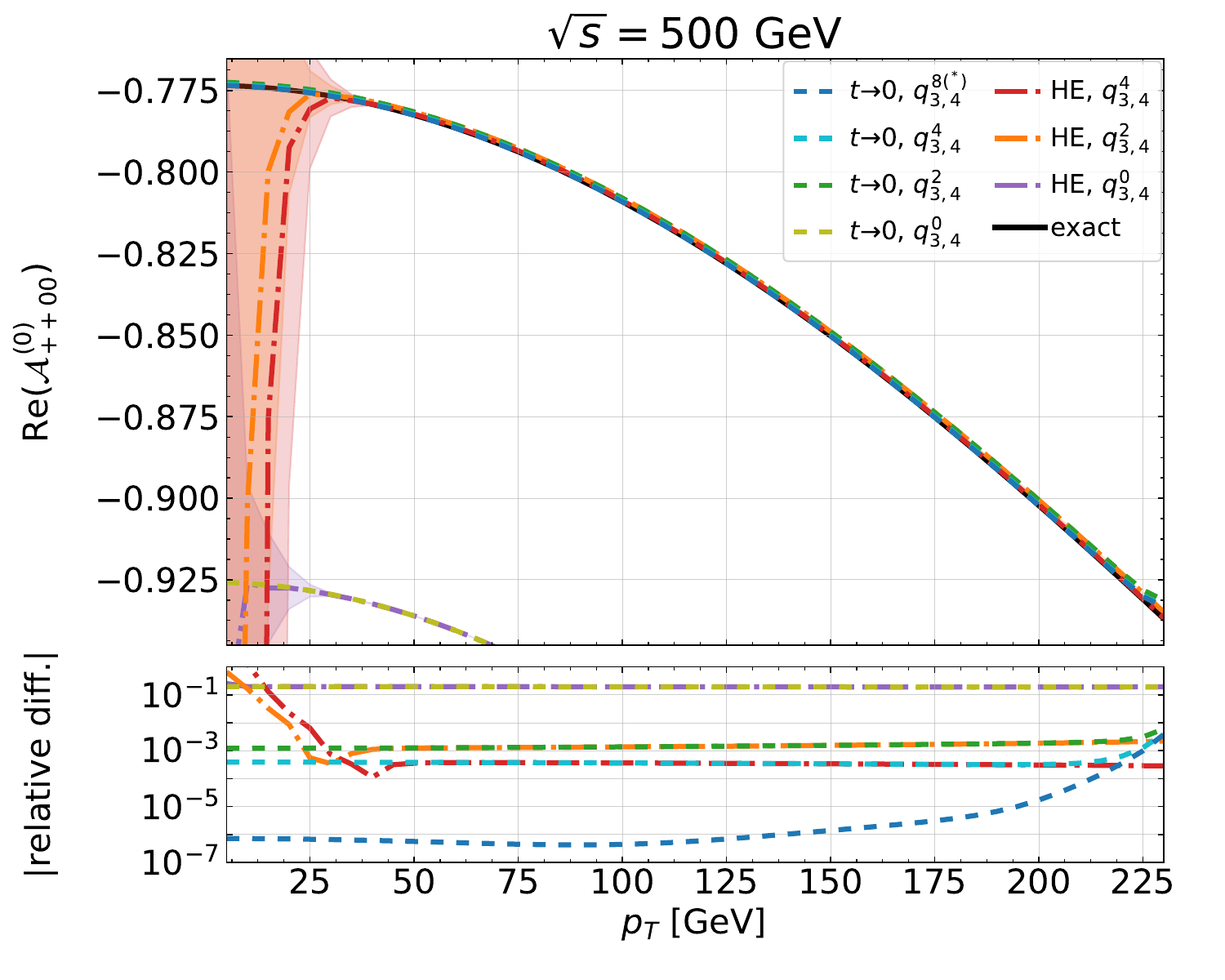} &
    \includegraphics[width=.45\textwidth]{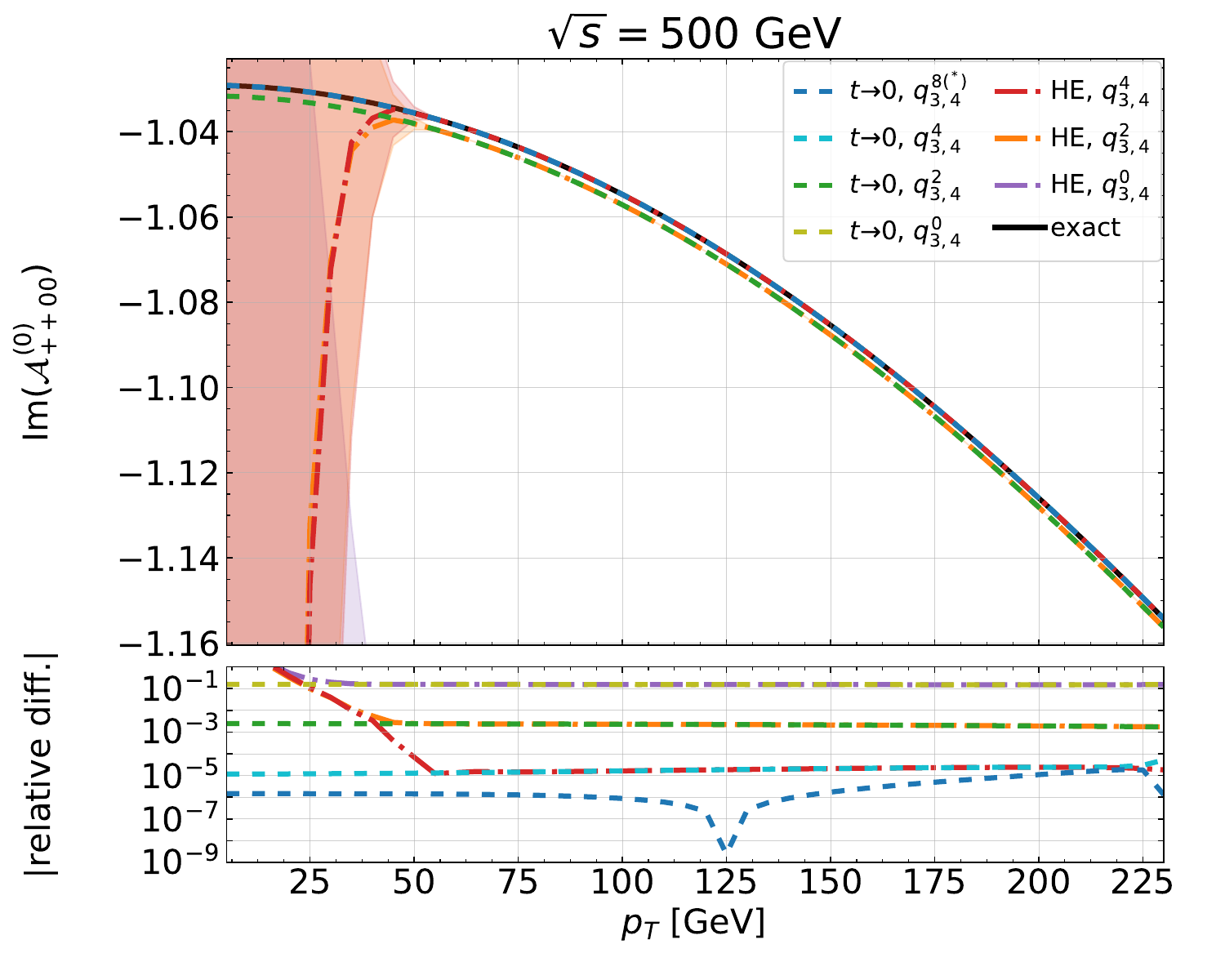}
    \end{tabular}
    \end{center}
  \caption{\label{fig::ggzz1l_sqrts500}
  Same as Figure \ref{fig::ggzz1l_pt150} but for fixed $\sqrt{s}=500$~GeV as a function of $p_T$.}
\end{figure}

In Fig.~\ref{fig::ggzz1l_sqrts500} we show the helicity amplitudes
${\cal A}^{(0)}_{++++}$ and ${\cal A}^{(0)}_{++00}$ for fixed $\sqrt{s}=500$~GeV as a function of $p_T$.
We use the same notation as in Fig.~\ref{fig::ggzz1l_pt150}.
These plots demonstrate that there is a relatively large
region for $p_T$ where both approximations agree with the exact results
below the per mille level, and for a large part of
the phase space even at the level of $10^{-6}$.
Furthermore, as expected, for small and large $p_T$ there is perfect agreement
between the exact results and the $t$ and high-energy expansions, respectively.
Note that for small values of $p_T$ the Pad\'e-improved high-energy expansion becomes unstable.

\begin{figure}[t]
    \begin{center}
    \begin{tabular}{cc}
    \includegraphics[width=.45\textwidth]{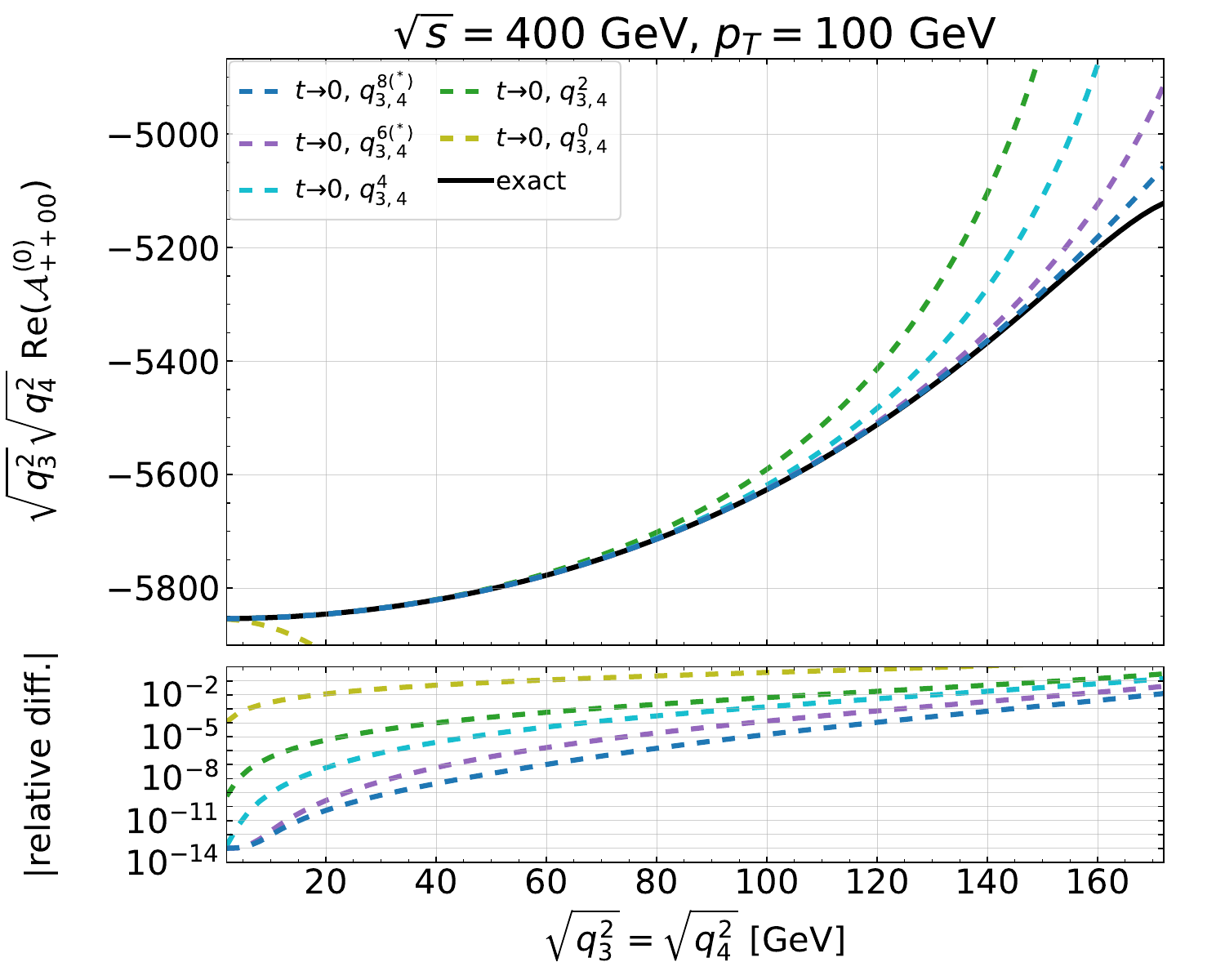} &
    \includegraphics[width=.45\textwidth]{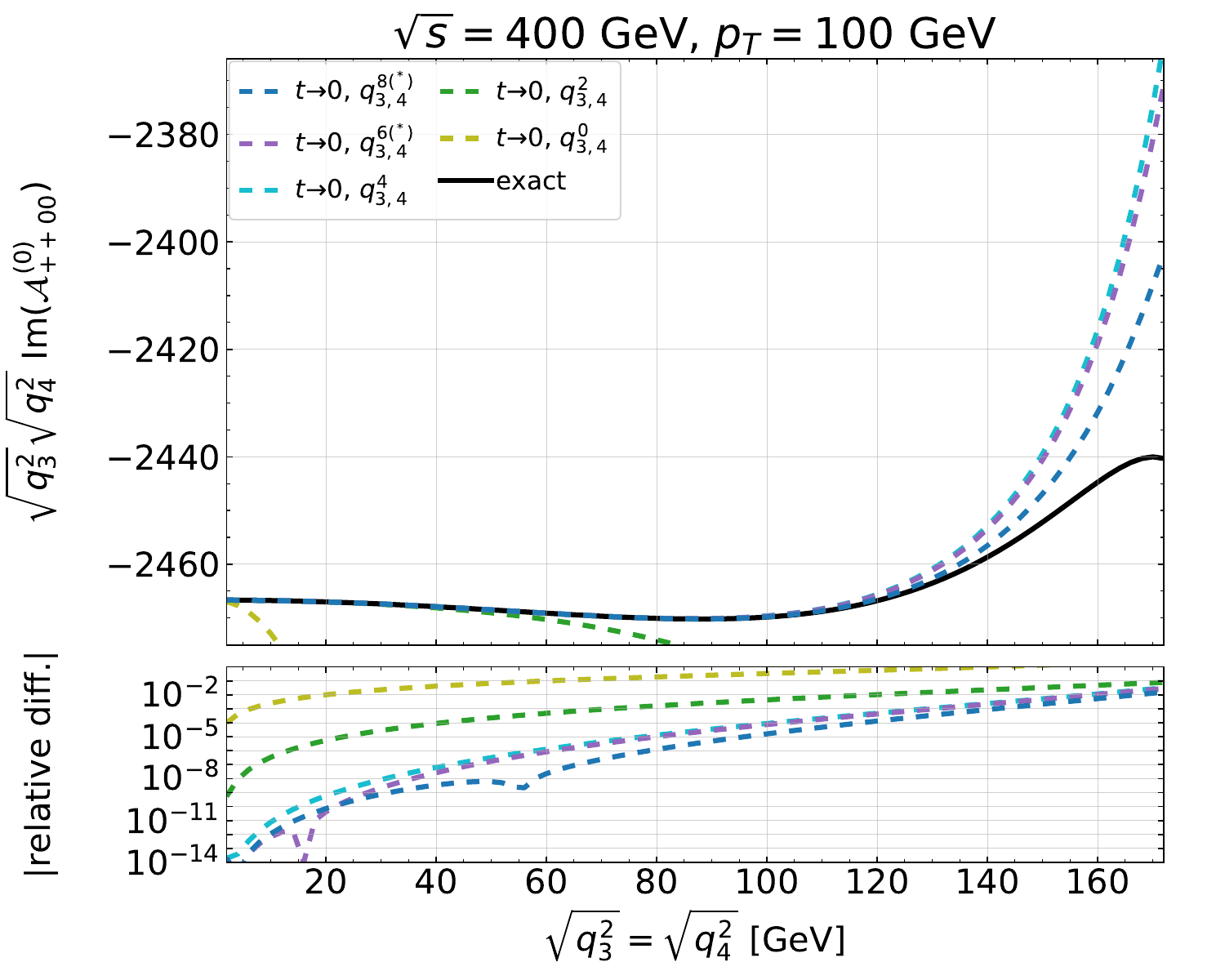}
    \end{tabular}
    \end{center}
  \caption{\label{fig::ggzz1l_q3sq4s}
  Same as Figure \ref{fig::ggzz1l_pt150} but for fixed $\sqrt{s}=400$~GeV and $p_T=100$~GeV as a function of $\sqrt{q_3^2}=\sqrt{q_4^2}$.}
\end{figure}

Finally, in Fig.~\ref{fig::ggzz1l_q3sq4s} we study the dependence of the expansion accuracy on the virtuality of
the final-state momenta $q_3$ and $q_4$. We fix $p_T=100$~GeV
and $\sqrt{s}=400$~GeV and show, for ${\cal A}^{(0)}_{++00}$,
the dependence on $q_3^2=q_4^2$. For the chosen values of
$p_T$ and $\sqrt{s}$ it is sufficient to consider only the $t$ expansion.
To avoid a divergent behaviour for small $q_3^2$ and $q_4^2$
we multiply the plotted helicity amplitude by
$\sqrt{q_3^2}\sqrt{q_4^2}$. The exact result is shown as black solid lines and the dashed curves include various expansion depths in $q_3^2$ and $q_4^2$. In the lower panel 
we show the relative differences between the expansions
and the exact result. It is impressive that for
 $q_3^2,q_4^2\to 0$ we reproduce more than 10 digits of the
 exact result after including quartic expansion terms. 
 For  $q_3^2=q_4^2=m_Z^2$ the agreement is at the level of $10^{-4}$
 and even for
 $q_3^2=q_4^2=(130~\mbox{GeV})^2$ the agreement with the exact result 
 is of the order of $0.1\%$ for the real
 and about a factor 10 better for the imaginary part.
 The agreement is similar for all other helicity amplitudes.

\subsection{\label{sec::ggZZ2l}Two-loop results to $gg \to Z^\star Z^\star$}

\begin{figure}[t]
    \begin{center}
    \begin{tabular}{cc}
    \includegraphics[width=.45\textwidth]{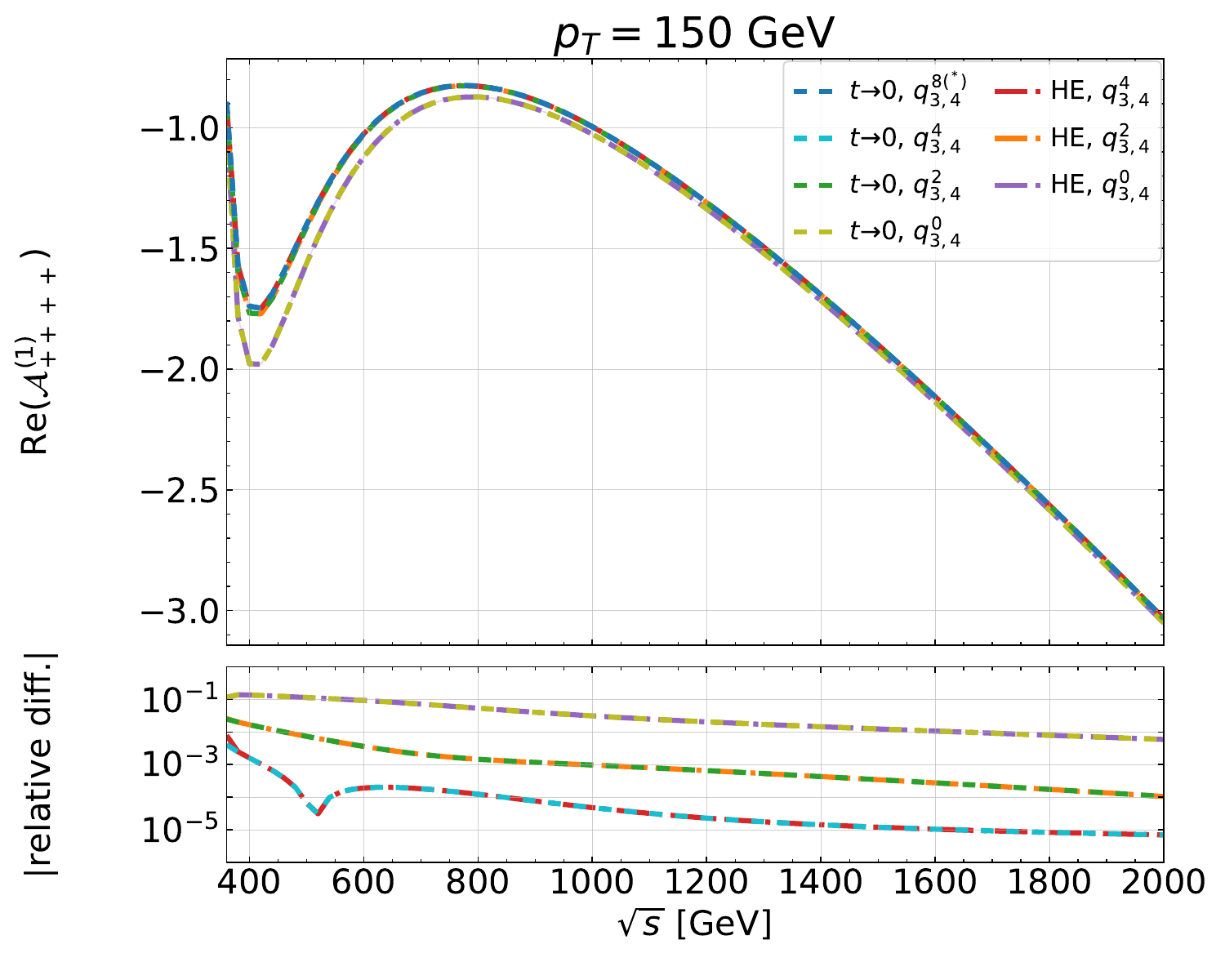} &
    \includegraphics[width=.45\textwidth]{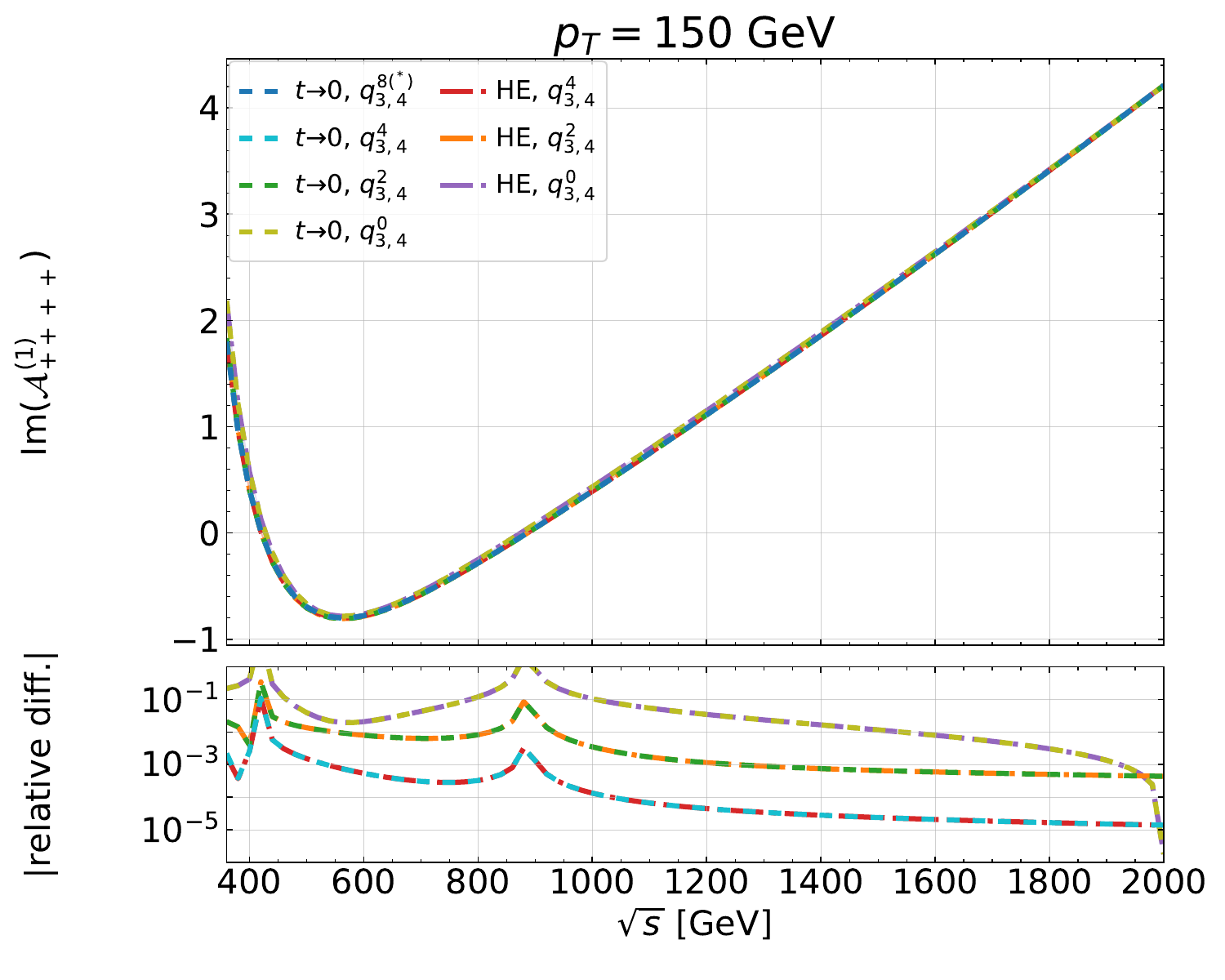} \\
    \includegraphics[width=.45\textwidth]{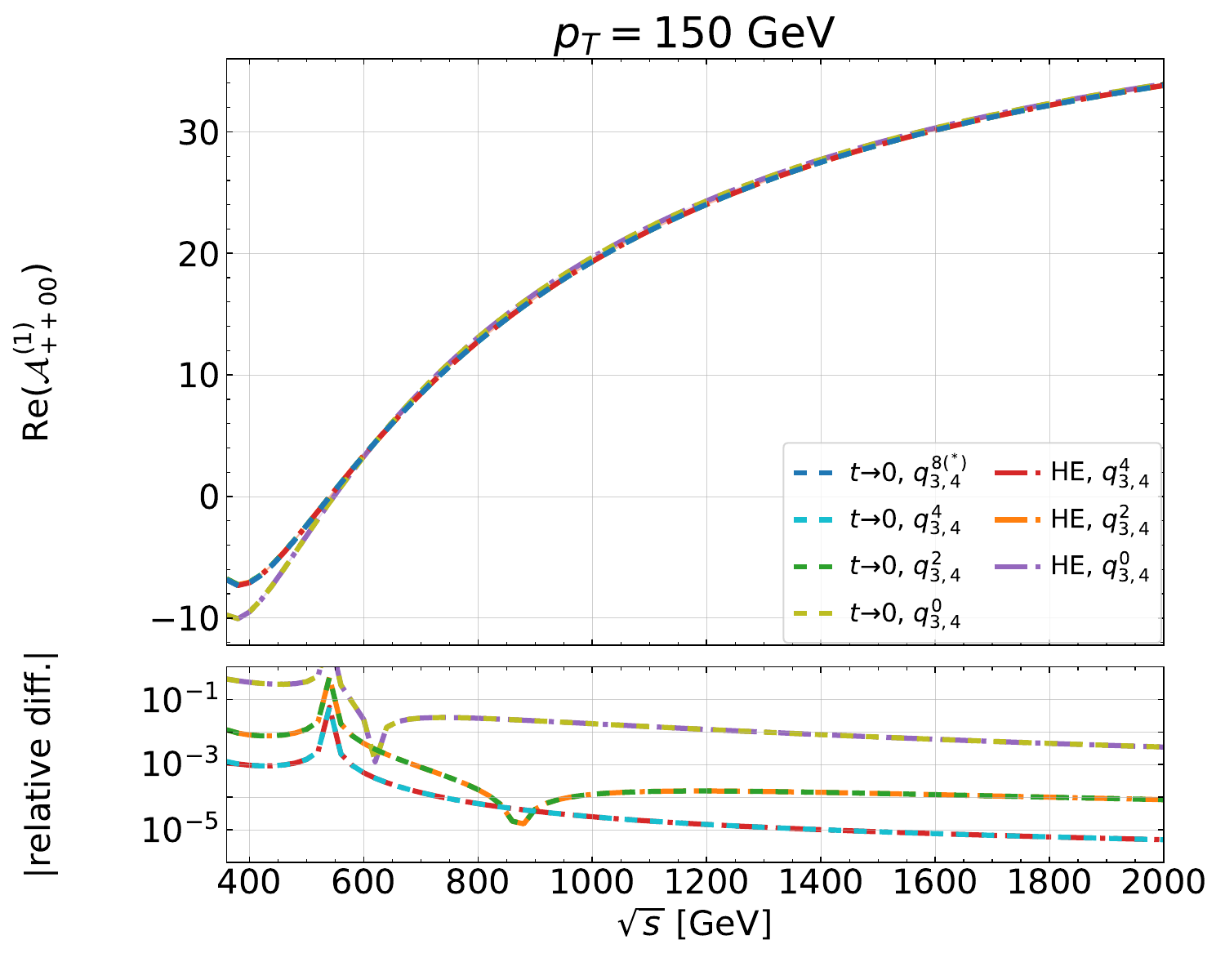} &
    \includegraphics[width=.45\textwidth]{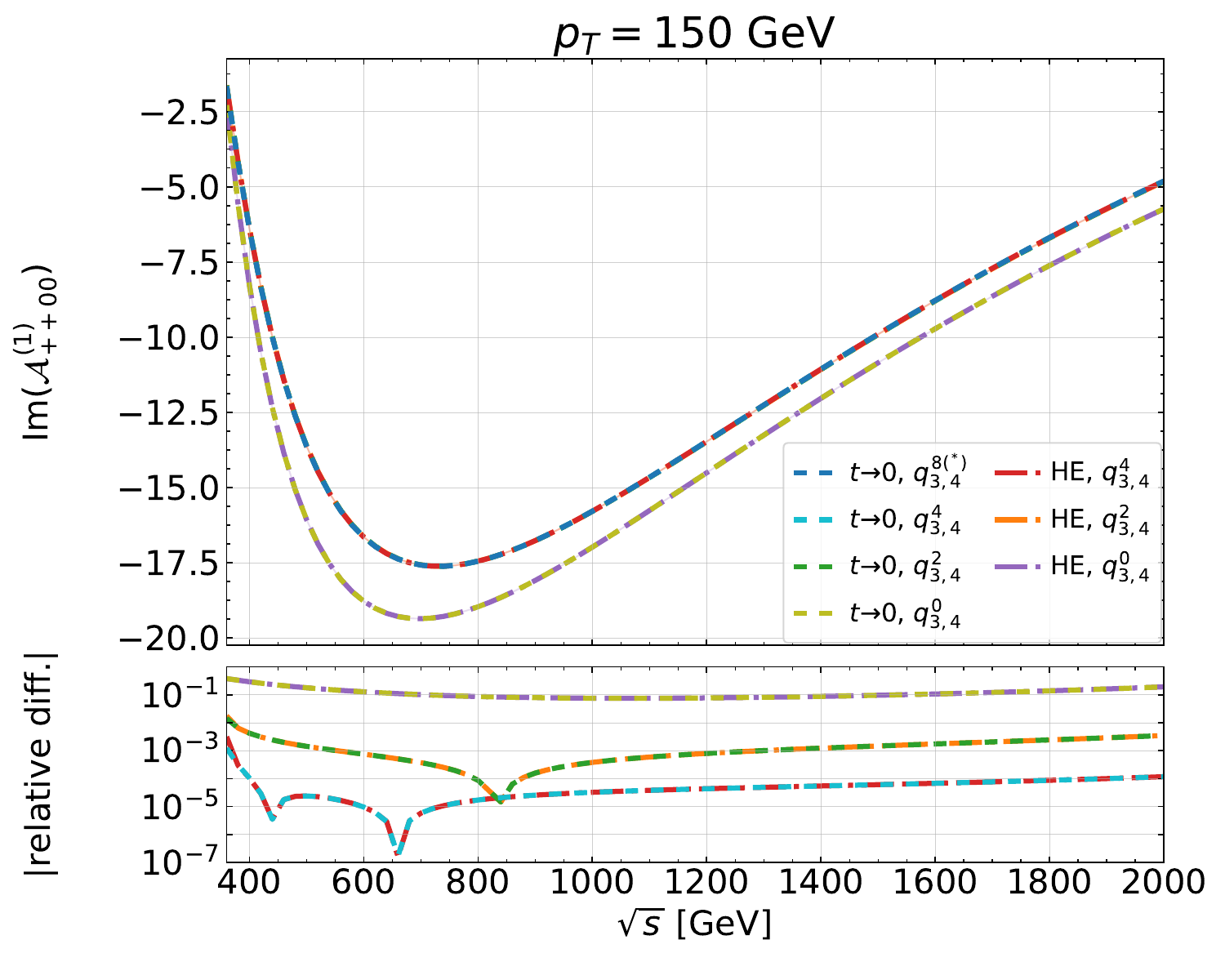}
    \end{tabular}
    \end{center}
  \caption{\label{fig::ggzz2l_pt150}
  Real and imaginary parts of ${\cal A}^{(1)}_{++++}$ and ${\cal A}^{(1)}_{++00}$ as a function of $\sqrt{s}$ for $p_T=150$~GeV and $q_3^2 = q_4^2 = m_Z^2$. High energy and $t\to 0$ expansions are shown including mass corrections up to $q_{3,4}^{\{0,2,4\}}$. Also shown are higher mass corrections up to $q_{3,4}^8$ for the $t\to 0$ expansion with fewer expansion terms in $t$ according to $t^{n_t}(q_3^2)^{n_3}(q_4^2)^{n_4}$ with $n_t+n_3+n_4\leq 4$ denoted by $\star$. Lower panels display the relative difference with respect to the best approximation of the $t\to 0$ expansion.}
\end{figure}

Next, we perform a similar comparison at the two-loop level where we study the differences between the forward and high-energy expansions and the convergence of the $q_{3,4}^2$ series.
In Fig.~\ref{fig::ggzz2l_pt150} we show the real and imaginary parts
of ${\cal A}^{(1)}_{++++}$ and ${\cal A}^{(1)}_{++00}$ for fixed $p_T=150$~GeV as a function of $\sqrt{s}$. The same notation as in the corresponding one-loop plots is used (see Fig.~\ref{fig::ggzz1l_pt150};
this time no exact results are shown). The lower panels show the relative
difference to the best available approximation from the  forward
expansion (shown as dark blue curve in the top panels). We observe a similar
pattern as at one-loop order. There is a 
good agreement between the $t\to0$ and high-energy expansion. Furthermore,
in general the quartic corrections in the final-state masses 
provide contributions in the sub-percent level which suggests
a good convergence.

\begin{figure}[t]
    \begin{center}
    \begin{tabular}{cc}
    \includegraphics[width=.45\textwidth]{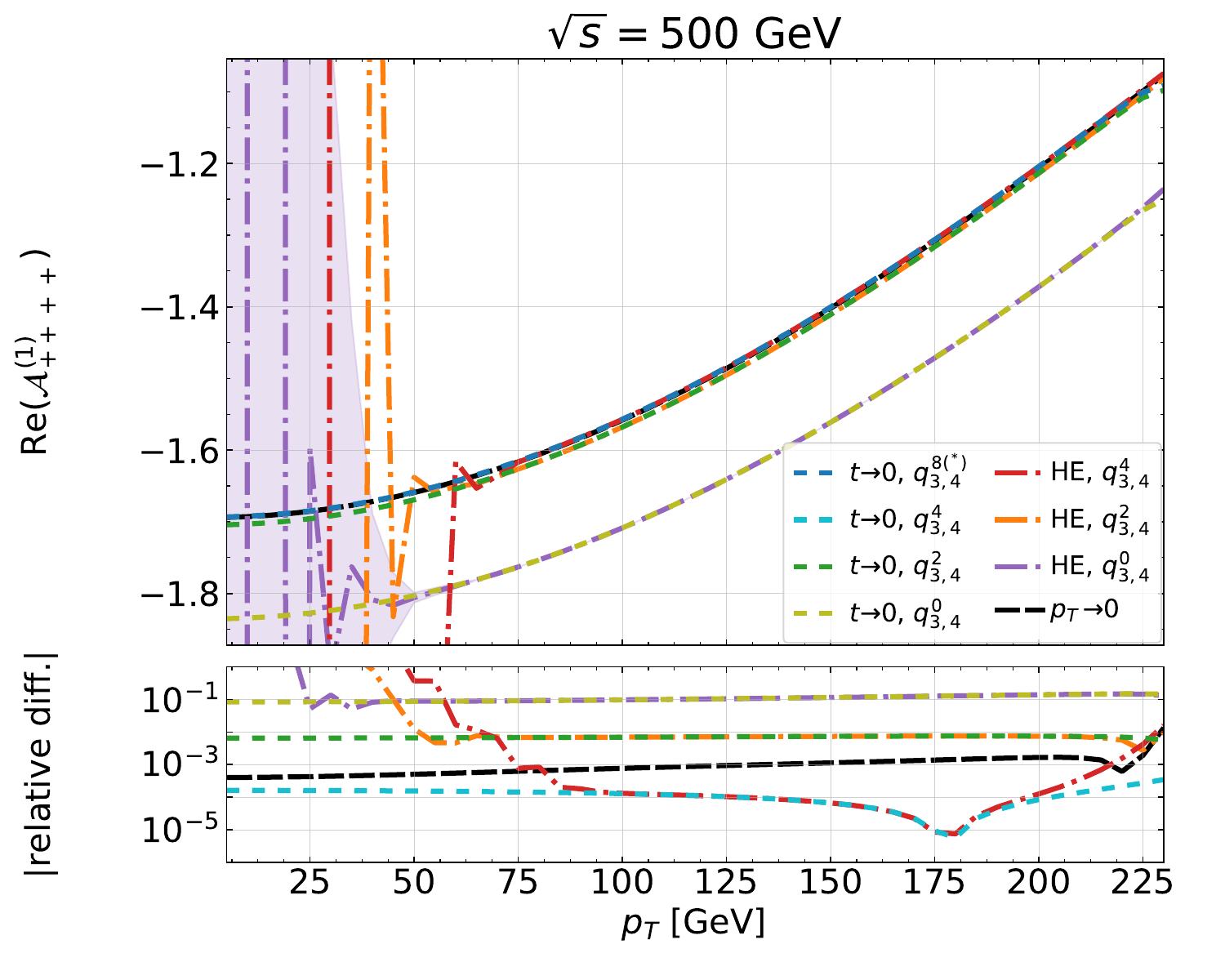} &
    \includegraphics[width=.45\textwidth]{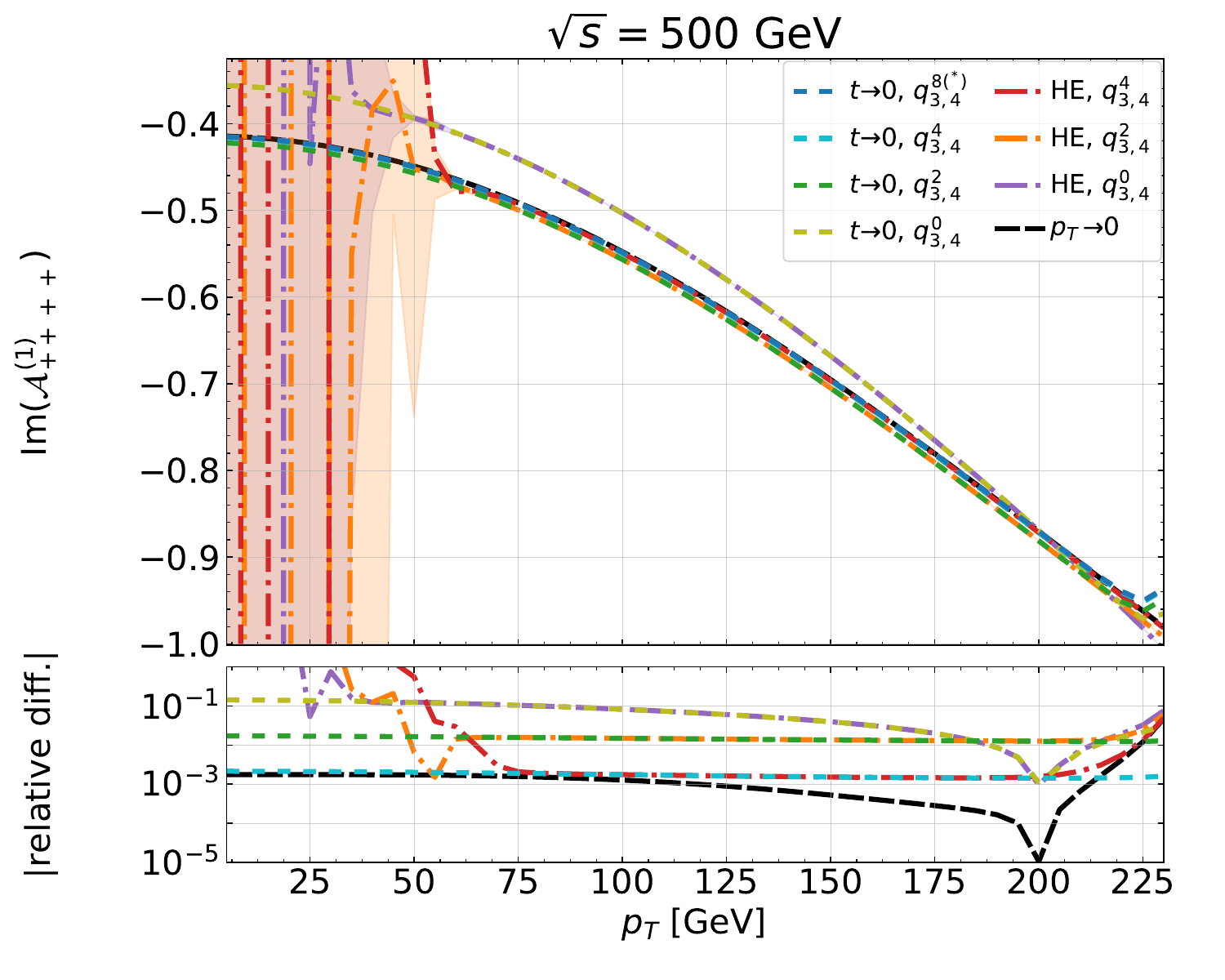} \\
    \includegraphics[width=.45\textwidth]{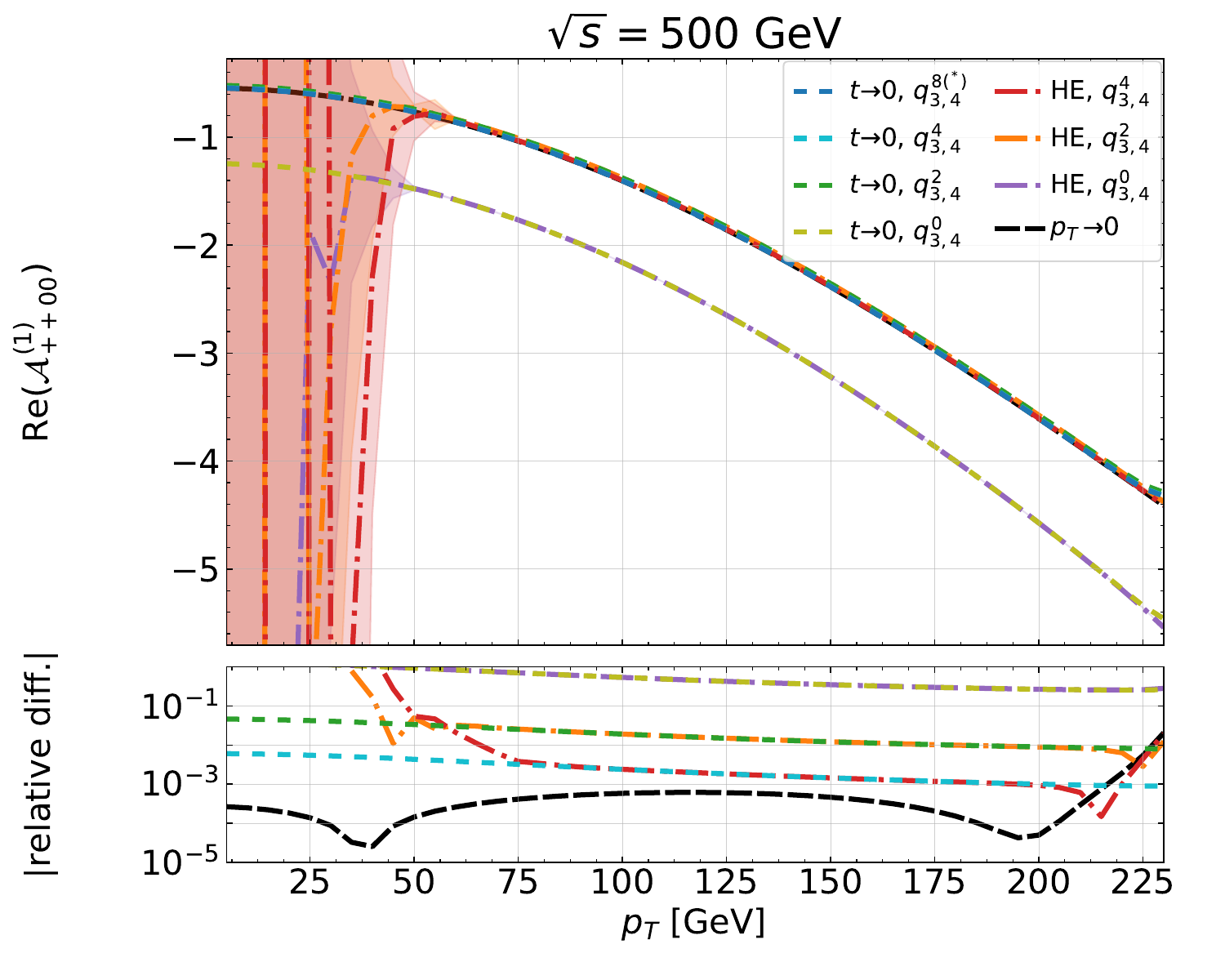} &
    \includegraphics[width=.45\textwidth]{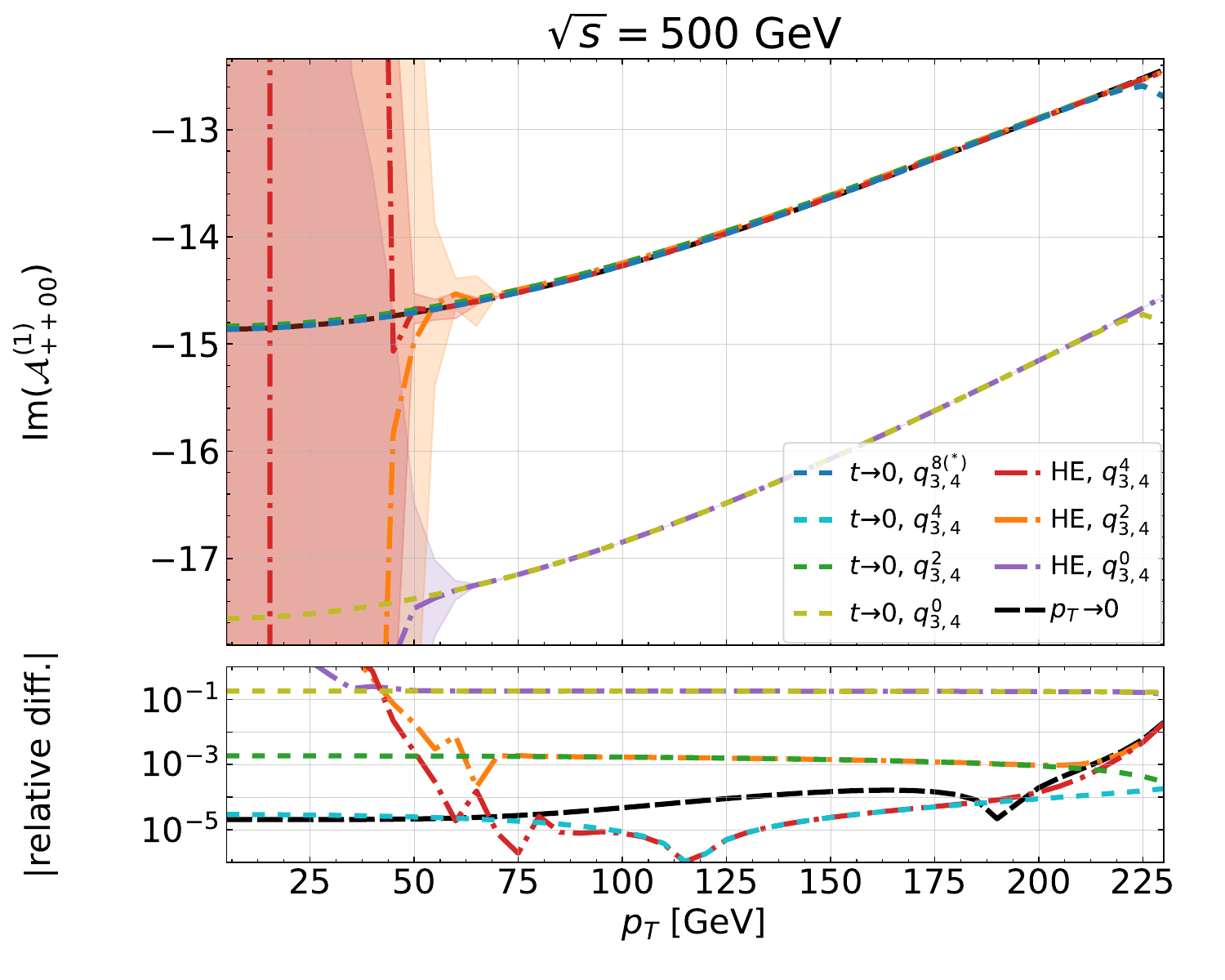}
    \end{tabular}
    \end{center}
  \caption{\label{fig::ggzz2l_ss500}
  Same as Figure \ref{fig::ggzz2l_pt150} but for fixed $\sqrt{s}=500$~GeV as a function of $p_T$.}
\end{figure}

In Fig.~\ref{fig::ggzz2l_ss500} we fix $\sqrt{s}=500$~GeV and vary $p_T$.
In all cases we observe an intermediate region for $p_T$ where both 
expansions agree at the sub-percent level. For small values of $p_T$ 
the Pad\'e procedure applied to the high-energy approximation produces large 
uncertainties. On the other hand, for large $p_T$ 
the forward expansion becomes unstable as can be seen on the
very right of the plots.

In Fig.~\ref{fig::ggzz2l_ss500} we also show the results
from the $p_T$ expansion computed in Ref.~\cite{Degrassi:2024fye}.
We observe good agreement at the level of $10^{-3}$ or below
between the $t$ and $p_T$ expansion.
For larger values of $\sqrt{s}$ (not shown in Fig.~\ref{fig::ggzz2l_ss500}), which allows for larger values of $p_T$,  the results from Ref.~\cite{Degrassi:2024fye} diverge earlier than the results obtained in the current
paper, which includes a much deeper expansion.

\begin{figure}[t]
    \begin{center}
    \begin{tabular}{cc}
    \includegraphics[width=.45\textwidth]{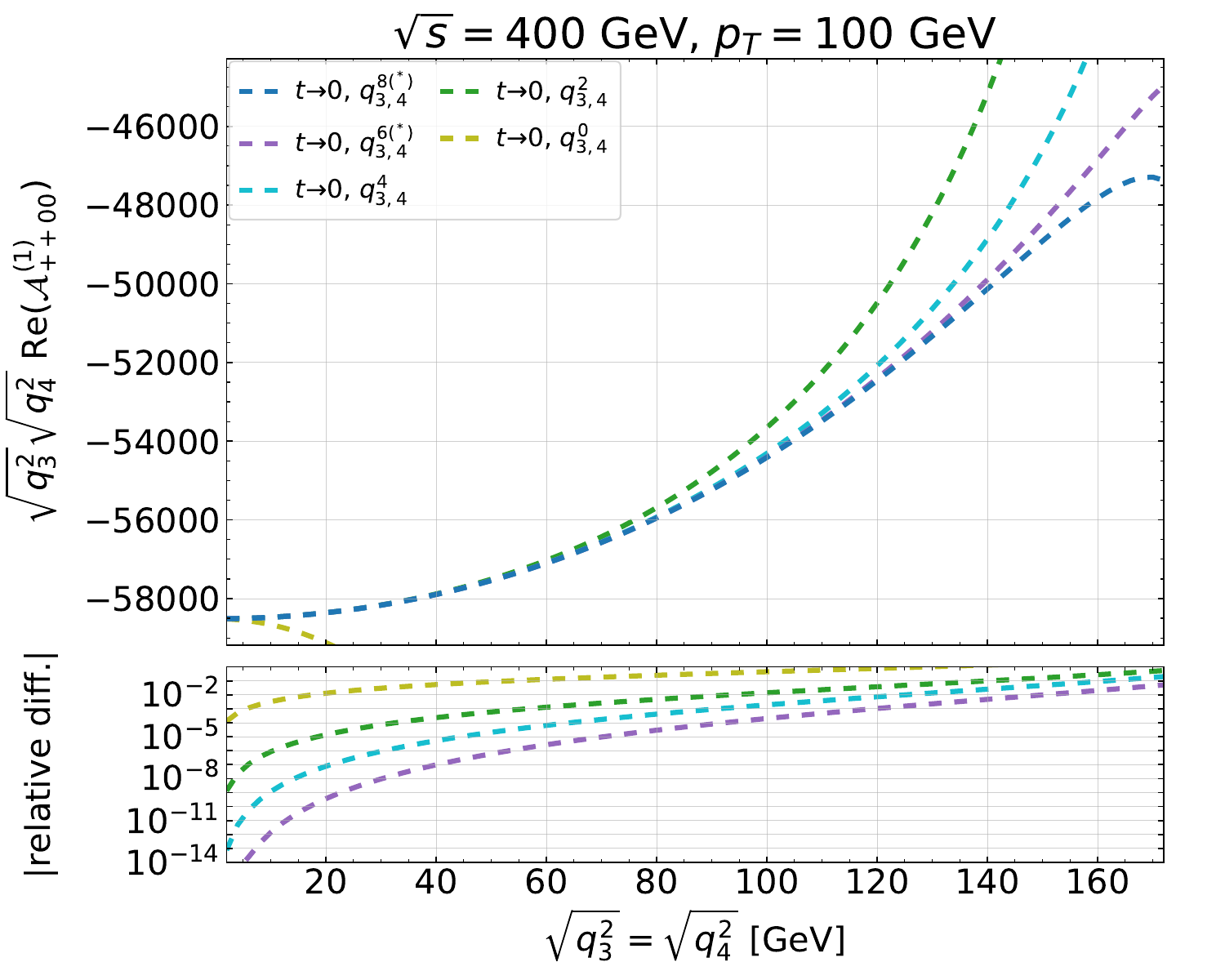} &
    \includegraphics[width=.45\textwidth]{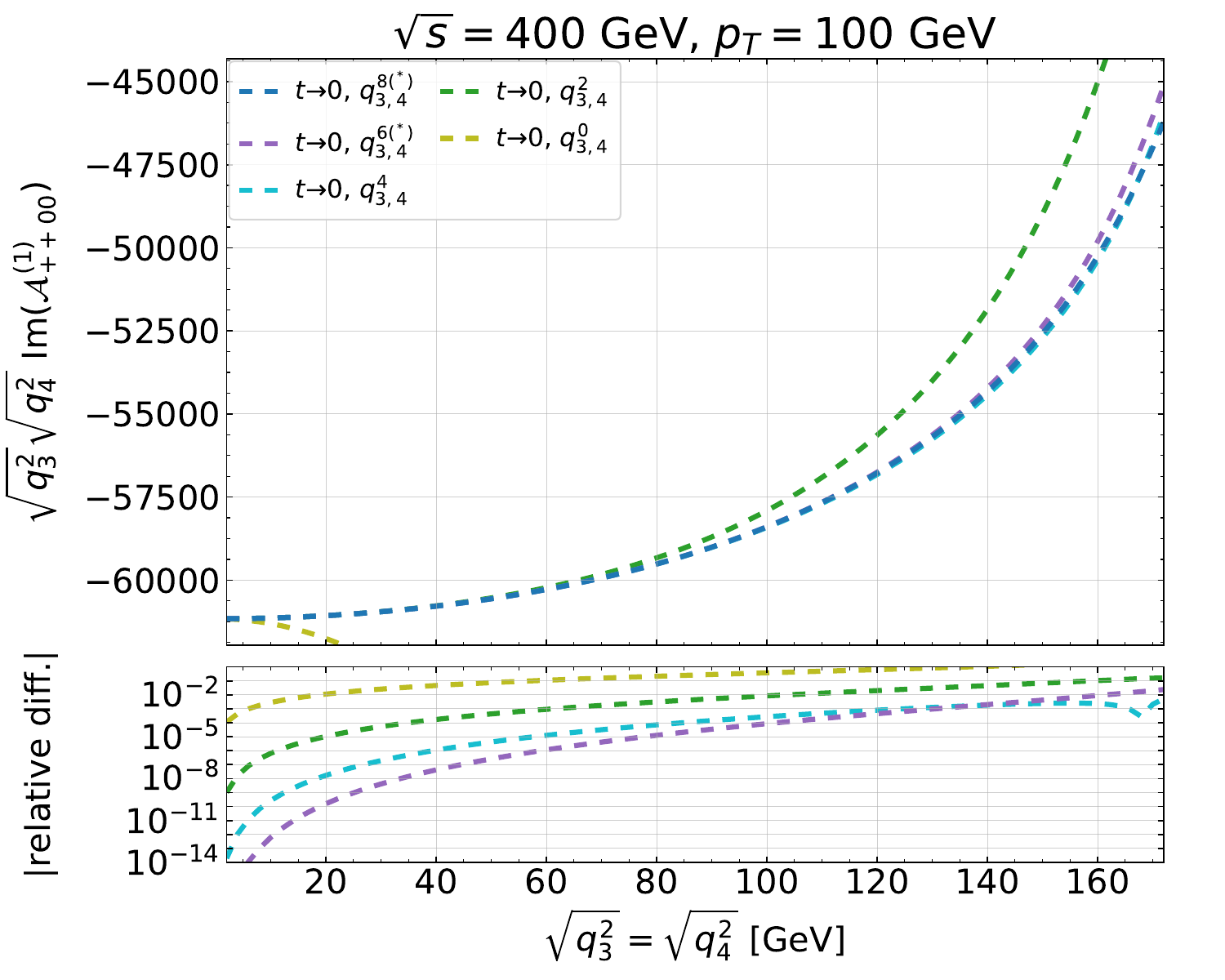}
    \end{tabular}
    \end{center}
  \caption{\label{fig::ggzz2l_q3sq4s}
  Same as Figure \ref{fig::ggzz1l_pt150} but for fixed $\sqrt{s}=400$~GeV and $p_T=100$~GeV as a function of $\sqrt{q_3^2}=\sqrt{q_4^2}$. Here, the relative difference is plotted with
  respect to the $q_{3,4}^{8(\star)}$ curve.}
\end{figure}

In Fig.~\ref{fig::ggzz2l_q3sq4s} we show for $\sqrt{q_3^2}\sqrt{q_4^2}{\cal A}_{++00}$ the
dependence on $\sqrt{q_3^2}=\sqrt{q_4^2}$ for 
$p_T=100$~GeV and $\sqrt{s}=400$~GeV.
We show results for the $t$ expansion (which for this phase-space point is the appropriate choice) including different orders in $q_3^2$ and $q_4^2$.
In the lower panel the relative difference to the best approximation is considered.
In the whole range of $\sqrt{q_3^2}=\sqrt{q_4^2}$ we observe a rapid convergence 
of our expansions which suggests that for small $\sqrt{q_3^2}=\sqrt{q_4^2}$
we have a precision of more than 10 digits which reduces to about
5 digits for $\sqrt{q_3^2}=\sqrt{q_4^2}=m_Z$
and 3~digits (or better) for 
$\sqrt{q_3^2}=\sqrt{q_4^2}=130$~GeV. Thus, our results
conveniently cover
the whole range including off-shell $Z$ boson pair production 
with a virtuality in the vicinity of the Higgs boson mass.

\subsection{\label{sec::ggZZ2lVfin}Finite virtual corrections for $gg \to Z^\star Z^\star$}

In analogy  to double Higgs boson production (see, e.g., 
Refs.~\cite{Heinrich:2017kxx,Davies:2019dfy}) we define
the finite, virtual contribution to the cross section as~\cite{Davies:2020lpf}
\begin{eqnarray}
  \tilde{\mathcal{V}}_{\textnormal{fin}}^{ZZ} &=&
  \frac{1}{64}\left(\frac{\alpha_s}{2\pi}\right)^2\sum_{\lambda_1,\lambda_2,\lambda_3,\lambda_4}
  \left[C_{\lambda_1\lambda_2\lambda_3\lambda_4} + \left(\tilde{\mathcal{A}}^{(0)*}_{\lambda_1\lambda_2\lambda_3\lambda_4} \tilde{\mathcal{A}}^{(1)}_{\lambda_1\lambda_2\lambda_3\lambda_4}+\tilde{\mathcal{A}}^{(0)}_{\lambda_1\lambda_2\lambda_3\lambda_4} \tilde{\mathcal{A}}^{(1)*}_{\lambda_1\lambda_2\lambda_3\lambda_4}\right)\right] \,,
  \nonumber\\
      \label{eq::Vtil}
\end{eqnarray}
where $C_{\lambda_1\lambda_2\lambda_3\lambda_4}$ is defined by
\begin{eqnarray}
C_{\lambda_1\lambda_2\lambda_3\lambda_4} &=& \left|\tilde{\mathcal{A}}^{(0)}_{\lambda_1\lambda_2\lambda_3\lambda_4}\right|^2 C_A
        \left(
        \pi^2 - \log^2\frac{\mu_r^2}{s}
        \right)
        \,
          \label{eq::Ci}
\end{eqnarray}
and $\tilde{\mathcal{A}}^{\{(0),(1)\}}_{\lambda_1\lambda_2\lambda_3\lambda_4}$ are the corresponding one-loop and two-loop ultra-violet finite and infra-red subtracted helicity amplitudes evaluated at $\mu^2=-s$ such that the $\mu_r^2$ dependence
of $\widetilde{\mathcal{V}}_{\textnormal{fin}}^{ZZ}$ is contained in 
$\alpha_s(\mu_r)\equiv\alpha_s^{(5)}(\mu_r)$
and the quantities $C_{\lambda_1\lambda_2\lambda_3\lambda_4}$. For convenience we introduce
the quantity
\begin{eqnarray}
  {\cal V}_{\rm fin}^{ZZ} &=& \frac{\tilde{{\cal V}}_{\rm fin}^{ZZ}}{\alpha_s^2}
                         \,.
\label{eq::Vfin_norm-ZZ}
\end{eqnarray}

\begin{figure}[t]
    \begin{center}
    \begin{tabular}{c}
    \includegraphics[width=.85\textwidth]{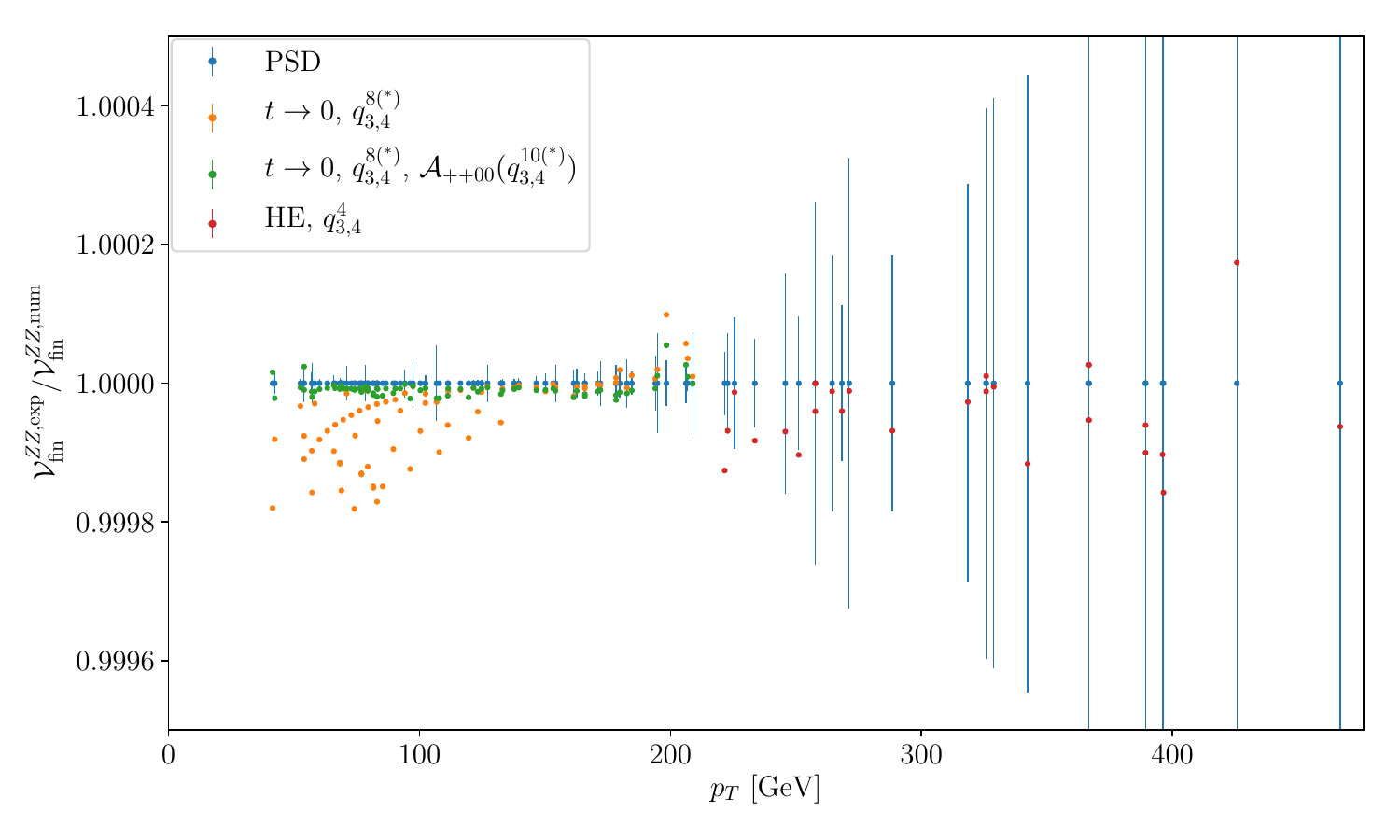}
    \end{tabular}
    \end{center}
    \vspace*{-2em}
  \caption{\label{fig::ggzz_vfin}
  Ratio ${\cal V}_{\rm fin}^{ZZ,\rm exp}/{\cal V}_{\rm fin}^{ZZ,\rm num}$ as a function of $p_T$. In the high-energy region terms up to order $m_t^{100}$ are included. For $t\to 0$ terms up to order $t^{10}$ for $q_{3,4}^{\{0,2,4\}}$ and
  $q_{3,4}^6$ and $q_{3,4}^8$ with a lower expansion depth in $t$ terms are included. In addition, results for $t\to 0$ are also shown with $\mathcal{A}_{++00}$ up to $q_{3,4}^{10}$ with a lower expansion depth in $t$.
  The reference values denoted by ``PSD'' are obtained from Refs.~\cite{Agarwal:2020dye,Agarwal:2024pod}.
  }
\end{figure}

\begin{figure}[t]
    \begin{center}
    \begin{tabular}{c}
    \includegraphics[width=.85\textwidth]{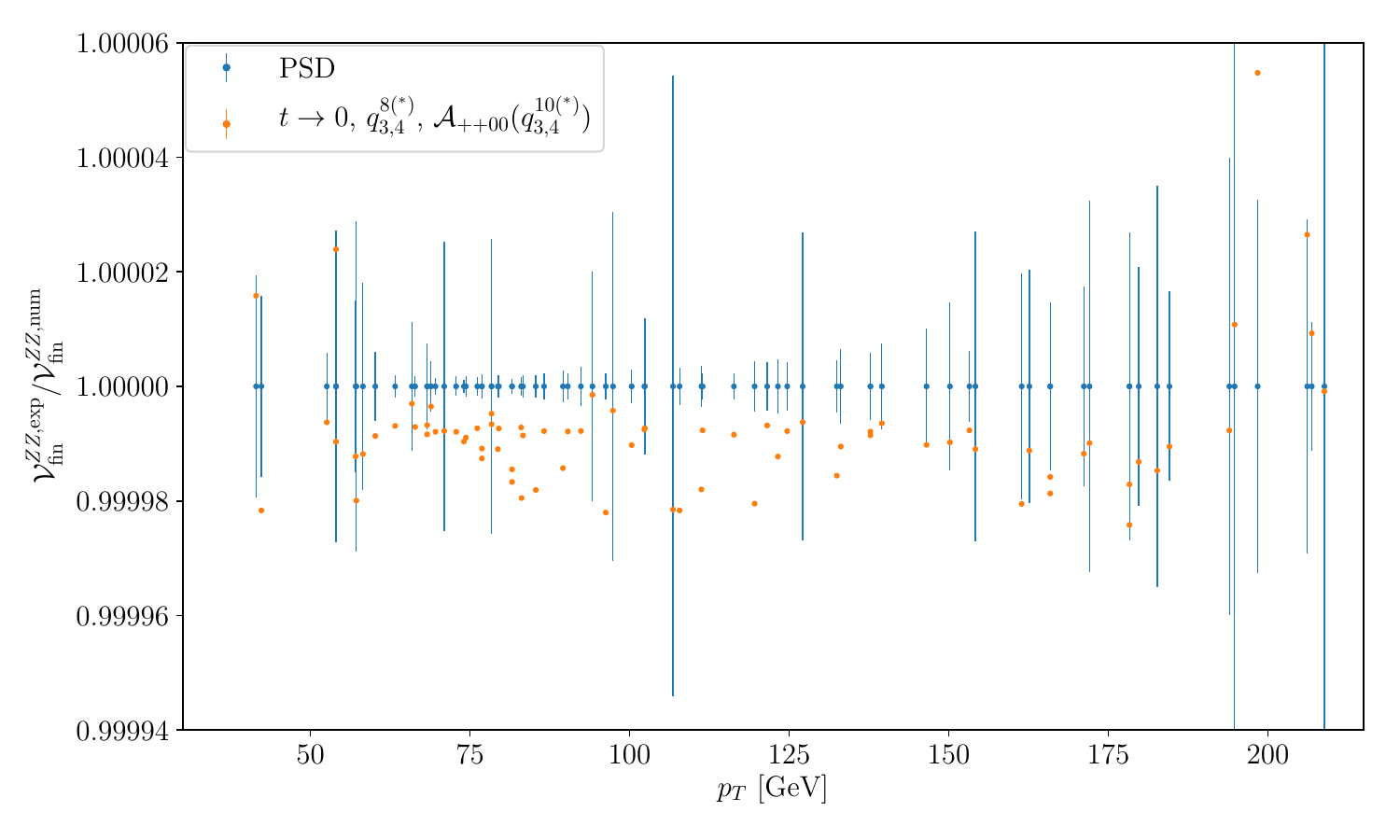}
    \end{tabular}
    \end{center}
    \vspace*{-2em}
  \caption{\label{fig::ggzz_vfin_mag} Magnification of data shown in Fig.~\ref{fig::ggzz_vfin}
  for $p_T\le 210$~GeV. Results are shown for the best approximation of the $t\to 0$ expansion.
  The reference values denoted by ``PSD'' are obtained from Refs.~\cite{Agarwal:2020dye,Agarwal:2024pod}.
  }
  \label{fig:zzlowpt}
\end{figure}

In Fig.~\ref{fig::ggzz_vfin} we compare 
results for ${\cal V}_{\rm fin}^{ZZ}$ with $\mu_r^2=s$
obtained on the basis of our expansions
to numerical results from Refs.~\cite{Agarwal:2020dye,Agarwal:2024pod} based on {\tt pySecDec}~\cite{Borowka:2017idc,Borowka:2018goh,Heinrich:2021dbf} as a ratio ${\cal V}_{\rm fin}^{ZZ,\rm exp}/{\cal V}_{\rm fin}^{ZZ,\rm num}$.
The plot contains $132$
data points over a broad range of $p_T$ and $\sqrt{s}$. For the construction of ${\cal V}_{\rm fin}^{ZZ,\rm exp}$ we use the exact one-loop amplitudes and all available expansion terms for the two-loop helicity amplitudes. The orange dots correspond to the $t$ expansion including the expansion depth
described in Section~\ref{subsub::texp_MI}.
To obtain the green dots 
we additionally include all terms with $n_t+n_3+n_4\leq 5$ for $\mathcal{A}_{++00}$, which provides the numerically dominant contribution. The improvement is clearly visible. The results of the high-energy expansion are shown as red dots. 
We find, for all data points, an agreement below $0.02\%$. Below $p_T=210~{\rm GeV}$, where we use the $t\to 0$ expansion, we find an agreement of at least $0.005\%$
as can be seen from Fig.~\ref{fig:zzlowpt}. From this plot one can see that ${\cal V}_{\rm fin}^{ZZ,\rm exp}$ approximates the exact result impressively well.

\clearpage

\subsection{\label{sec::ggaa} Two-loop results for $gg\to \gamma\gamma$}

Our amplitudes for $gg \to Z^* Z^*$ can also be used to obtain 
results for the process $gg\to \gamma\gamma$ by setting
\begin{align}
    v_t &\to e Q_t~, & a_t &\to 0 ~, & q_3^2&=q_4^2=0 \,.
\end{align}
The triangle and double triangle contributions vanish in this case. 
For on-shell photon production the helicity amplitudes in Eq.~\eqref{eq:ZZhelicity}
with $\lambda_i=0$ vanish and we are left with the eight 
helicity amplitudes, which can be reduced to four linearly 
independent ones using the symmetries in Eq.~(\ref{eq:ZZsymmetry}).

The two-loop amplitudes have been calculated before with numerical methods 
in Refs.~\cite{Maltoni:2018zvp,Becchetti:2023wev,Becchetti:2023yat}. More
recently, analytic results have also become available~\cite{Becchetti:2025rrz,Ahmed:2025osb}.
The full process of di-photon production involving an internal heavy top quark 
also has contributions from $q \bar{q}$ initial states, for which we do not provide 
amplitudes.

\begin{figure}[t]
    \begin{center}
    \begin{tabular}{cc}
    \includegraphics[width=.45\textwidth]{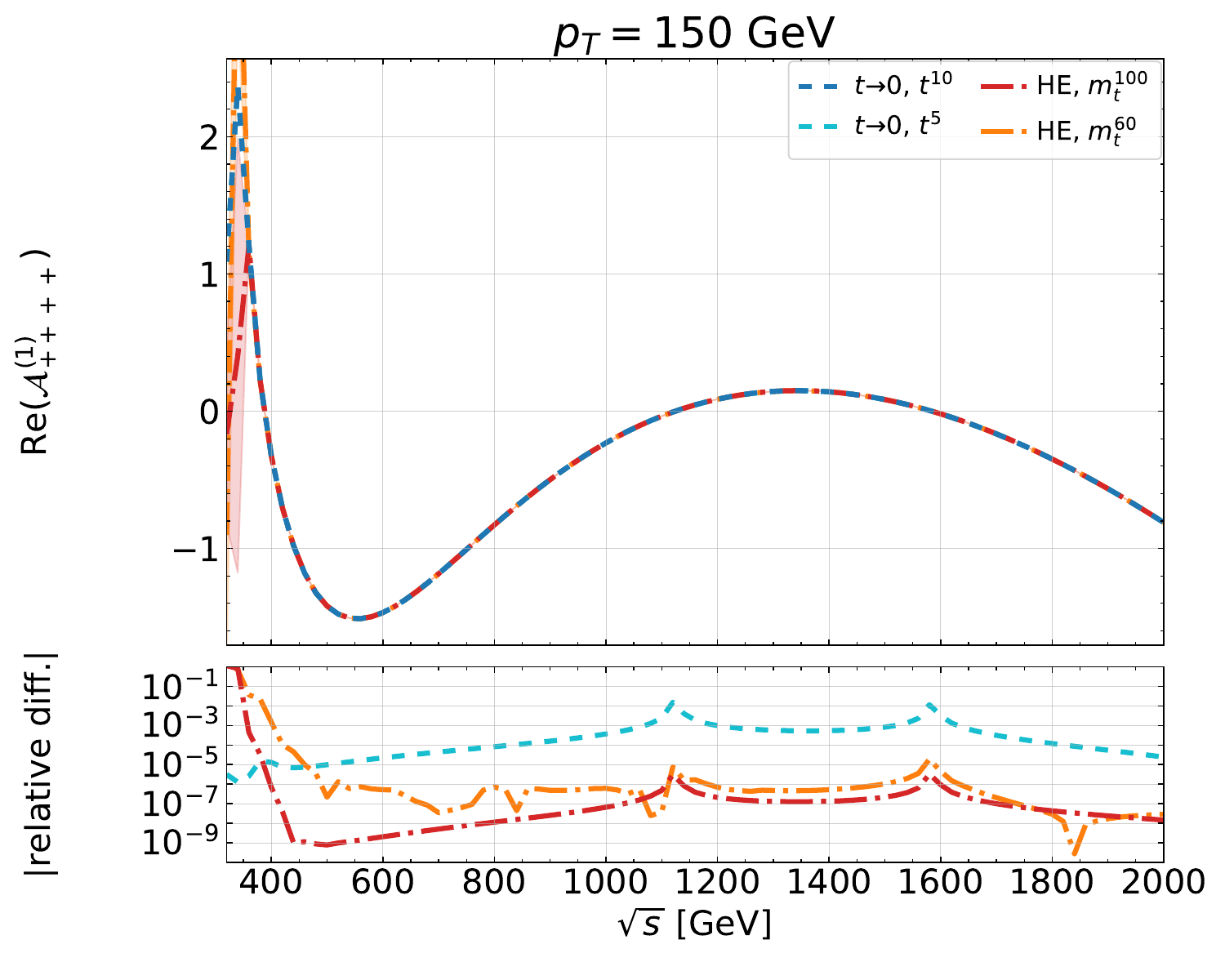} &
    \includegraphics[width=.45\textwidth]{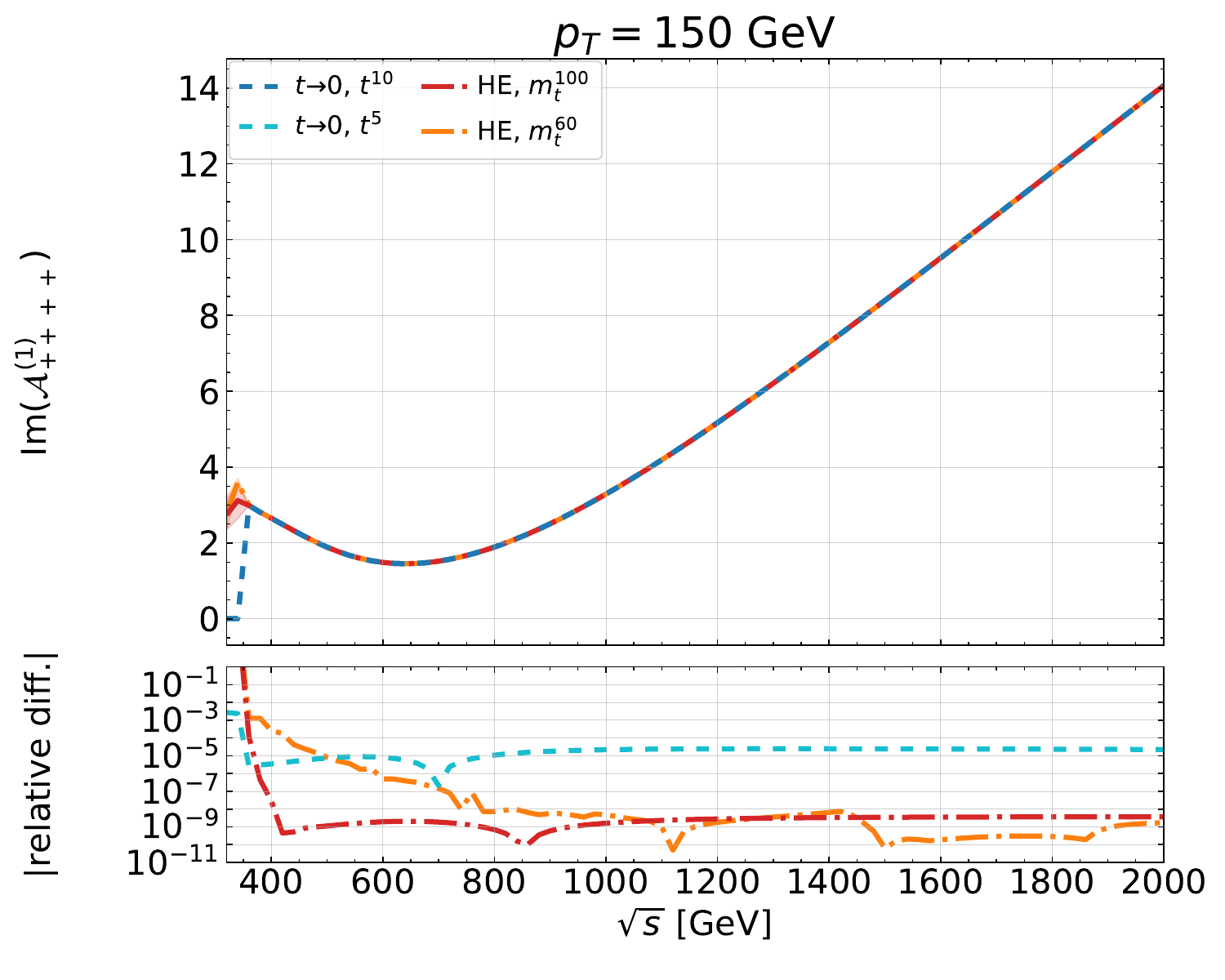} \\
    \includegraphics[width=.45\textwidth]{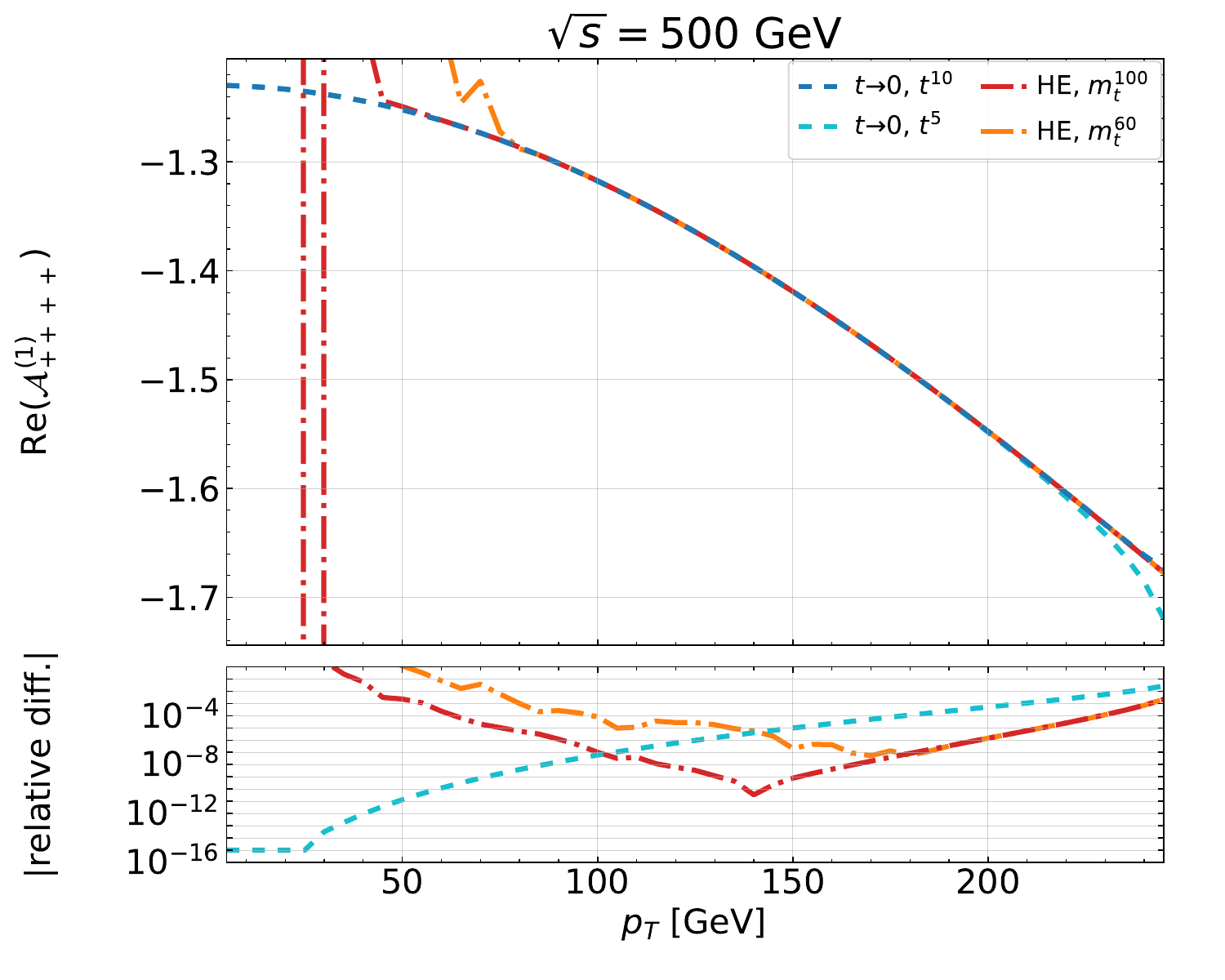} &
    \includegraphics[width=.45\textwidth]{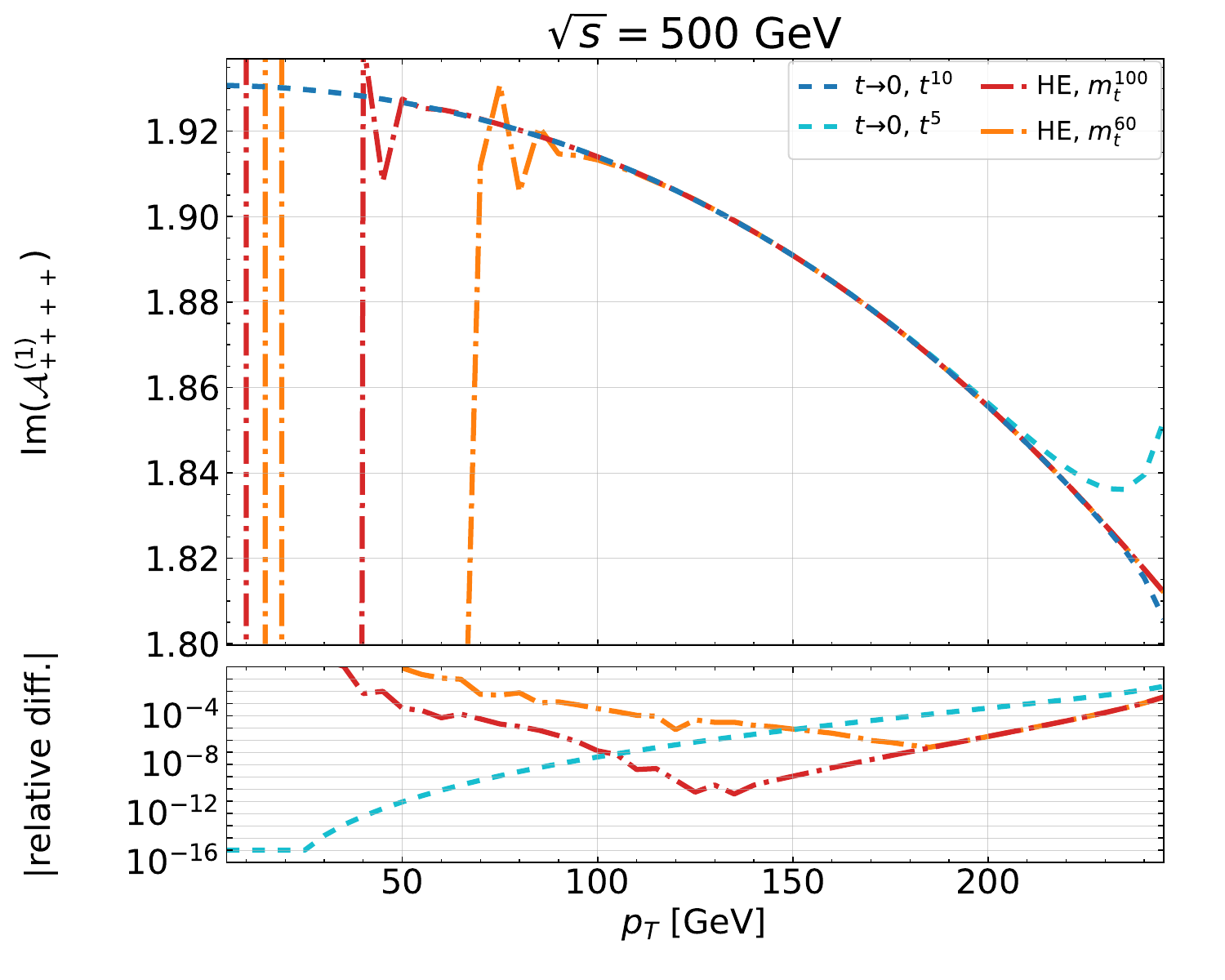}
    \end{tabular}
    \end{center}
  \caption{\label{fig::ggaa2l}
  Real and imaginary parts of ${\cal A}^{(1)}_{++++}$ for $gg\to\gamma\gamma$ for $p_T=150$~GeV as a function of $\sqrt{s}$ and for $\sqrt{s}=500$~GeV as a function of $p_T$. High-energy and $t\to 0$ expansions are shown with expansions terms up to $m_t^{\{60,100\}}$ and $t^{\{5,10\}}$, respectively. Lower panels display the relative difference with respect to the best approximation of the $t\to 0$ expansion.}
\end{figure}

For illustration we show in Fig.~\ref{fig::ggaa2l} the two-loop results for the 
helicity amplitude ${\cal A}_{++++}$ for fixed-$p_T$ (top panels)
and fixed-$\sqrt{s}$ (bottom panels). They show a similar behaviour as
the corresponding plots for $gg\to ZZ$ (see Figs.~\ref{fig::ggzz2l_pt150}
and~\ref{fig::ggzz2l_ss500}). Note, however, that the agreement between
the high-energy and $t\to0$ expansion is significantly better and
reaches in some case about 10~digits as can be seen in the plots
showing the relative difference. This improvement is due to the fact that the expansion in $q_3^2$ and $q_4^2$, which limits the accuracy of the approximation, is not required in this case.

We note that our results can readily be used to consider off-shell photon production 
by considering $q_3^2, q_4^2 \neq 0$.
Since we have a joint expansion in $q_3^2$ and $q_4^2$ the photons cannot be arbitrarily off-shell. 
However, we know that the expansion provides precise results at least up to $q_{3}^2,q_{4}^2 \sim (125~\mbox{GeV})^2$ from the $gg \to HH$~\cite{Davies:2023vmj}
and $gg\to Z^\star Z^\star$ (see Fig.~\ref{fig::ggzz2l_q3sq4s}) amplitudes.
We expect the quality of the expansion to behave similarly
to the one shown in Figs.~\ref{fig::ggzz1l_q3sq4s} and \ref{fig::ggzz2l_q3sq4s}, which suggest that we, 
approximately, loose about one digit of accuracy for every $20$\,GeV increase in $q_{3}^2$ and
$q_{4}^2$.
In this case the full set of helicity amplitudes
(see Eq.~(\ref{eq::ggZZ_hel})) contributes.

\subsection{\label{sec::ggZa} Two-loop results for $gg\to Z\gamma$}

The amplitudes for $gg \to Z^* Z^*$ can additionally be used to obtain 
results for the process $gg\to Z \gamma$, where 
one or both final state bosons are either on- or off-shell.
Here we have to set 
\begin{align}
    v_t^2 &\to \frac{e^2 Q_t}{2 s_w c_w}(I_t^3 -2 Q_t s_w^2) ~, & a_t^2 &\to 0 ~.
\end{align}
The triangle and double triangle contributions also vanish in this case.
For on-shell photon production the helicity amplitudes in Eq.~\eqref{eq:ZZhelicity}
with $\lambda_4=0$ vanish.

The process mediated by massless quarks has been calculated in Refs.~\cite{Gehrmann:2013vga,Grazzini:2015nwa,Buccioni:2023qnt} and matched to parton showers in Refs.~\cite{Lombardi:2020wju,Wiesemann:2020gbm}. 
In the latter case the gluon fusion channel was found to be small, cf.~Ref.~\cite{Grazzini:2015nwa}, and neglected for the numerical studies. 
It would be interesting to study the impact of massive quarks on the differential distributions. 

\begin{figure}[t]
    \begin{center}
    \begin{tabular}{cc}
    \includegraphics[width=.45\textwidth]{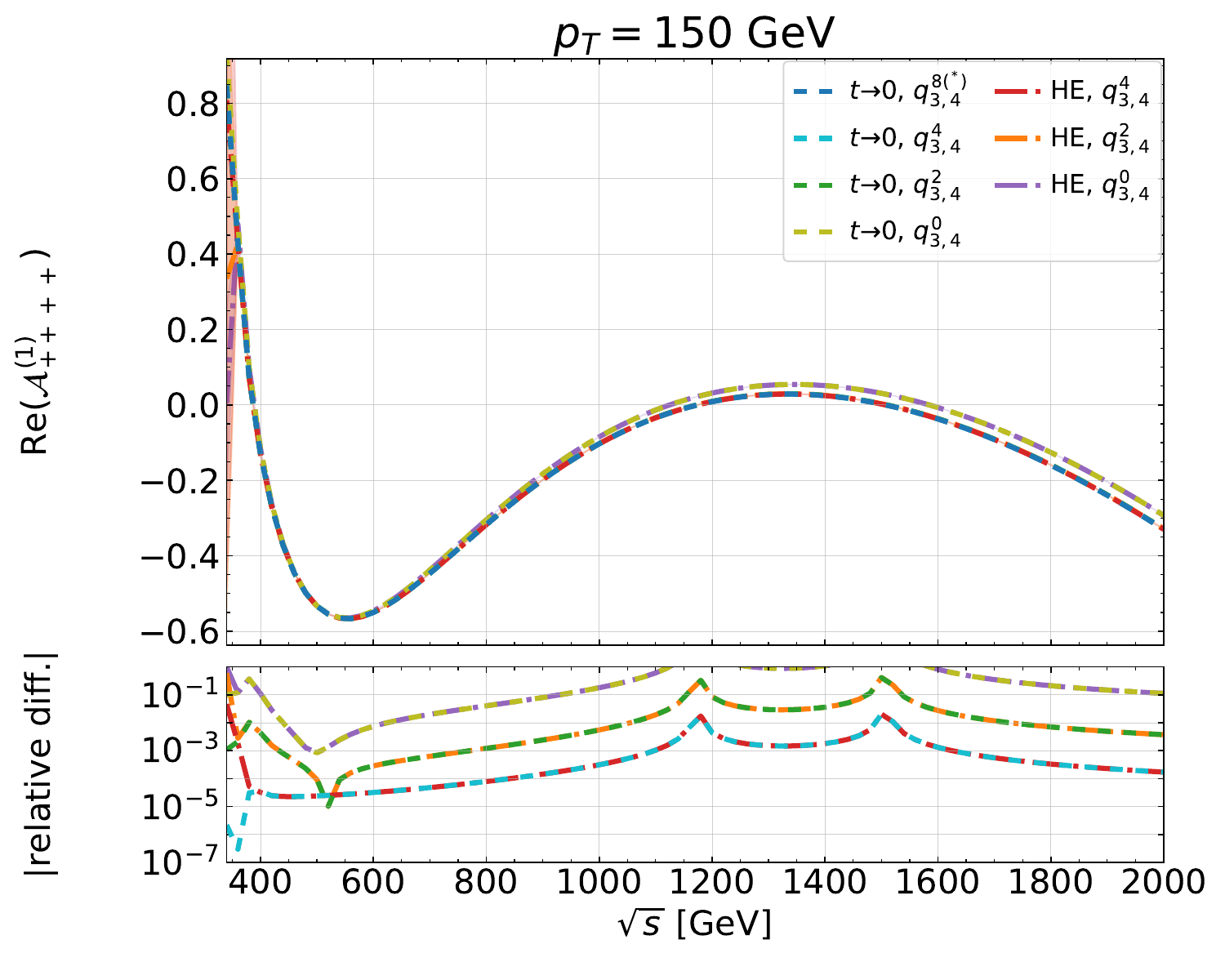} &
    \includegraphics[width=.45\textwidth]{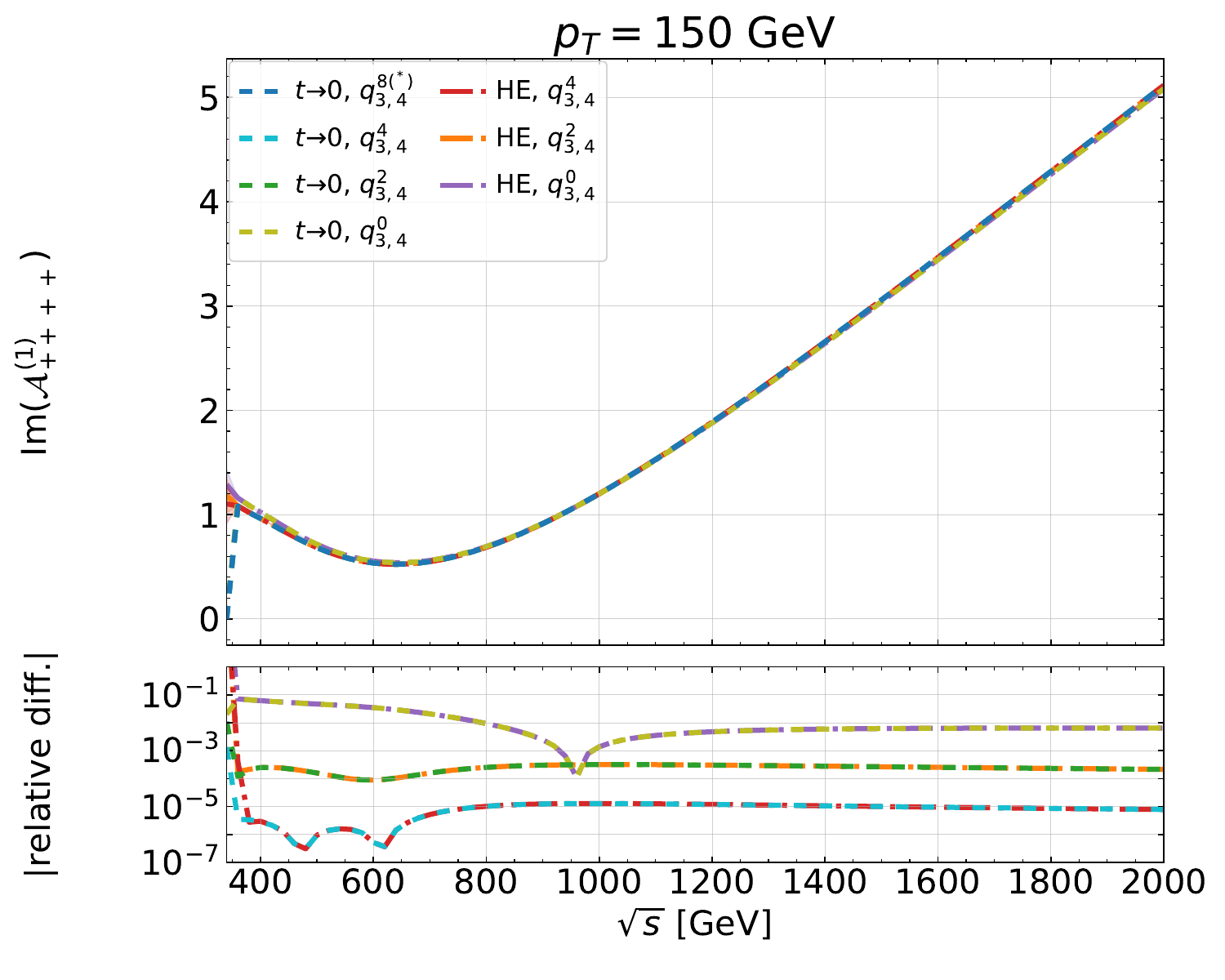} \\
    \includegraphics[width=.45\textwidth]{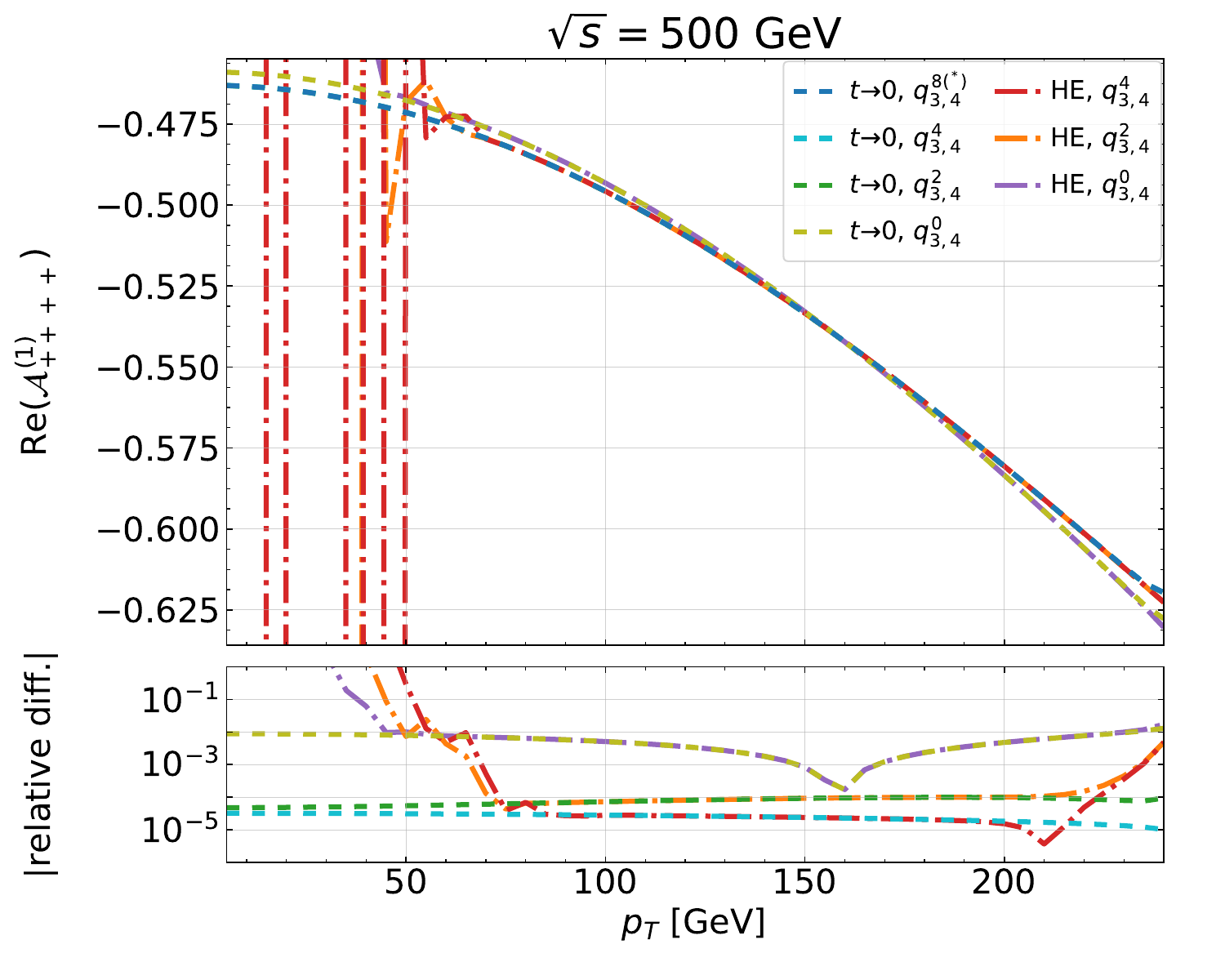} &
    \includegraphics[width=.45\textwidth]{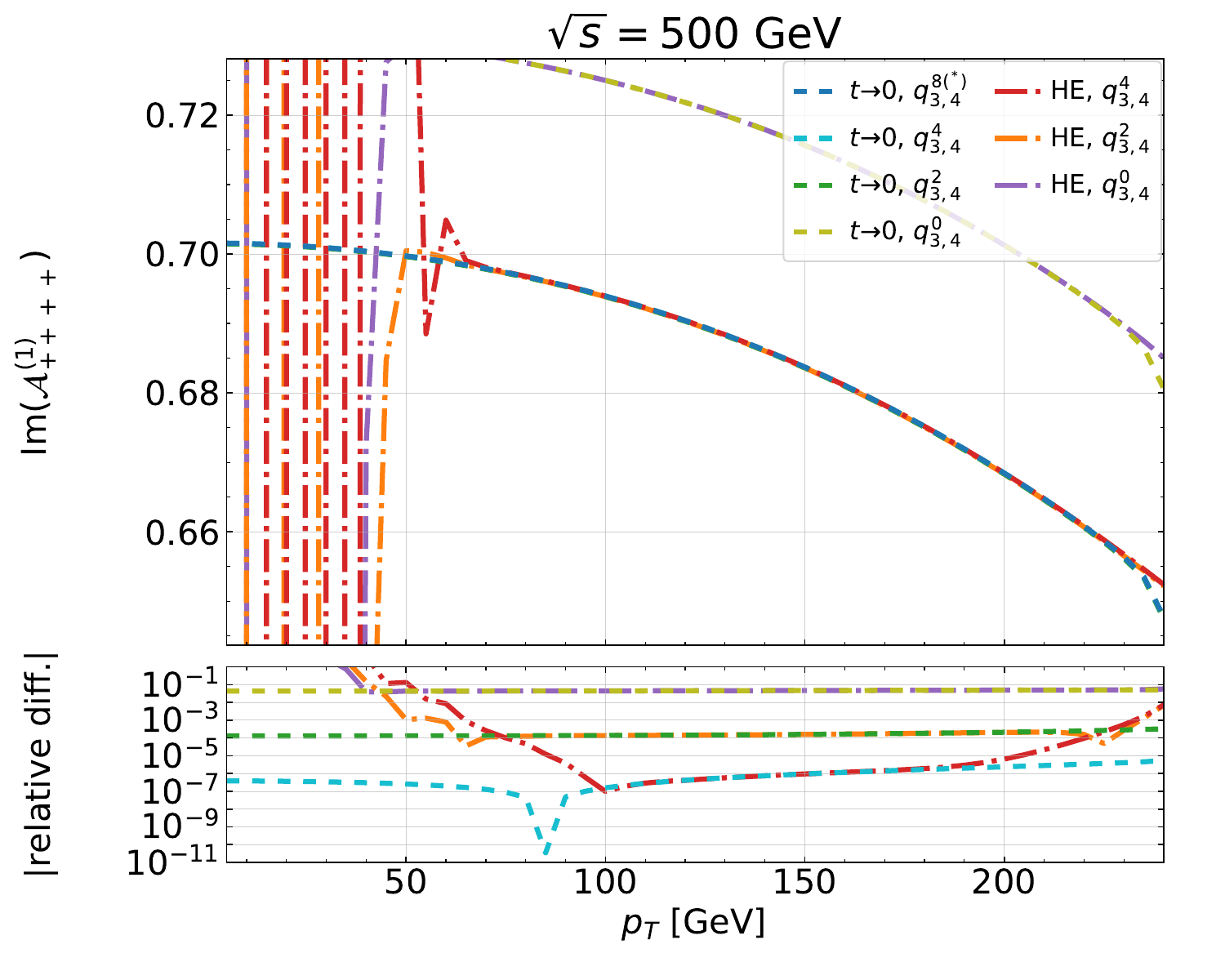}
    \end{tabular}
    \end{center}
  \caption{\label{fig::ggza2l}
  Real and imaginary parts of ${\cal A}^{(1)}_{++++}$ for $gg\to Z\gamma$ for $p_T=150$~GeV as a function of $\sqrt{s}$ and for $\sqrt{s}=500$~GeV as a function of $p_T$. High-energy and $t\to 0$ expansions are shown including mass corrections up to $q_{3,4}^{\{0,2,4\}}$. Also shown are higher mass corrections up to $q_{3,4}^8$ for the $t\to 0$ expansion with fewer expansion terms in $t$ according to $t^{n_t}(q_3^2)^{n_3}(q_4^2)^{n_4}$ with $n_t+n_3+n_4\leq 4$ denoted by $\star$. Lower panels display the relative difference with respect to the best approximation of the $t\to 0$ expansion.}
\end{figure}

We show the result for the two-loop contributions to the helicity amplitude ${\cal A}_{++++}$ in Fig.~\ref{fig::ggza2l} for illustration.

\clearpage

\section{\label{sec::ggZH}$gg\to ZH$}

\subsection{Amplitude}

We decompose the amplitude for $g(q_1)g(q_2) \to Z(q_3)H(q_4)$ as a linear combination of form factors~\cite{Kniehl:1990iva},
\begin{align}
    A &(q_1,q_2,q_3) = i \delta_{a b} \frac{\sqrt{2} G_F m_Z}{s} \frac{\alpha_s (\mu)}{\pi} A^{\mu \nu \rho} (q_1, q_2, q_3) \epsilon_\mu^a(q_1) \epsilon_\nu^b(q_2) \epsilon_\rho^*(q_3) , \nonumber\\
   &A^{\mu \nu \rho}(q_1,q_2,q_3) = \nonumber\\
     &\Biggl\{   
    \left( \frac{s}{2} \varepsilon^{\mu \nu \rho \alpha} q_{2 \alpha}  - q_2^\mu \varepsilon^{\nu \rho \alpha \beta} q_{1 \alpha} q_{2 \beta}\right) F_1(t,u) - 
    \left( \frac{s}{2} \varepsilon^{\mu \nu \rho \alpha} q_{1 \alpha} - q_1^\nu \varepsilon^{\mu \rho \alpha \beta} q_{1 \alpha} q_{2 \beta}\right) F_1(u,t) \nonumber\\
    & + \left(q_3^\mu + \frac{m_Z^2-t}{s} q_2^\mu \right) \varepsilon^{\nu \rho \alpha \beta} q_{2 \alpha} \left[ q_{1 \beta} F_2(t,u) + q_{3\beta} F_3 (t,u)\right] \nonumber\\
    & + \left(q_3^\nu + \frac{m_Z^2-u}{s} q_1^\nu \right) \varepsilon^{\mu \rho \alpha \beta} q_{1 \alpha} \left[ q_{2 \beta} F_2(u,t) + q_{3\beta} F_3 (u,t)\right] \nonumber\\
    & + \left(\frac{s}{2} \varepsilon^{\mu \nu \rho \alpha} q_{3 \alpha} - q_2^\mu \varepsilon^{\nu \rho \alpha \beta} q_{1 \alpha} q_{3 \beta} + q_1^\nu \varepsilon^{\mu \rho \alpha \beta}q_{2\alpha} q_{3 \beta} + g^{\mu \nu }\varepsilon^{\rho \alpha \beta \gamma} q_{1 \alpha} q_{2 \beta} q_{3 \gamma} \right) F_4(t,u) \Biggr\}.
    \label{eq:ggzhamp}
\end{align}
and use on-shell kinematics\footnote{Our results
for the form factors have an explicit dependence 
on $m_Z^2$ and $m_H^2$ and thus they
can immediately be applied to the production of off-shell
$Z$ and Higgs bosons.}
specified in Section~\ref{sub::kin}.

We follow Ref.~\cite{Davies:2020drs} and discuss the results for the six independent form factors 
\begin{align}
    F_{12}^+(t,u),  F_{12}^-(t,u), F_2^-(t,u), F_3^+(t,u), F_3^-(t,u), F_4(t,u)
\end{align}
that can be obtained after taking linear combinations of those of Eq.~\eqref{eq:ggzhamp}:
\begin{align}
    F_2^-(t,u) =& F_2 (t,u) - F_2(u,t), &\quad F_2^+(t,u) =& F_2 (t,u) + F_2(u,t),\nonumber\\
    F_{12}^-(t,u) =& F_{12} (t,u) - F_{12}(u,t), &\quad F_{12}^+(t,u) =& F_{12} (t,u) + F_{12}(u,t), \nonumber\\
    F_3^-(t,u) =& F_3 (t,u) - F_3(u,t), &\quad F_3^+(t,u) =& F_3 (t,u) + F_3(u,t)\,,
\end{align}
where
\begin{align}
    F_{12} (t,u) &= F_1(t,u) - \frac{t-m_Z^2}{s} F_2(t,u)\,, \nonumber\\
    F_{12} (u,t) &= F_1(u,t) - \frac{u-m_Z^2}{s} F_2(u,t)\,.
\end{align}
In the following, we only discuss the results for
box diagrams. The triangle diagrams only contribute to $F_{12}^+$.
The contribution from double-triangle diagrams
can be cast into one form factor which is given in the Appendix of Ref.~\cite{Davies:2020drs}.

\subsection{One-loop results for $gg\to ZH$}

\begin{figure}[t]
\begin{subfigure}{0.5\textwidth}
\centering    
\includegraphics[width=1\textwidth,keepaspectratio]{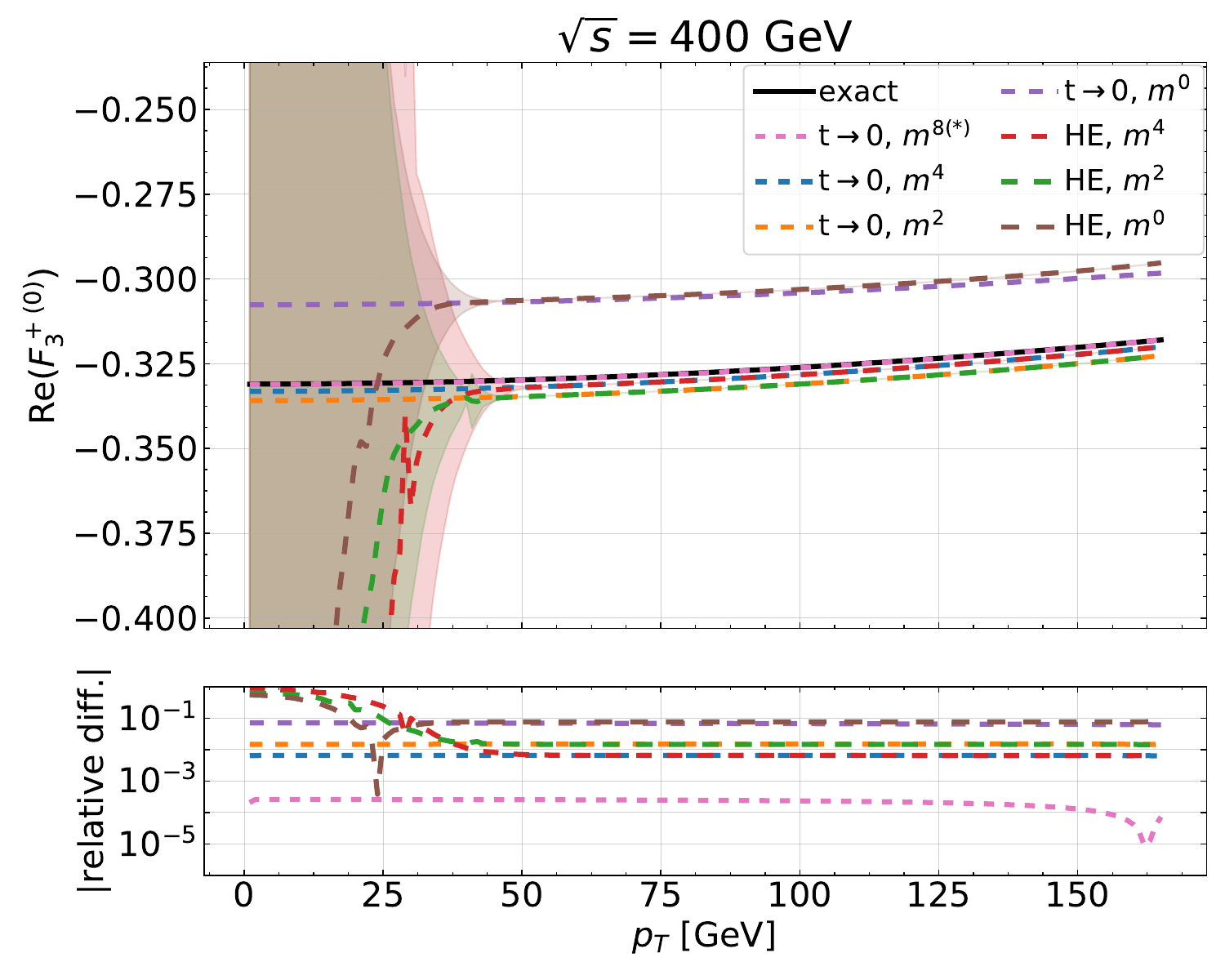}
\end{subfigure}
\begin{subfigure}{0.5\textwidth}
    \centering
    \includegraphics[width=1\textwidth,keepaspectratio]{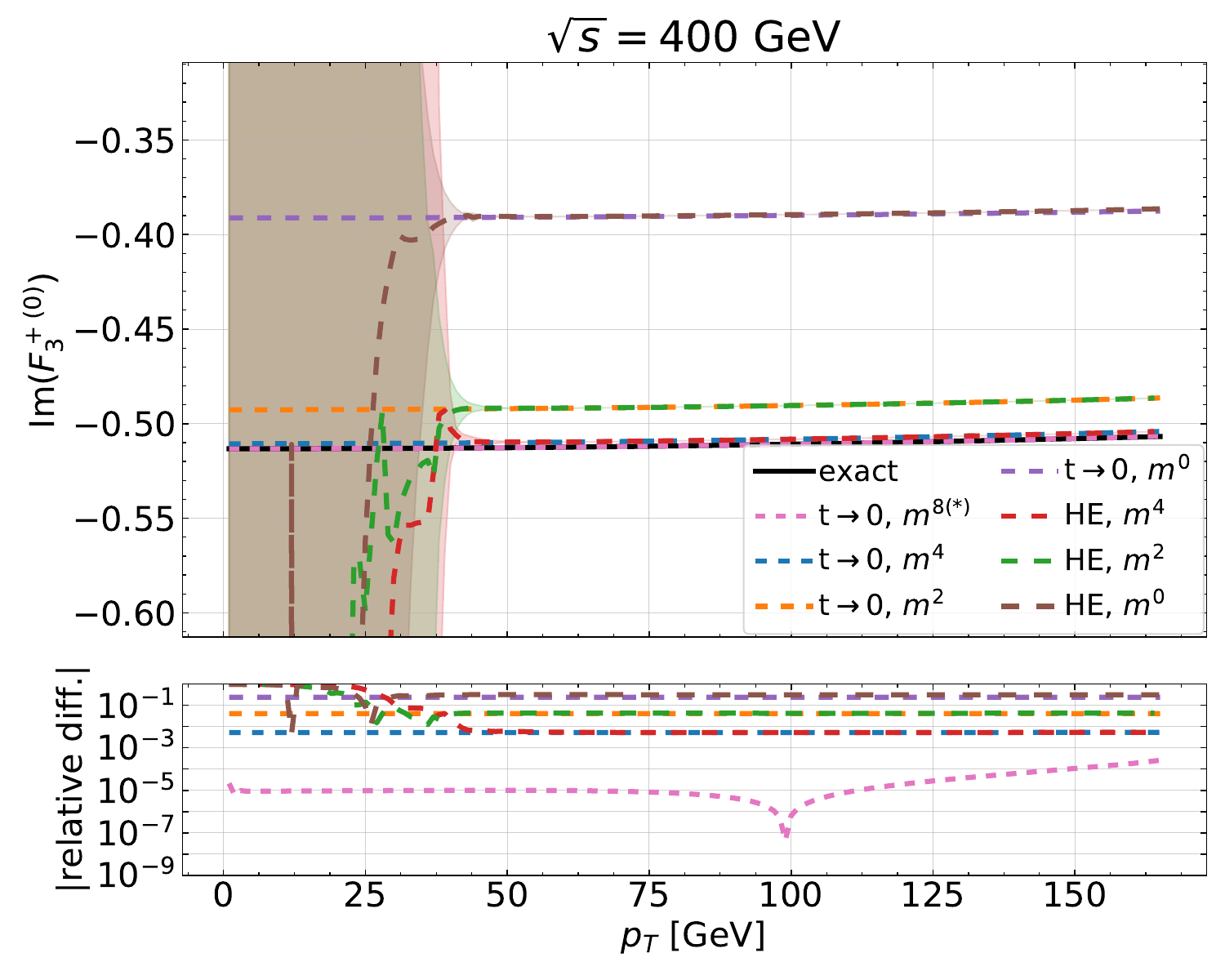}
\end{subfigure}

\begin{subfigure}{0.5\textwidth}
\centering    
\includegraphics[width=1\textwidth,keepaspectratio]{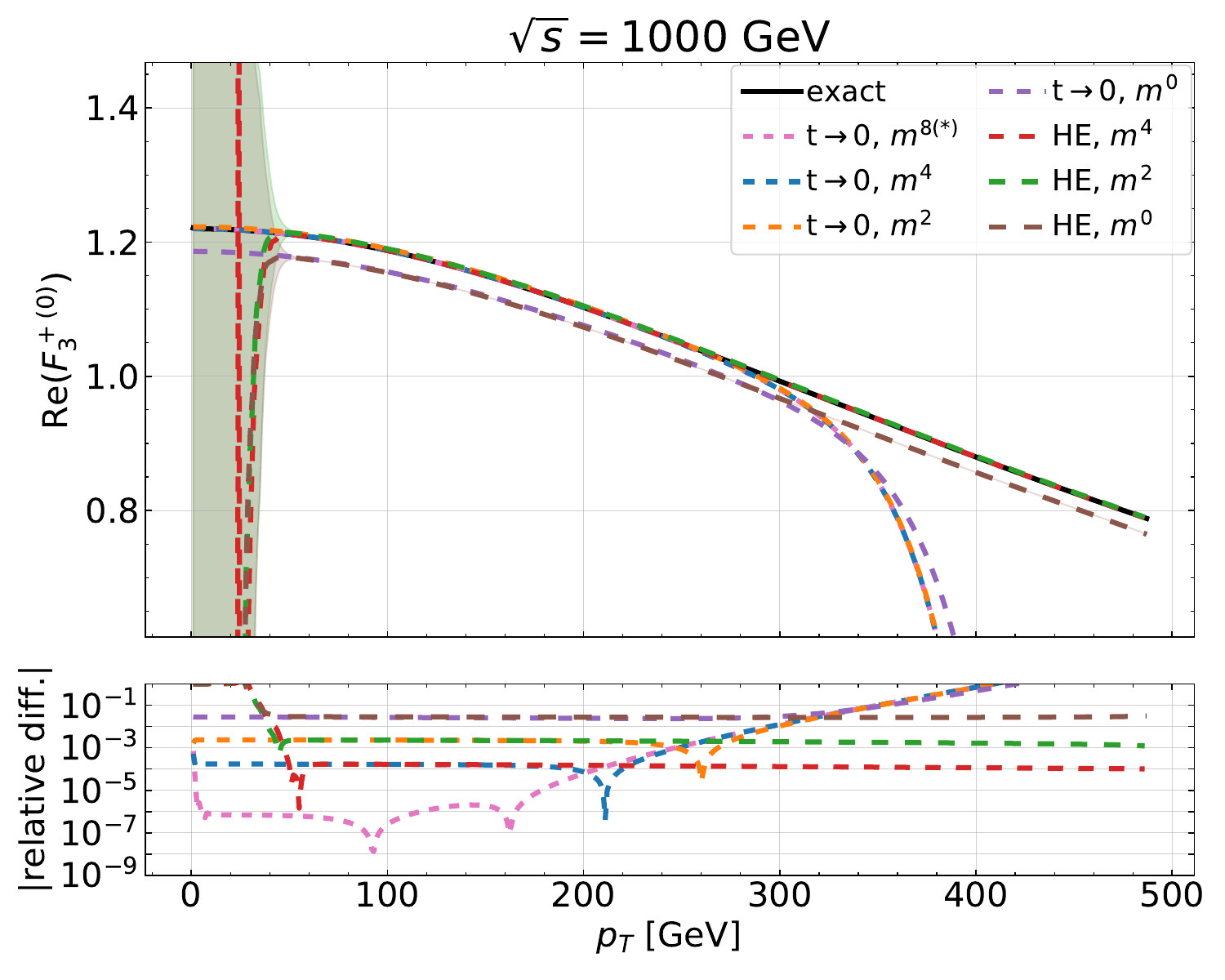}
\end{subfigure}
\begin{subfigure}{0.5\textwidth}
    \centering
    \includegraphics[width=1\textwidth,keepaspectratio]{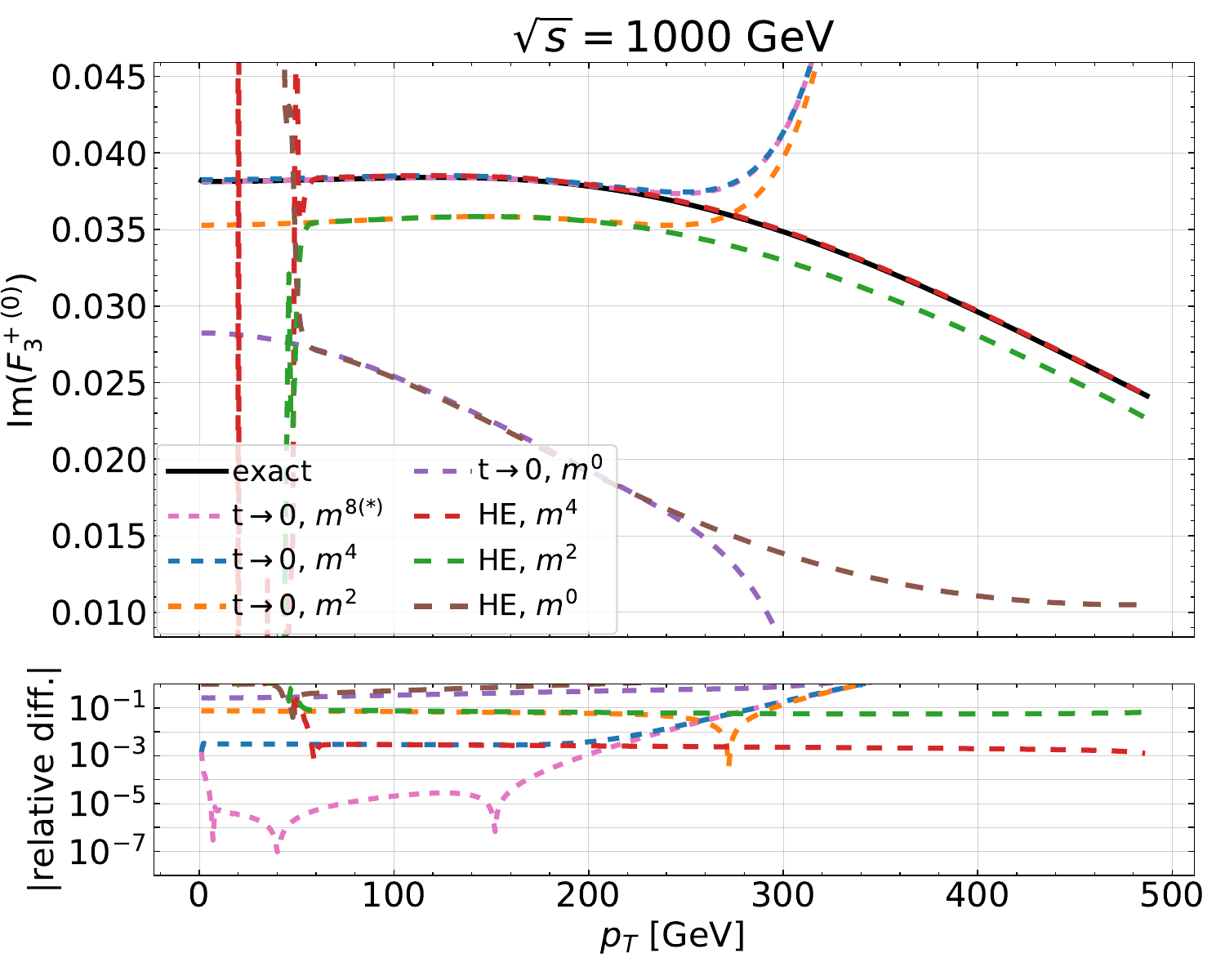}
\end{subfigure}
\caption{
	\label{fig:zhlocomparison}
    Real and imaginary parts of $F_3^{+(0)}$ as a function of $p_T$ for $\sqrt{s}=\{400,1000\}$~GeV. High-energy and $t\to 0$ expansions are shown including mass corrections up to $m_{Z,H}^{\{0,2,4\}}$. Also shown are higher mass corrections up to $m_{Z,H}^8$ for the $t\to 0$ expansion with fewer expansion terms in $t$ according to $t^{n_t}(m_Z^2)^{n_3}(m_H^2)^{n_4}$ with $n_t+n_3+n_4\leq 4$ denoted by $\star$. The lower panels display the relative difference with respect to the exact result.    
     }
\end{figure}

\begin{figure}[t]
\begin{subfigure}{0.5\textwidth}
\centering    
\includegraphics[width=1\textwidth,keepaspectratio]{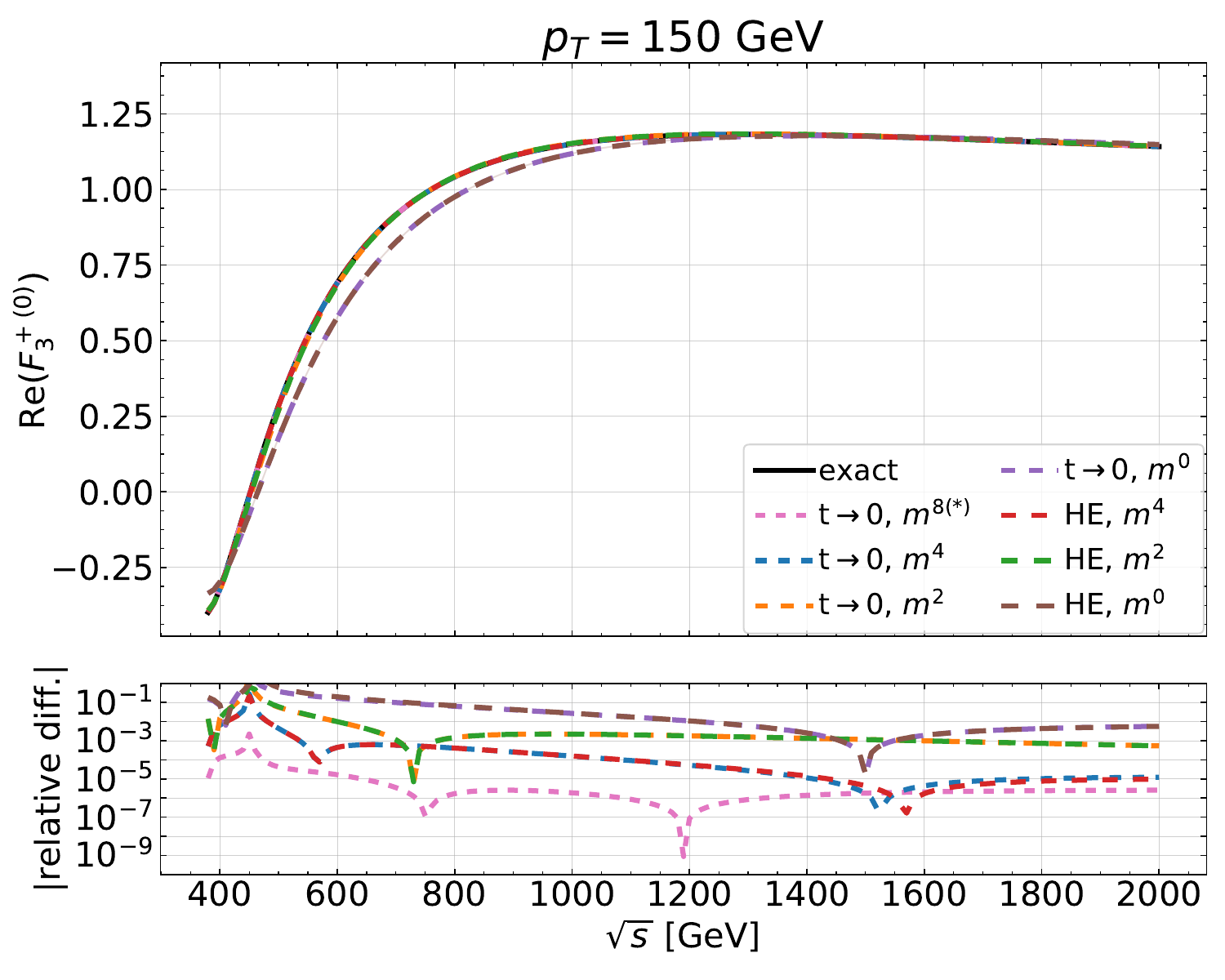}
\end{subfigure}
\begin{subfigure}{0.5\textwidth}
    \centering
    \includegraphics[width=1\textwidth,keepaspectratio]{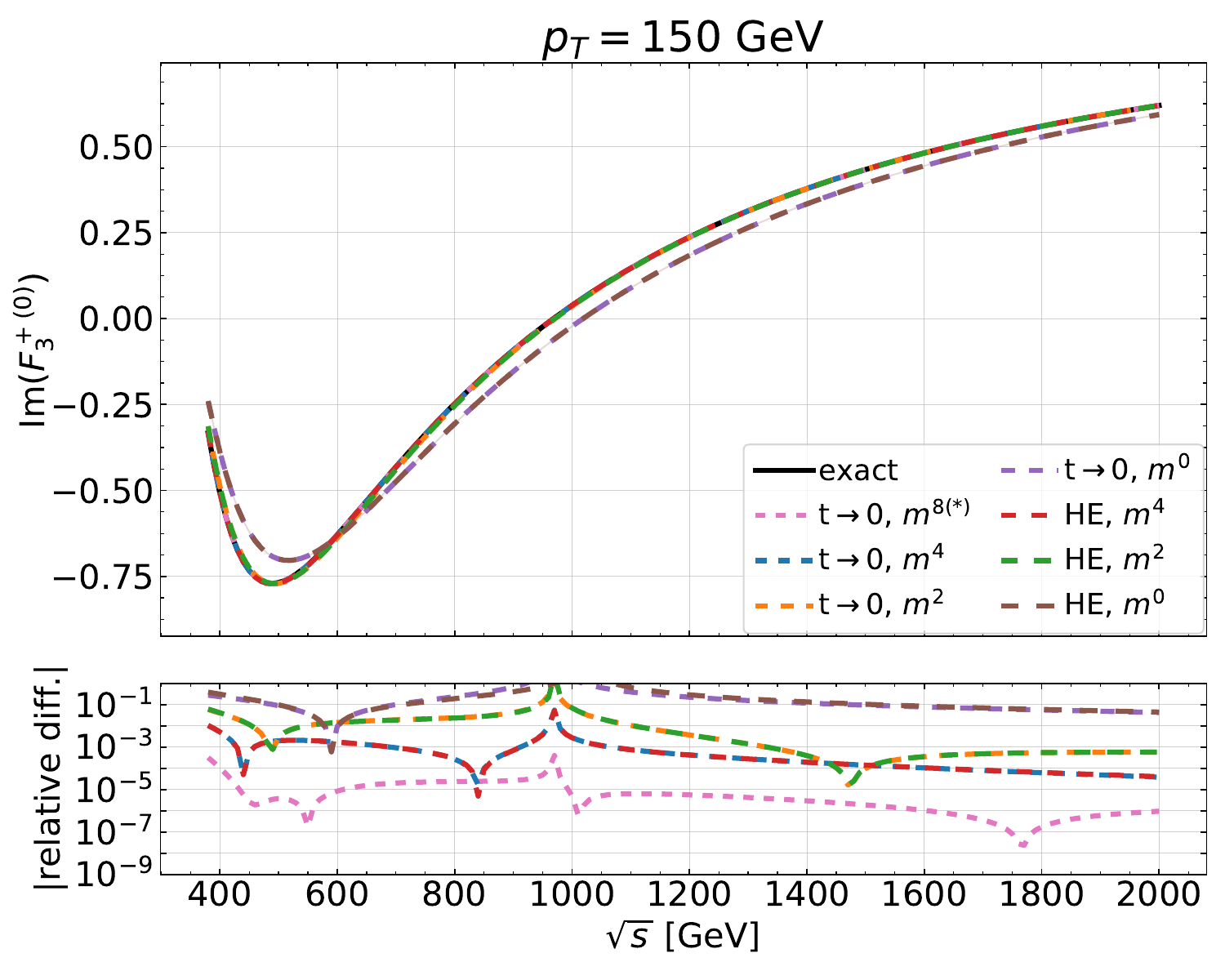}
\end{subfigure}
\caption{
    \label{fig:zhlocomparison_fixpT}
    Same as Figure \ref{fig:zhlocomparison} but for fixed $p_T=150$~GeV as a function of $\sqrt{s}$.
    }
\end{figure}

In Figs.~\ref{fig:zhlocomparison} and~\ref{fig:zhlocomparison_fixpT}
we compare our approximations to the exact one-loop results for
the real and imaginary parts of the form factor $F_3^+$.  
Similar results are also obtained for all other form factors.
The exact one-loop results have been calculated with the \texttt{calc} setup without 
applying any expansions and are expressed in terms of scalar one-loop functions. 
We use \texttt{LoopTools}~\cite{Hahn:1998yk} for their numerical evaluation.

Both in the forward-limit and high-energy approximation we include up to (at least)
quartic expansion terms in the final-state Higgs and $Z$ boson masses.  Furthermore we
incorporate expansions up to order $t^5$ and $m_t^{112}$,
respectively. The high-energy expansion is supplemented by Pad\'e
approximations following the procedure outlined in
Ref.~\cite{Davies:2020lpf}. For the construction of the
Pad\'e approximants we use expansions in $m_t$ between
$m_t^{98}$ and $m_t^{112}$.
It provides a central value and an
uncertainty for each phase-space point. In the plots, the uncertainty is only visible for low
energies; here the $t$ expansion provides precise predictions.

In each plot we show the form factors in the upper panel and in the
lower panel the absolute value of the relative difference to the exact result. Both for
fixed-$\sqrt{s}$ (cf.~Fig.~\ref{fig:zhlocomparison}) and fixed-$p_T$
(cf.~Fig.~\ref{fig:zhlocomparison_fixpT}) we observe an agreement
with the exact results for the form factor at 0.1\% or better.  The residual
difference mainly comes from the limited expansion depth in $m_H$ and $m_Z$; the
agreement between the high-energy and $t$ expansion is far below the
per mille level in the intermediate regions of
$p_T$ and $\sqrt{s}$, respectively. We remark that the relative difference
increases by about a factor 10 in case we drop the quartic terms and only
keep the quadratic terms in $m_H$ and $m_Z$.

From Figs.~\ref{fig:zhlocomparison} and \ref{fig:zhlocomparison_fixpT}
it is clear that one can simply switch between the two approximations
at some intermediate value for $p_T$ or $\sqrt{s}$.
In practice we will switch for $p_T=150$~GeV where we find 
a six-digit agreement between the $t$-expansion and high-energy results. Note that for $p_T=150$~GeV 
the minimal value for  $\sqrt{s}$
is sufficiently high such that no additional
criterion for $\sqrt{s}$ is needed.

\subsection{Two-loop results for $gg\to ZH$}

\begin{figure}[t]

\begin{subfigure}{0.5\textwidth}
\centering    
\includegraphics[width=1\textwidth,keepaspectratio]{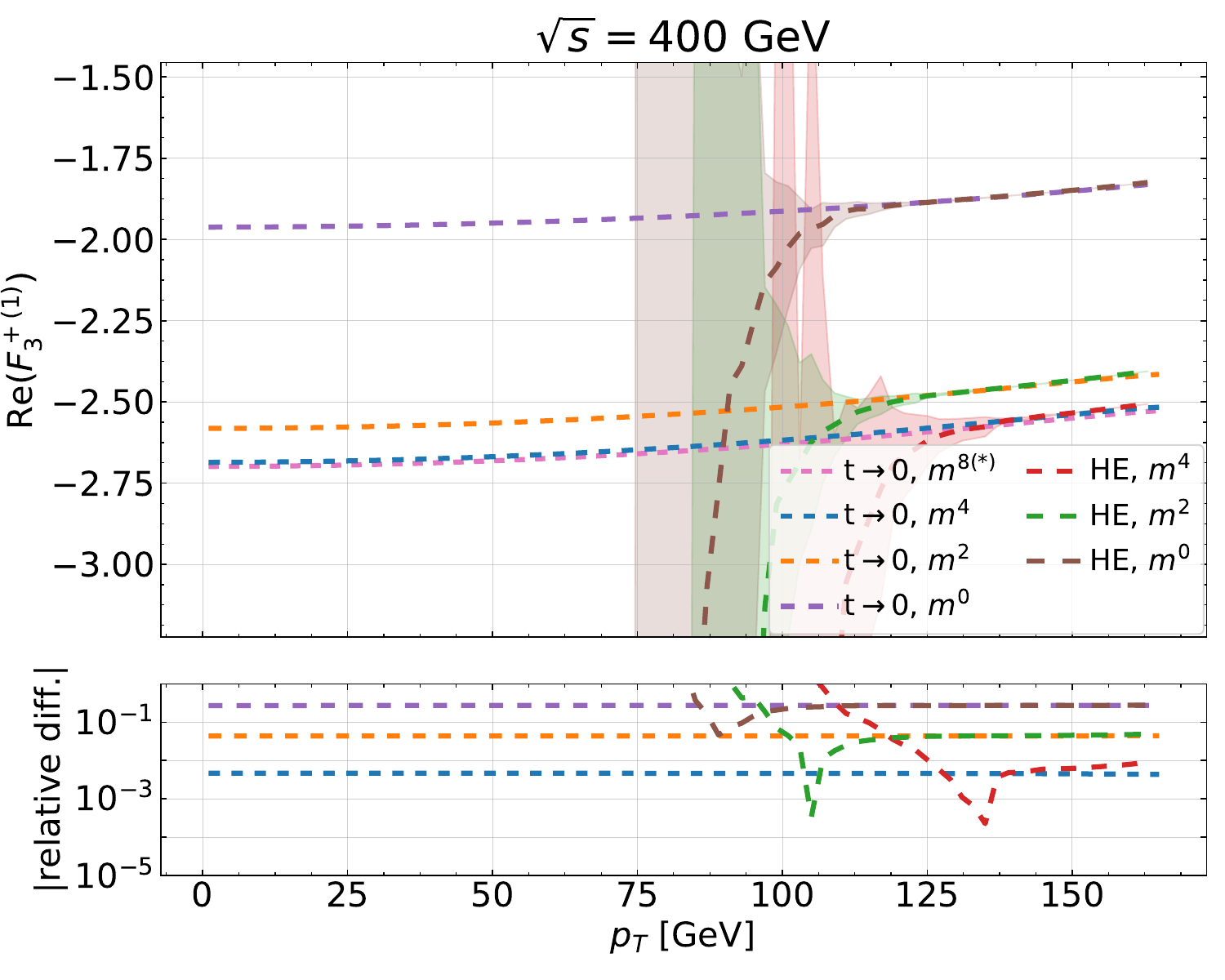}
\end{subfigure}
\begin{subfigure}{0.5\textwidth}
    \centering
    \includegraphics[width=1\textwidth,keepaspectratio]{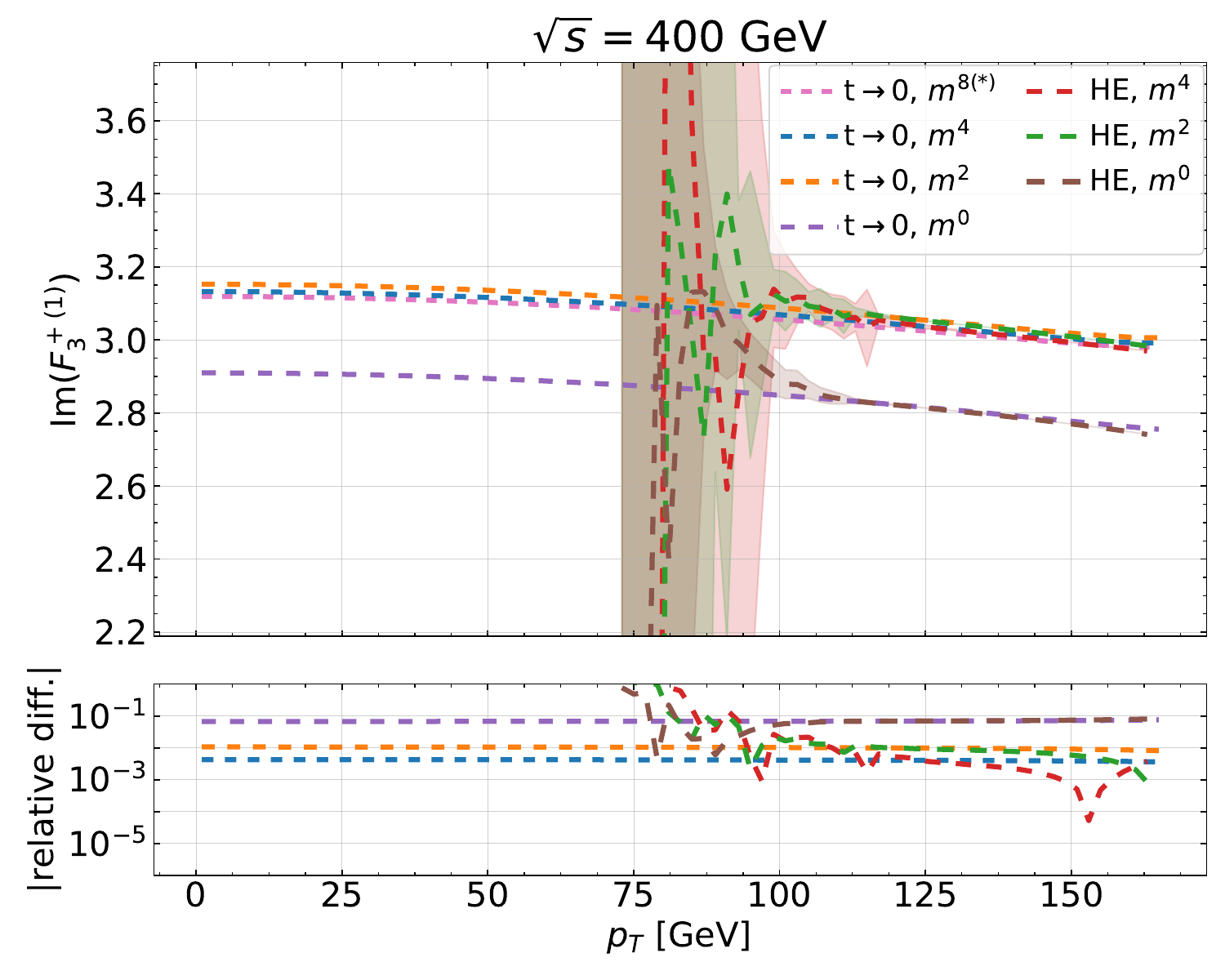}
\end{subfigure}

\begin{subfigure}{0.5\textwidth}
\centering    
\includegraphics[width=1\textwidth,keepaspectratio]{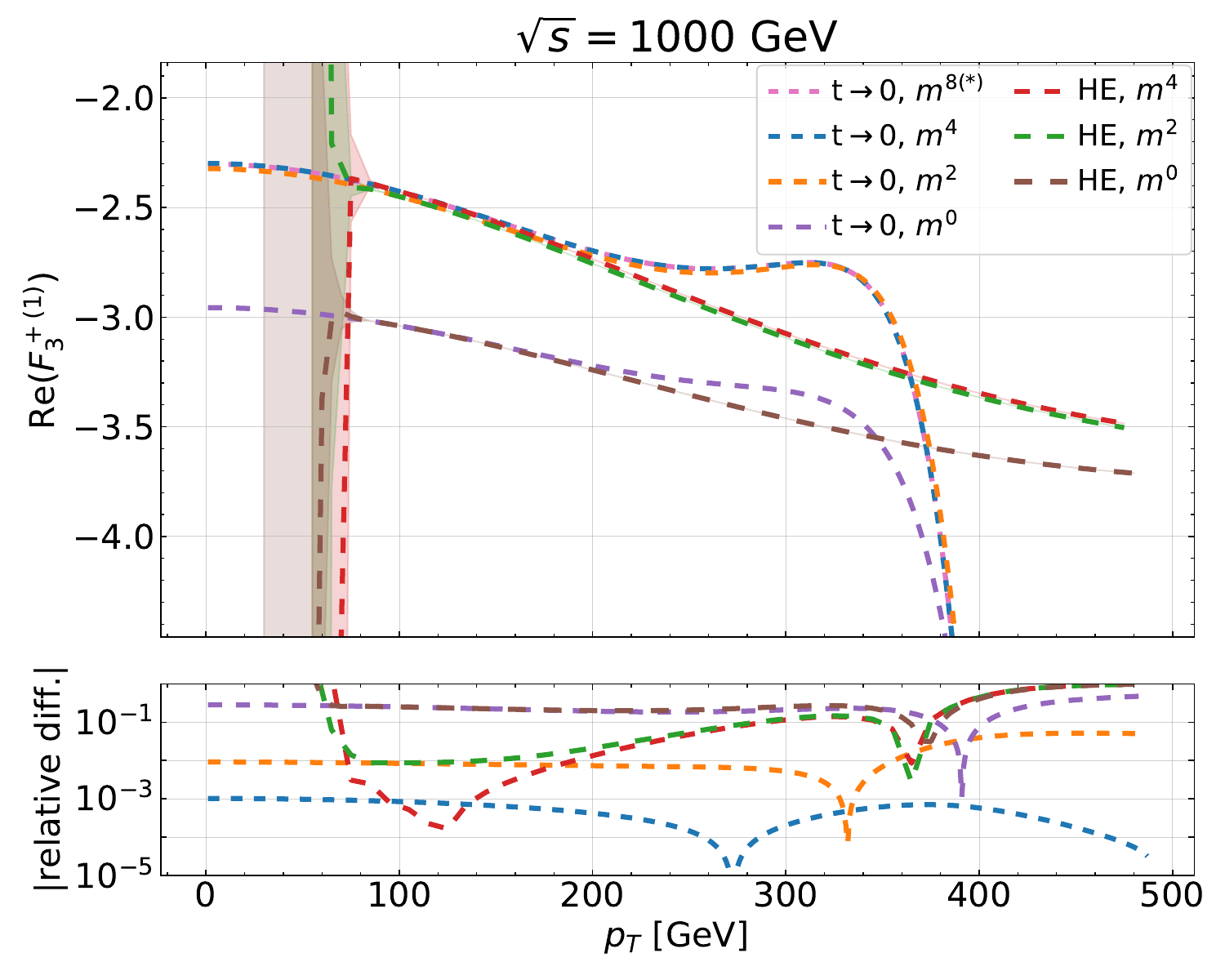}
\end{subfigure}
\begin{subfigure}{0.5\textwidth}
    \centering
    \includegraphics[width=1\textwidth,keepaspectratio]{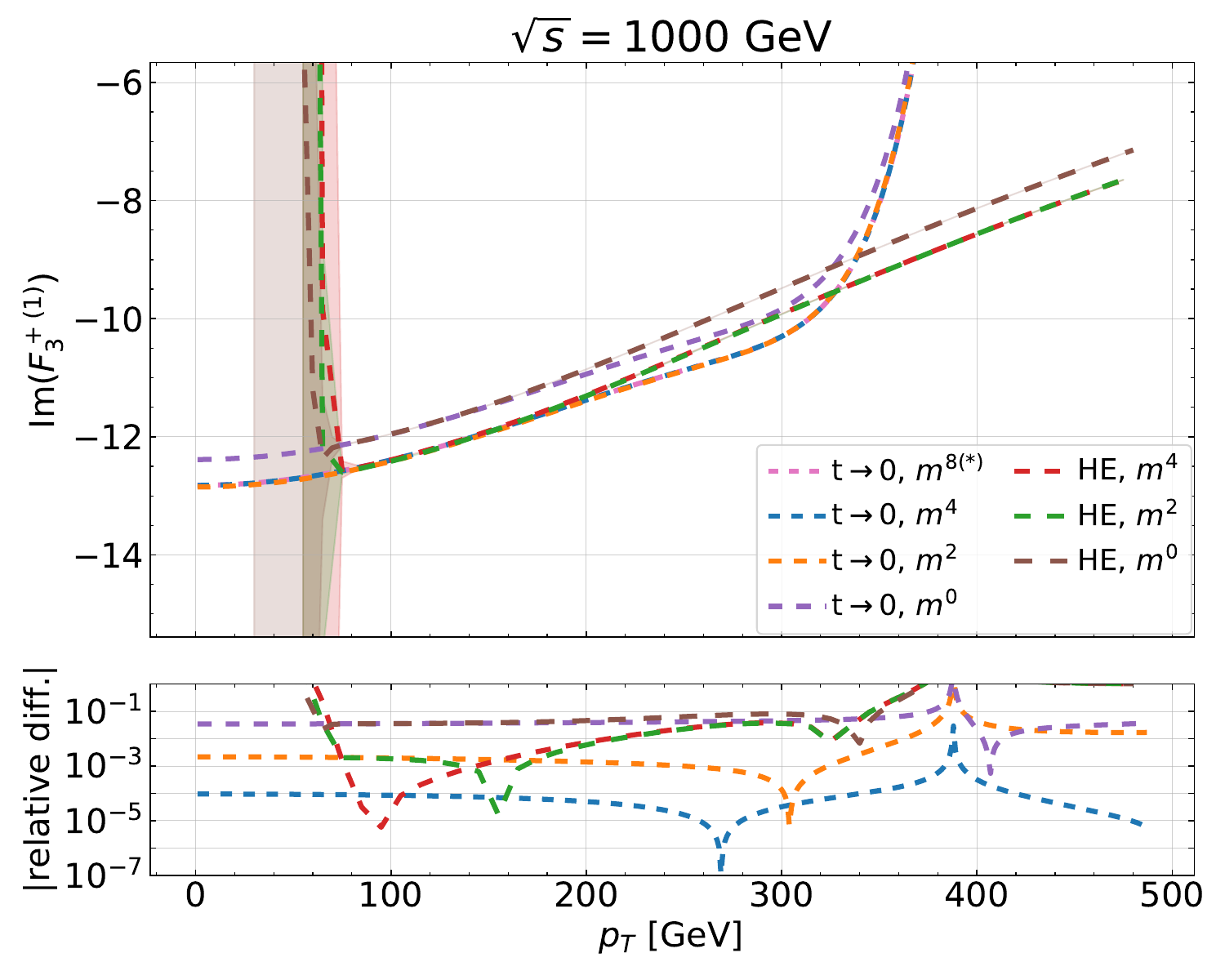}
\end{subfigure}

\caption{\label{fig::zh_nlo_fixsqrts}
      Real and imaginary parts of $F_3^{+(1)}$ as a function of $p_T$ for $\sqrt{s}=\{400,1000\}$~GeV. High-energy and $t\to 0$ expansions are shown including mass corrections up to $m_{Z,H}^{\{0,2,4\}}$. Also shown are higher mass corrections up to $m_{Z,H}^8$ for the $t\to 0$ expansion with fewer expansion terms in $t$ according to $t^{n_t}(m_Z^2)^{n_3}(m_H^2)^{n_4}$ with $n_t+n_3+n_4\leq 4$ denoted by $\star$. The lower panels display the relative difference with respect to the best approximation of the $t\to 0$ expansion.
	}
\end{figure}

\begin{figure}[t]
\begin{subfigure}{0.5\textwidth}
\centering    
\includegraphics[width=1\textwidth,keepaspectratio]{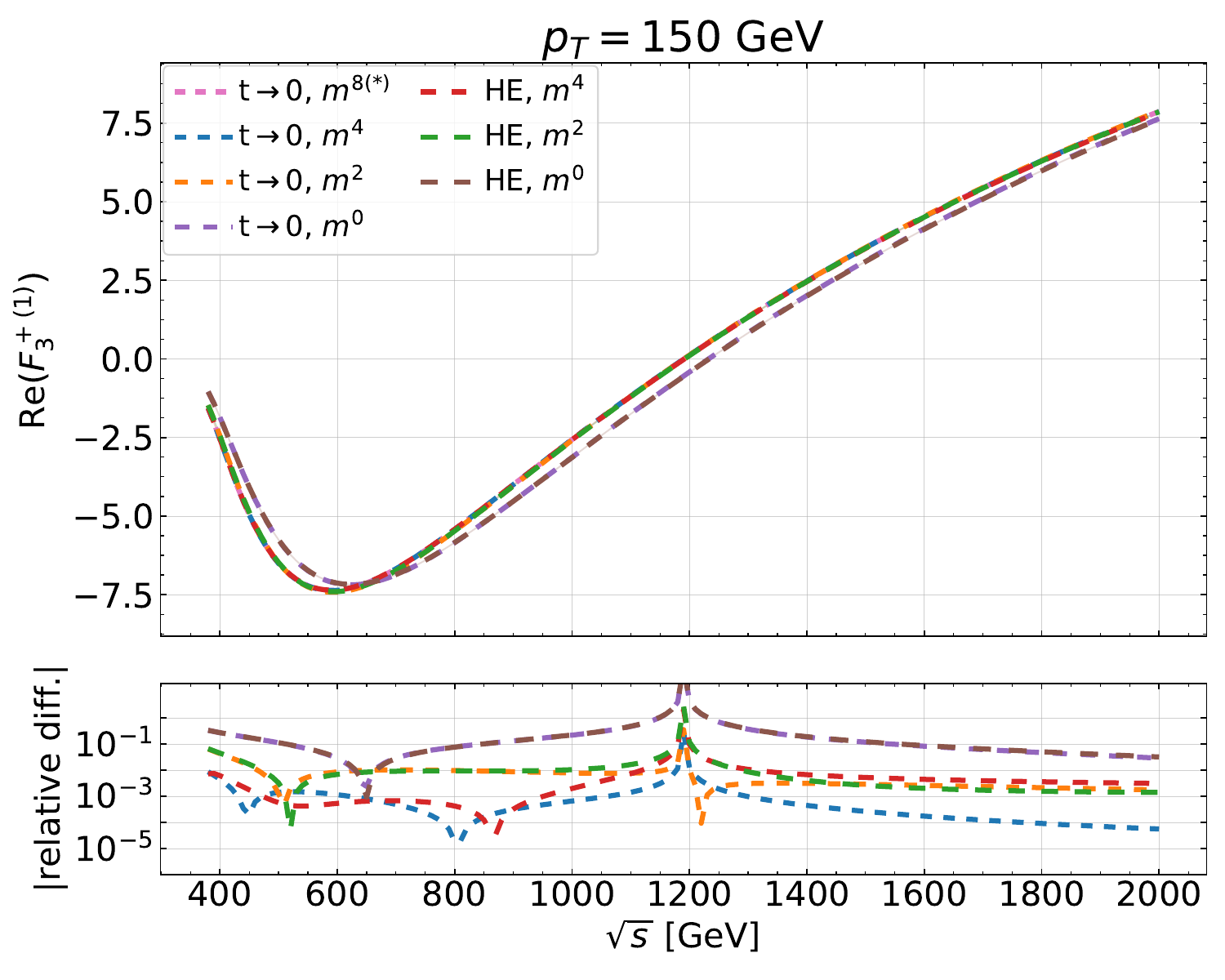}
\end{subfigure}
\begin{subfigure}{0.5\textwidth}
    \centering
    \includegraphics[width=1\textwidth,keepaspectratio]{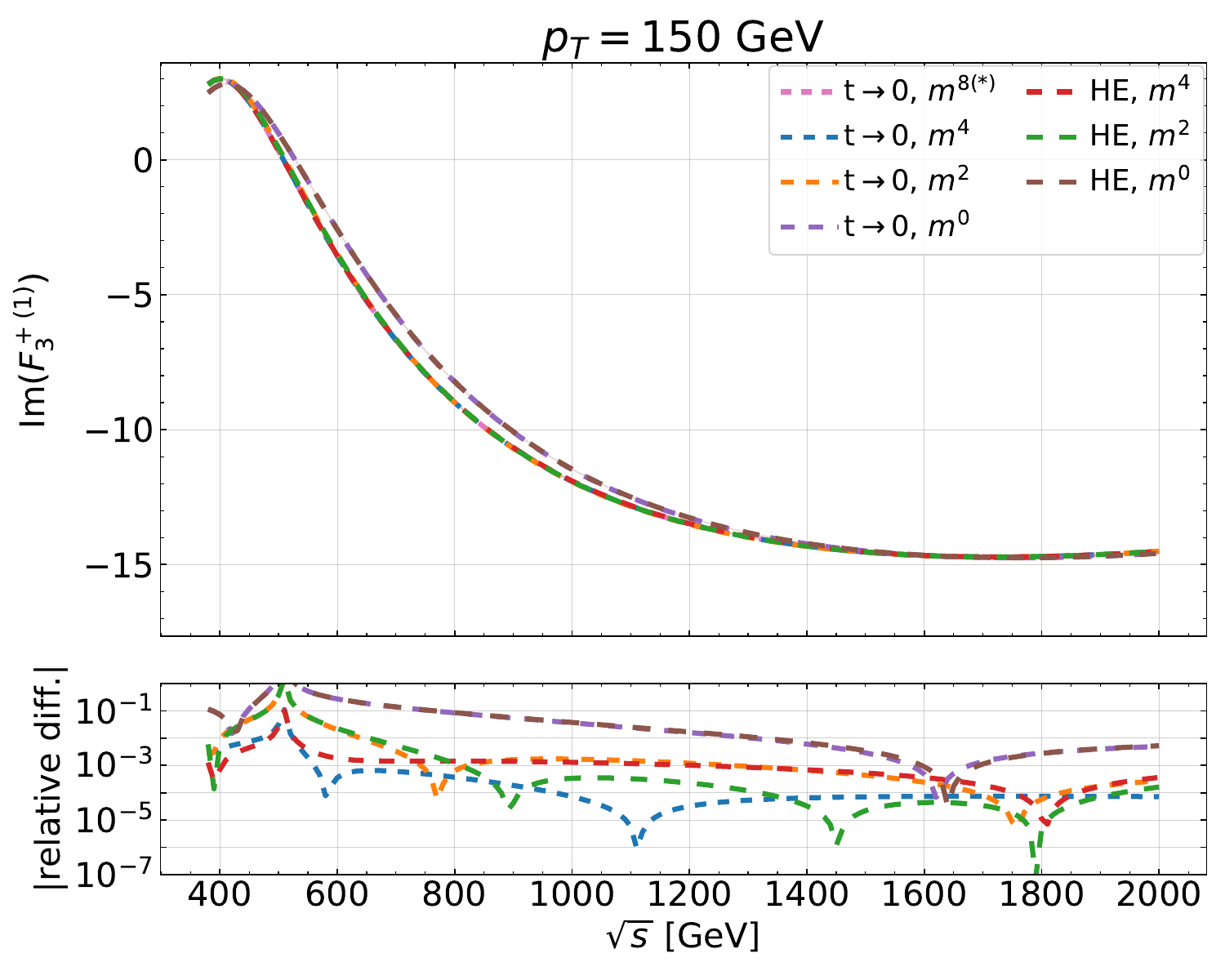}
\end{subfigure}
    \caption{\label{fig::zh_nlo_fixpT}
      Same as Figure \ref{fig::zh_nlo_fixsqrts} but for fixed $p_T=150$~GeV as a function of $\sqrt{s}$.
  }
\end{figure}

At two loops, exact results are not available to us for the form
factors. Thus, we estimate the quality of our approximation
by comparing the forward-limit and high-energy expansions
and by varying the expansion depths in $t$, $m_t$ and the external masses $m_H$ and $m_Z$. 

In Figs.~\ref{fig::zh_nlo_fixsqrts} and \ref{fig::zh_nlo_fixpT} we
show results for $F_3^{+,(1)}$ for fixed-$\sqrt{s}$ and fixed-$p_T$,
respectively, using the same parameters as in the one-loop plots in
the previous subsection. 
We use $t^5$ and $m_t^{112}$ terms in the construction
of the approximations and show results for massless final-state
particles and with quadratic and quartic terms included.
One observes a rapid convergence: the difference between
the quadratic and quartic results is below 1\%.
In Fig.~\ref{fig::zh_nlo_fixsqrts} we observe that for
$p_T\lesssim 125$~GeV the high-energy expansion does not converge.
Similarly, for $\sqrt{s}=1000$~GeV the $t$ expansion is only
valid for $p_T\lesssim 300$~GeV. However, for all
values of $\sqrt{s}$ both expansions show good agreement
for $p_T$ values around $150$~GeV, which we use for the transition
between the $t$ and high-energy expansions. This is supported by Fig.~\ref{fig::zh_nlo_fixpT} which shows an agreement at the level of $10^{-5}$. The upwards-oriented kinks originate from zero-crossings of the form factors.\footnote{The downwards-oriented kinks in the lower panels of Fig.~\ref{fig::zh_nlo_fixpT} are at $\sqrt{s}$ values where the expansions cross each other.}
Figs.~\ref{fig::zh_nlo_fixsqrts} and \ref{fig::zh_nlo_fixpT} show that we are able to cover the whole
phase-space by combining our expansions.

\subsection{Virtual NLO corrections}

\begin{figure}[t]
    \centering
    \includegraphics[width=0.98\linewidth]{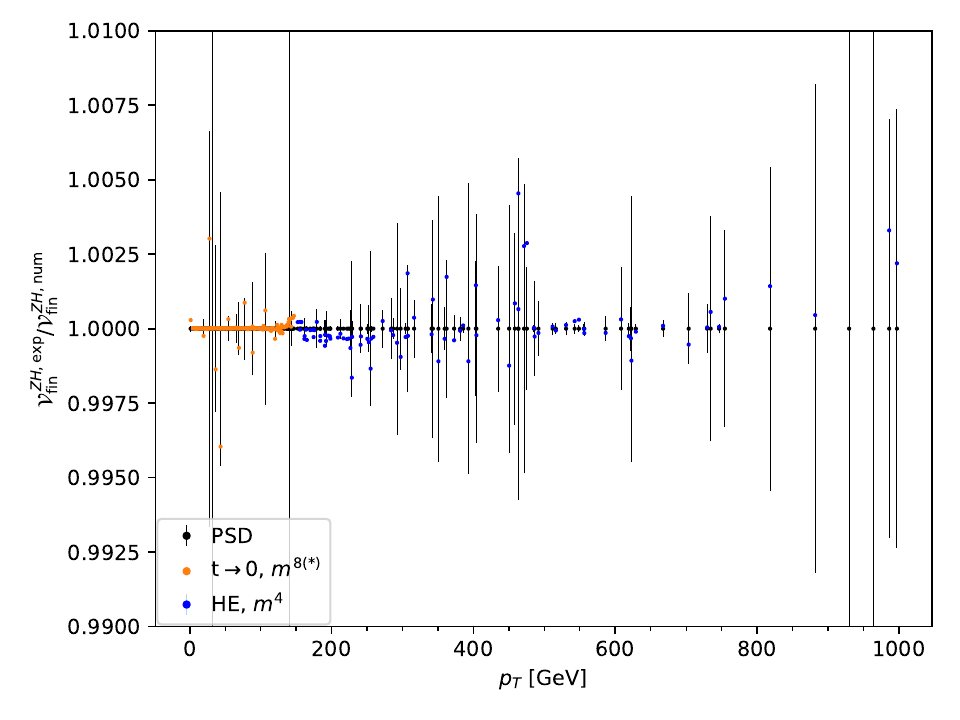}
    \caption{\label{fig::VfinZH_ratio}Ratio of ${\cal
    V}_{\rm fin}^{ZH,\rm exp}$ and ${\cal V}_{\rm fin}^{ZH, \rm num}$. In the legend ``$m^4$'' refers to the inclusion of 
      quartic terms in $m_Z$ and $m_H$ and
      in ``$m^{8\star}$'' all available mass terms are included; ``HE'' stands for
      the high-energy expansion and ``PSD'' refers to
      the results from Ref.~\cite{Chen:2020gae} based on {\tt
        pySecDec}. 
    }
\end{figure}

\begin{figure}[t]
    \centering
    \includegraphics[width=0.98\linewidth]{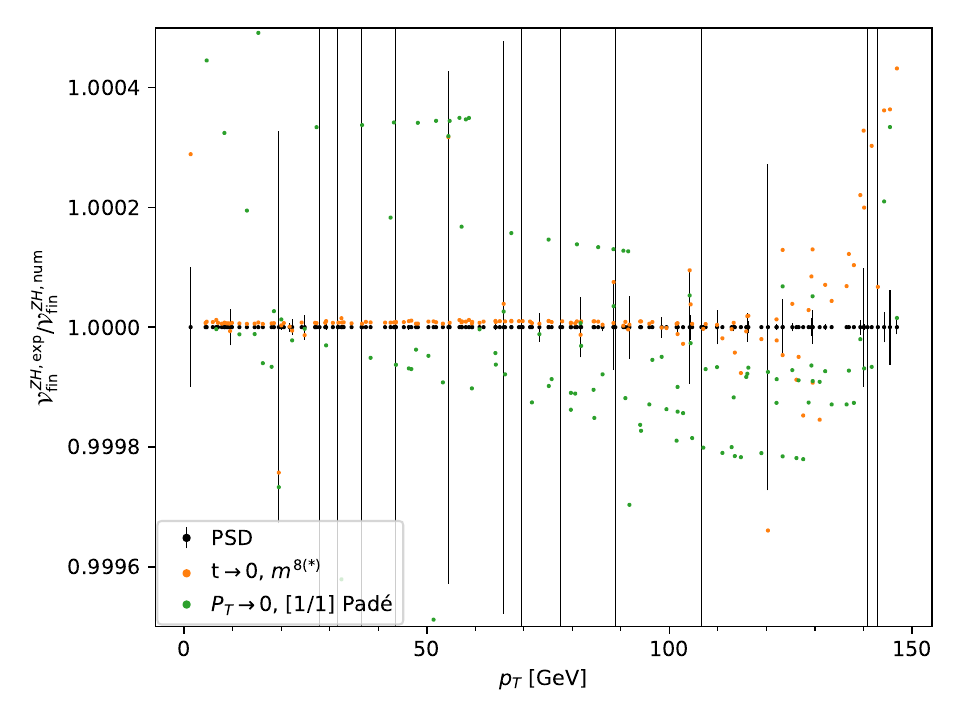}
    \caption{\label{fig::Vfin_zh_mag}
    Magnification of Fig.~\ref{fig::VfinZH_ratio} for $p_T\le 150$~GeV. Note that
    as compared to Fig.~\ref{fig::VfinZH_ratio} the $y$ axis is blown up by a factor of 20.
    We also show the results of the $p_T$ expansion
    from Ref.~\cite{Degrassi:2022mro} as green dots. Note that for some phase-space points the $p_T$ expansion results are outside the frame.}
\end{figure}

\begin{figure}[t]
    \centering
    \includegraphics[width=0.98\linewidth]{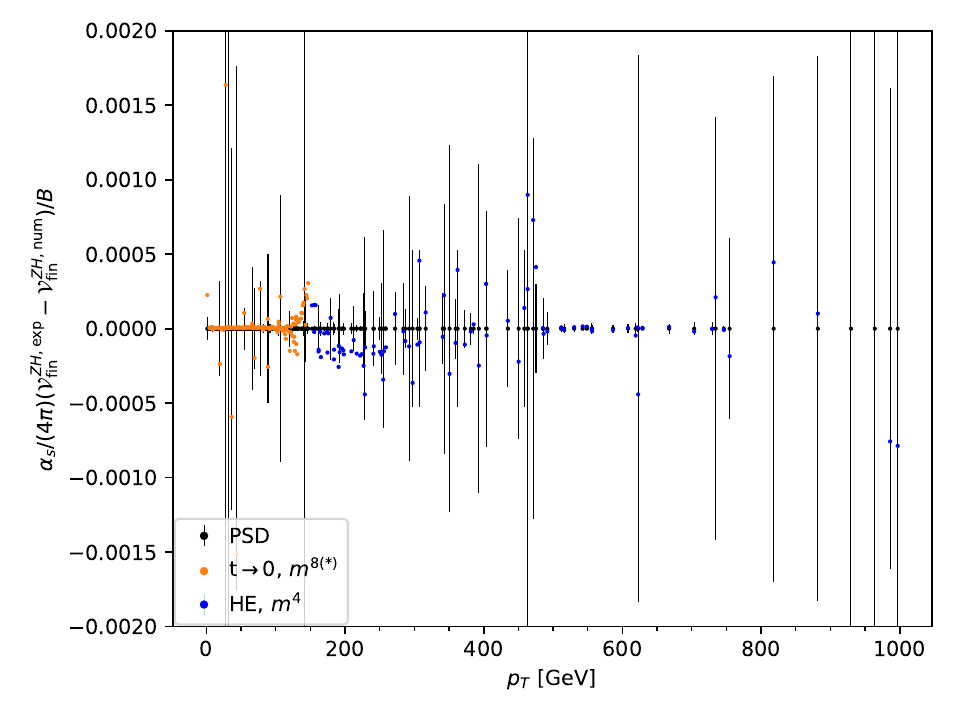}
    \caption{\label{fig::Vfin}
      Difference of ${\cal V}_{\rm fin}^{ZH}$ computed from analytic
      expansion and using a numerical approach. The quantity on the
      $y$ axis represents the uncertainty at NLO relative to the Born contribution.
      In the legend ``$m^4$'' refers to the inclusion of 
      quartic terms in $m_Z$ and $m_H$, ``HE'' stands for the
      high-energy expansion and ``PSD'' refers to
      the results from Ref.~\cite{Chen:2020gae} based on {\tt pySecDec}.
    }
\end{figure}

For convenience we repeat the formula derived in Ref.~\cite{Davies:2020drs}
for the finite virtual corrections as constructed from the form factors.
It is given by\footnote{As compared to Ref.~\cite{Davies:2020drs} the formula
presented here is multiplied by a factor of four to match the conventions of Ref.~\cite{Chen:2020gae}.}
\begin{eqnarray}
  \tilde{{\cal V}}_{\rm fin}^{ZH}
  \!\!&\!=\!&\!\!\frac{G_F^2 m_Z^2}{4s^2} \left(\frac{\alpha_s}{\pi}\right)^2\!
      \sum_{\lambda_1,\lambda_2,\lambda_3}
      \Bigg\{
      \left[ \tilde{A}_{\rm sub}^{\mu\nu\rho} \tilde{A}_{\rm
      sub}^{\star,\mu^\prime\nu^\prime\rho^\prime} \right]^{(1)}
      + \frac{C_A}{2}\left(\pi^2 - \log^2{\frac{\mu_r^2}{s}}\right) \left[ \tilde{A}_{\rm sub}^{\mu\nu\rho} \tilde{A}_{\rm
      sub}^{\star,\mu^\prime\nu^\prime\rho^\prime} \right]^{(0)} \Bigg\} 
      \nonumber\\&&\mbox{}\qquad \times 
      \varepsilon_{\lambda_1,\mu}(q_1)\:
      \varepsilon^\star_{\lambda_1,\mu^\prime}(q_1)\:
      \varepsilon_{\lambda_2,\nu}(q_2)\:
      \varepsilon^\star_{\lambda_2,\nu^\prime}(q_2)\:
      \varepsilon_{\lambda_3,\rho}(q_3)\:
      \varepsilon^\star_{\lambda_3,\rho^\prime}(q_3)\,,
      \label{eq::Vfin_ZH}
\end{eqnarray}
where the form factors entering the amplitude $\tilde{A}_{\rm
  sub}^{\mu\nu\rho}$ are the infrared-subtracted finite form factors
constructed according to our prescription outlined in
Section~\ref{sub::UVIR}, evaluated at $\mu^2=-s$.  The superscripts ``(0)'' and ``(1)'' after
the square brackets in Eq.~(\ref{eq::Vfin_ZH}) indicate that we take
the coefficients of $(\alpha_s/\pi)^0$ and $(\alpha_s/\pi)^1$,
respectively, of the squared amplitude.  For the following discussion
it is convenient to introduce the $\alpha_s$-independent quantity
\begin{eqnarray}
  {\cal V}_{\rm fin}^{ZH} &=& \frac{\tilde{{\cal V}}_{\rm fin}^{ZH}}{\alpha_s^2}
                         \,.
                         \label{eq::Vfin_norm-ZH}
\end{eqnarray}

${\cal V}_{\rm fin}^{ZH}$ has been computed in the high-energy limit in
Ref.~\cite{Davies:2020drs} and using a numerical approach in
Ref.~\cite{Chen:2020gae}. A first comparison is available in Fig.~2 of
Ref.~\cite{Chen:2022rua}.  We are now in a position to update this plot
in two ways. First, we considerably extend the depth
of the high-energy expansion from 32 to over 100 terms in
$m_t$. Additionally, we also use the $t$ expansion for the form factors
as discussed in the previous subsections. 
For the construction of ${\cal V}_{\rm fin}^{ZH}$ we use the exact one loop amplitude and all available expansion terms of our analytical approximations for the NLO contribution to the form factors.

In Fig.~\ref{fig::VfinZH_ratio} we compare our predictions for ${\cal V}_{\rm
  fin}^{ZH}$ with $\mu_r^2=s$ based on the $t$ and high-energy expansions (shown as orange and
blue dots) with results obtained in Ref.~\cite{Chen:2020gae} with a
purely numerical approach (black dots and uncertainty bars) which is
based on {\tt pySecDec}~\cite{Borowka:2017idc,Borowka:2018goh,Heinrich:2021dbf}. 
We observe that for most of the data points
our analytic expansion approximates the numerical result
far below the percent level. In fact, larger deviations
are only observed in those cases where the numerical
uncertainty is large.

To compare in more detail, in Fig.~\ref{fig::Vfin_zh_mag} we show
a magnification of Fig.~\ref{fig::VfinZH_ratio} for $p_T\le 150$~GeV.
Note that the $y$ axis has been re-scaled by a factor
of 20. 
The results of the Pad\'e-improved $p_T$ expansion
from Ref.~\cite{Degrassi:2022mro} are shown as green dots.
They are less precise; in fact, some of the data
points lie outside of the plot area.
Note, however, that for phenomenological applications the approximation provided
by these points is sufficiently precise.

Finally, we are in the position to present Fig.~\ref{fig::Vfin}, which is an update of
Fig.~2 of Ref.~\cite{Chen:2022rua}
  where the numerical results were
compared to the high-energy results of Ref.~\cite{Davies:2020drs}.
The comparison included quartic terms in the final-state masses $m_Z$ and
$m_H$.  However instead of $112$, only $32$ expansion terms
in $m_t$ were included. Furthermore, at the time Fig.~2 of
Ref.~\cite{Chen:2022rua} was made no $t$ expansion was available.
In Fig.~\ref{fig::Vfin} we use an expansion up to $t^5$.

On the $y$ axis of Fig.~\ref{fig::Vfin} we show the quantity
\begin{eqnarray}
    \frac{\alpha_s}{4\pi} 
    \frac{{\cal
    V}_{\rm fin}^{ZH,\rm exp} - {\cal V}_{\rm fin}^{ZH, \rm num}}{{\cal B}} 
    \label{eq::DeltaVin_ZH}
\end{eqnarray}
which quantifies the difference in terms of a typical NLO contribution.
Furthermore, it is
independent of the infrared subtraction scheme used for removing the infrared
singularities.  Here, ${\cal B}$ is the exact Born contribution and ${\cal
  V}_{\rm fin}^{ZH,\rm exp}$ and ${\cal V}_{\rm fin}^{ZH,\rm num}$ stand for ${\cal V}_{\rm
  fin}^{ZH}$ obtained with the help of analytic approximations and a
numerical approach, respectively. 
Over the whole $p_T$ range we observe good agreement with
the numerical results.
Fig.~\ref{fig::Vfin} represents an improvement of the results shown in Fig.~2 of Ref.~\cite{Chen:2022rua} and Fig.~1 of Ref.~\cite{CampilloAveleira:2025rbh}.

In practical applications our expansions have the advantage that
they can be evaluated very quickly. We will soon implement them in \texttt{ggxy}~\cite{Davies:2025qjr}
and expect an evaluation time of a few milli-seconds per
phase-space point.

\section{\label{sec::concl}Conclusions}

In this paper, deep expansions 
around the forward limit and the 
high-energy limit have been computed 
for the processes $gg\to ZH$ and the top quark contribution to $gg\to Z^\star Z^\star$.
We have shown that the combination of both expansions covers the whole phase space with very good accuracy. 
Our analytic results retain
explicit dependence on all parameters and it is straightforward to change them and
to change the renormalization scheme.
This makes them very suitable for the study of off-shell production, 
since the invariant mass of the final state particles can be changed 
on-the-fly and no grid has to be computed.
Furthermore, their numerical evaluation is very fast. 
For these reasons, they provide a convenient and practically equivalent alternative to the results obtained with purely numerical methods. In fact, no results for $gg \to Z^\star Z^\star$ have been calculated in the literature so far, not even with a numerical approach.
As a by-product to the calculation
of the top-quark contribution to $gg\to Z^\star Z^\star$ we obtain the corresponding results
for $gg\to\gamma^\star \gamma^\star $ and $gg\to Z^\star \gamma^\star$.

As compared to  previously
known 
expansion based results for on-shell $gg \to ZZ$~\cite{Degrassi:2024fye} and $gg \to ZH$~\cite{Degrassi:2022mro} 
our expansions are much deeper and thus
more precise.
We compare the virtual NLO corrections to numerical results~\cite{Chen:2020gae,Agarwal:2020dye,Agarwal:2024pod}
and observe agreement
below the per mille level, and for small values of $t$ even better. 
We will implement the results obtained in this paper into the \texttt{C++} library
\texttt{ggxy}~\cite{Davies:2025qjr}. This allows for a
convenient and flexible way to evaluate
the form factors, helicity 
amplitudes and virtual NLO corrections.

\section*{Acknowledgements}  

This research was supported by the Deutsche Forschungsgemeinschaft (DFG,
German Research Foundation) under grant 396021762 --- TRR 257 ``Particle
Physics Phenomenology after the Higgs Discovery''. 
The work of K.~S.~was supported by the European Research Council (ERC)
under the European Union’s Horizon 2020 research and innovation programme
grant agreement 101019620 (ERC Advanced Grant TOPUP) and the UZH Postdoc Grant,
grant no.~[FK-24-115].
The work of J.~D.~was supported by STFC Consolidated Grant ST/X000699/1.
We thank Stephen Jones and Matthias Kerner for providing numerical results for the virtual-finite
cross sections.

\begin{appendix}

\section{\label{app::proj}Projectors for $gg \to Z^\star Z^\star$}

For the helicity amplitude $\mathcal{A}_{++00}$ the coefficients
$a^{(i)}_{\lambda_1\lambda_2\lambda_3\lambda_4}$
in Eq.~(\ref{eq::polproj}) are given by
\begin{align*}
a^{(1)}_{++00} &=
a^{(2)}_{++00} =
a^{(3)}_{++00} =
a^{(8)}_{++00} =
a^{(9)}_{++00} = 
a^{(10)}_{++00} = 
0 ~, \\
a^{(11)}_{++00} &= 
a^{(12)}_{++00} = 
a^{(13)}_{++00} = 
a^{(14)}_{++00} = 
a^{(15)}_{++00} = 
a^{(16)}_{++00} 
= 0 ~,\\
a^{(4)}_{++00} &= 
a^{(5)}_{++00}  =
a^{(6)}_{++00}  =
a^{(7)}_{++00}  =
\frac{2 q_3^2 q_4^2}{\beta ^2 (d-3) s^3 \sqrt{q_3^2} \sqrt{q_4^2}} ~,\\
a^{(17)}_{++00} &= 
a^{(18)}_{++00}  =
a^{(19)}_{++00}  =
a^{(20)}_{++00}  =
\frac{2 (d-4) q_3^2 q_4^2}{\beta ^2 (d-3) s \sqrt{q_3^2} \sqrt{q_4^2} \left(q_3^2 \left(q_4^2-t\right)+t \left(-q_4^2+s+t\right)\right)} ~,
\end{align*}
with $d=4-2\epsilon$.
The corresponding coefficients for all helicity amplitudes can be
found in Ref.~\cite{progdata}.

\end{appendix}

\bibliographystyle{jhep}
\bibliography{inspire.bib,extra.bib}

%

\end{document}